\documentclass[3p,12pt]{elsarticle}

\pdfoutput=1
\usepackage{indentfirst}
\usepackage{color,graphicx}
\usepackage{subfigure}
\usepackage{threeparttable}
\usepackage{array}
\usepackage{multirow}
\usepackage{amsmath}
\usepackage{amssymb}
\usepackage{bm}
\usepackage{hyperref}
\usepackage{lineno}

\begin{document}


\title{An efficient topology optimization method for steady gas flows in all flow regimes}

\author[sustech]{Ruifeng Yuan}

\author[sustech,GHM]{Lei Wu~\corref{cor1}}
\cortext[cor1]{yuanrf@sustech.edu.cn (Ruifeng Yuan)\\
	Corresponding author: wul@sustech.edu.cn (Lei Wu). }

\address[sustech]{Department of Mechanics and Aerospace Engineering, Southern University of Science and Technology, Shenzhen 518055, China }

\address[GHM]{Guangdong-Hong Kong-Macao Joint Laboratory for Data-Driven Fluid Mechanics and Engineering Applications, Southern University of Science and Technology, Shenzhen 518055, China.}


\begin{abstract}
	
An efficient topology optimization method applicable to both continuum and rarefied gas flows is proposed in the framework of gas-kinetic theory. 
The areas of gas and solid are marked by the material density, based on which a fictitious porosity model is used to reflect the effect of the solid on the gas and mimic the diffuse boundary condition on the gas-solid interface.
The formula of this fictitious porosity model is modified to make the model work well in all flow regimes, i.e., from the continuum to free-molecular flow regimes.
To find the optimized material density, a gradient-based optimizer is adopted and the design sensitivity is obtained by the continuous adjoint method. 
To solve the primal kinetic equation and the corresponding adjoint equation, the numerical schemes efficient and accurate in all flow regimes are constructed.
Several airfoil optimization problems are solved to demonstrate the good performance and high efficiency of the present topology optimization method, covering the flow conditions from continuum to rarefied, and from subsonic to supersonic.
\end{abstract}

\begin{keyword}
	topology optimization, adjoint method, rarefied gas flow, multiscale flow, discrete velocity method
\end{keyword}

\maketitle



\section{Introduction}\label{sec:intro}

In modern industry, many problems involve the multiscale gas flows which include the continuum (described by the traditional Navier-Stokes (NS) equation) and the rarefied (described by the gas-kinetic equation) flows at different positions. These problems appear in a wide range of fields such as aerospace, micro-electromechanical system, precision manufacturing, and even nuclear industry. For instances, the near-space/re-entry vehicles \cite{reed2010investigation,li2021kinetic} which travel between the low- and high-altitude atmospheres can experience the continuum flow and rarefied flow during flight; the micro components and devices used in the vacuum environment, such as the molecular pump \cite{hablanian1997high,sharipov2005numerical},  lithography \cite{bakshi2009euv} and nuclear fusion \cite{tantos2020deterministic}, often have gas flows undergoing the transition from the continuum to rarefied flow regimes \cite{tsien1946superaerodynamics}. Obviously, the aerodynamic geometries of these vehicles and devices greatly influence their performance, which should be carefully designed. To this end, the optimization method applicable to gas flows in all flow regimes is highly demanded. For gas dynamics, this can only be considered by the gas-kinetic equations that work in all flow regimes~\cite{Cercignanibook1988,Lei2022}.

Among various optimization methods, in the present work we consider the method of topology optimization, which is initially developed to optimize solid structures. As the famous French architect Robert le Ricolais said, ``the art of structure is where to put the holes'', the topology optimization can be viewed as a method to design the number, location and shape of holes on a structure to optimize the structural performance under certain constraints. Typically, for  classic method topology optimization \cite{bendsoe2003topology}, a material density field is introduced to distinguish the solid and void regions; then the derivative of the objective function with respect to the material density is calculated by the adjoint method; finally the material density distribution is optimized by the gradient-based optimization algorithm. Due to the vast design space and no reliance on intuition or experience, the topology optimization method is able to create novel designs of high performance, and thus is well developed and widely applied in the structural optimization \cite{rozvany2009critical,huang2010evolutionary,sigmund2013topology}.

Borrvall and Petersson first introduced the topology optimization into the design problems in fluid mechanics~\cite{borrvall2003topology}. They established the theory from the problem of the Stokes flow between two surfaces with a short distance, and obtained the model in which the fluid acts as a porous medium; the design variable, which can be physically interpreted as the distance between surfaces, is related to the permeability of the porous medium, and the solid region is represented by the area with very small permeability. This laid the basis for the density-based (or more precisely, porosity-based) fluid topology optimization. Since then, various fluid topology optimization methods based on the NS equation were proposed, such as the density-based method of Gersborg-Hansen \textit{et al.}~\cite{gersborg2005topology},
the level set approach with the embedded boundary condition of Kreissl and Maute~\cite{kreissl2012levelset}, and  the level set approach with the immersed boundary method of Kubo \textit{et al.}~\cite{kubo2021level}. Besides, the topology optimization methods for the lattice Boltzmann method (which is an alternative solver of the weakly compressible NS equation) have also been developed, such as Pingen \textit{et al.}~and Liu \textit{et al.}'s methods based on the fictitious porosity model \cite{pingen2007topology,liu2014discrete}, Kreissl \textit{et al.}'s method combining the level set approach and the interpolation bounce-back scheme~\cite{kreissl2011explicit}, N{\o}rgaard \textit{et al.}'s method based on the partial bounce-back model~\cite{norgaard2016topology}.

Apart from these methods for the optimization of continuum gas flows which are essentially governed by the Stokes and NS equations, recently there were also researches (but only a few) on the topology optimization of rarefied gas flows. Based on the Bhatnagar-Gross-Krook (BGK)~\cite{bhatnagar1954model} gas-kinetic equation which is able to describe the rarefied gas dynamics, Sato \textit{et al.}~\cite{sato2019topology} modified the fictitious porosity model and developed the topology optimization method. In their work, the BGK equation with the fictitious porosity model was solved by the conventional discrete velocity method (DVM) that adopts the advection-collision splitting. They applied their method to the 2D optimization problem of a thermally driven pump and took 230 hours on a 80-core computer to finish the computation; the optimal result, however, seemed to suffer from a large area of abnormal gray material density distributions. In addition to the deterministic DVM method, the rarefied gas flow can also be solved by the stochastic method, i.e., the prevailing direct simulation Monte Carlo (DSMC) method~\cite{Bird1994Molecular}. Therefore, Caflisch \textit{et al.}~\cite{caflisch2021adjoint} and Yang \textit{et al.}~\cite{yang2023adjoint} proposed the adjoint DSMC for the spatially-homogeneous Boltzmann equation, which makes it possible to do the sensitivity analysis in the stochastic particle method. Based on the works, Guan \textit{et al.}~\cite{guan2023topology} established the topology optimization method using the information preservation DSMC method for the rarefied flow problems. It should be noted that, due to the intrinsic unsteady property of the Monte Carlo algorithm, these adjoint DSMC methods are memory and time consuming. For example, in the two-dimensional (2D) optimization of a bent pipe by Guan \textit{et al.}, the computer memory requirement is 75 GB~\cite{guan2023topology}, and the computational complexity is proportional to such a memory cost according to their analysis, though the exact computation time cost was not published.

It is worth mentioning that, existing researches on topology optimization of gas dynamics mainly consider flow problems which are only continuum flows or only rarefied flows, and due to the different physical properties of gas flows in different flow regimes, often different theoretical models (e.g., NS or gas-kinetic equations) and numerical methods are used to handle the continuum or rarefied gas flows. For example, for continuum flows typically the methods based on the NS equation or lattice Boltzmann method are used, while for rarefied flows the conventional DVM or DSMC are used. To the best of our knowledge, so far there is no research on the unified method of fluid topology optimization which can handle both continuum and rarefied gas flows.
On the other hand, for the rarefied gas simulation, in recent years we have seen much progress in the multiscale numerical methods applicable to gas flows in all flow regimes. One category of these methods is the deterministic method based on the DVM framework, such as the (discrete) unified gas-kinetic scheme~\cite{Xu2010A,guo2013discrete,guo2015discrete}. The other category is the stochastic particle method, such as the unified gas-kinetic wave-particle method~\cite{liu2020unified,zhu2019unified} and the unified stochastic particle BGK method~\cite{fei2020unified}. Furthermore, various of acceleration algorithms for the multiscale methods have also been developed,
 such as the macroscopic-prediction implicit method~\cite{Zhu2016Implicit,yuan2021multi} and the general synthetic iterative scheme~\cite{su2020can}.
Therefore, it will be of great importance and advantage if the topology optimization methods are coupled with these efficient multiscale methods to deal with the continuum and rarefied gas flows simultaneously.

The present work is dedicated to realizing the above goal. To this end, in Section \ref{sec:formula}, we first introduce the theoretical basis of the gas-kinetic equation and the modification of the fictitious porosity model, then the formulations of the primal problem and the adjoint problem are presented. 
In Section \ref{sec:method}, the overall computation procedure of the optimization method is described, and then the efficient multiscale numerical methods to solve the primal and adjoint equations are proposed. 
In Section~\ref{sec:test}, several tests are carried out to verify the performance and efficiency of the present method.
The summary is given in Section~\ref{sec:conc}.

\section{Formulation}\label{sec:formula}

\subsection{Gas kinetic theory}\label{sec:kinetictheory}

Since the NS equation is only applicable in the continuum flow regime, the gas-kinetic theory should be used to describe the gas dynamics from the continuum to free-molecular flow regimes~\cite{Cercignanibook1988,Lei2022}. In kinetic theory, the state of the gas is described by the velocity distribution function $f$, from which the macroscopic information can be obtained by integration in the molecular velocity space, such as
\begin{equation}
\bm W = \int {\bm \psi fd\Xi }.
\end{equation}
Here, $\bm W=(\rho,\rho\bm u,\rho E)^T$ is the macroscopic conservative variables,  with $\rho$ being the mass density, $\bm u$ the flow velocity, and the energy $E=|{\bm u}|^2/2 + {RT}/({\gamma  - 1})$, where $T$ is the gas temperature, $R$ is the specific gas constant and $\gamma$ is the specific heat ratio. $\bm \psi$ is the vector of moments $\bm \psi  = {\left( {1,\bm v,\frac{1}{2}{{\bm v}^2}} \right)^T}$, $\bm v$ is the molecular velocity and $d\Xi  = dv_1dv_2dv_3$ is the molecular velocity space element. 

To describe the evolution of $f$, we consider the BGK equation \cite{bhatnagar1954model} which is a widely used Boltzmann model equation:
\begin{equation}\label{eqn:bgk_org}
\frac{{\partial f}}{{\partial t}}{\rm{ + }}\bm v \cdot \nabla f = \frac{{g - f}}{\tau },
\end{equation}
where $\tau$ is the relaxation time calculated by the dynamical viscosity coefficient $\mu$ (we also consider the gas molecules with the hard-sphere model \cite{Bird1994Molecular} so $\mu  \propto \sqrt T $) and the pressure $p=\rho RT$ by
\begin{equation}
\tau  = \frac{\mu}{p}.
\end{equation}
The equilibrium state $g$ follows the Maxwellian distribution
\begin{equation}
g =g_{\rm M}(\rho,\bm{u},T)= \rho {\left( {\frac{1}{{2\pi RT}}} \right)^{\frac{3}{2}}}{e^{ - \frac{{{{(\bm v - \bm u)}^2}}}{{2RT}}}}.
\end{equation}

Take moments of the BGK equation \eqref{eqn:bgk_org} about $\bm \psi$ will yield the transport equation for the macroscopic conserved quantities
\begin{equation}
\frac{{\partial \bm W}}{{\partial t}} + \nabla  \cdot {\bf{F}} = 0,
\end{equation}
with the flux tensor ${\bf{F}} = \int {\bm v\bm \psi fd\Xi } $ and the conservation condition $\int {\bm \psi \left( {g - f} \right)d\Xi }  = \bm{0}$.

To fully describe the rarefied gas dynamics, the gas-solid boundary condition is needed, i.e., giving the velocity distribution of the gas molecules hitting the solid wall, the velocity distribution of the gas molecules bouncing back from the wall should be determined. Here we consider the widely used diffuse boundary condition, where the bounce-back molecules follows the Maxwellian distribution:
\begin{equation}\label{eqn:formula_bgk_fdw0}
f = {g_{\rm{M}}}({\rho _{\rm{w}}},{\bm u_{\rm{w}}},T_{\rm{w}})\quad {\rm for}\quad \bm v \cdot \bm n < 0,
\end{equation}
where $\bm n$ is the normal unit vector pointing outward from the gas field; $T_{\rm{w}}$ is the prescribed wall temperature; $\bm u_{\rm{w}}$ is the prescribed wall velocity and since we don't consider the wall motion in the normal direction it satisfies ${\bm u_{\rm{w}}} \cdot \bm n = 0$; $\rho _{\rm{w}}$ is determined by the zero mass flux at the solid wall as
\begin{equation}
{\rho _{\rm{w}}} = \sqrt {\frac{{2\pi }}{{R{T_{\rm{w}}}}}} \int_{{\bm v} \cdot \bm n \ge 0} {\bm v \cdot \bm n{f}d\Xi } .
\end{equation}

An important parameter to indicate the rarefaction effect of the gas flow is the Knudsen (Kn) number, which is the ratio of the molecular mean free path to the characteristic length $l_{\rm ref}$ of the flow. In the present study we use the hard-sphere model \cite{Bird1994Molecular} to determine the mean free path and the Kn number can be calculated as
\begin{equation}\label{eqn:kndefine}
{\rm{Kn}} = \frac{{16}}{5}\frac{\tau }{{{l_{{\rm{ref}}}}}}\sqrt {\frac{{RT}}{{2\pi }}} .
\end{equation}
According to the magnitude of Kn number, gas flows are usually classified into four flow regimes \cite{tsien1946superaerodynamics} qualitatively: the continuum flow ($\rm{Kn}<0.01$), the slip flow ($0.01<\rm{Kn}<0.1$), the transition flow ($0.1<\rm{Kn}<10$) and the free-molecular flow ($\rm{Kn}>10$). In the continuum flow regime the gas can be viewed as the continuous medium, and the NS equation with the no-slip boundary condition holds true; in the slip flow regime the NS equation still applies to the bulk flow region, but the velocity slip and the temperature jump will occur on the gas-solid interface; in the transition flow regime the NS equation is no longer valid, and the gas flow should be described by the molecular transport and collision; in the free-molecular flow regime the gas flow dynamics falls into the free transport of gas molecules with negligible intermolecular collisions, and the system is very hard to describe from the macroscopic perspective. For all flow regimes the Boltzmann model equation \eqref{eqn:bgk_org} is applicable. Specially, in the continuum flow regime where Kn is very small, intensive molecular collisions drive $f$ very close to the equilibrium state $g$, and the NS equation can be recovered from the kinetic equation \eqref{eqn:bgk_org} by the Chapman-Enskog expansion \cite{chapman1990mathematical}. Moreover, the diffuse boundary condition \eqref{eqn:formula_bgk_fdw0} automatically recovers the no-slip boundary condition for the continuum flow. Nevertheless, in the continuum flow regime, because the spacial/temporal scales of the molecular collision are much smaller than the characteristic scales of the gas flow, the numerical scheme for the gas-kinetic equation  needs to be carefully constructed to guarantee the accuracy and efficiency. 

It is also worth noting that in the field of fluid mechanics, the Mach (Ma) number and the Reynolds (Re) number are two important parameters indicating the properties of the flow, which are defined as
\begin{equation}\label{eqn:maredefine}
{\rm{Ma}} = \frac{\left| \bm u \right|}{a},\;\;\;\;{\rm{Re}} = \frac{{\rho \left| \bm u \right|{l_{{\rm{ref}}}}}}{\mu },
\end{equation}
where $a=\sqrt{\gamma RT}$ is the acoustic speed. Ma indicates the compressibility of the flow, and Re is conventionally used in the continuum regime to indicate the viscous effect of the flow. According to Eqs.~\eqref{eqn:kndefine} and \eqref{eqn:maredefine}, Re is linked to Kn by
\begin{equation}
{\rm{Kn}} = \frac{{16}}{5}\sqrt {\frac{\gamma }{{2\pi }}} \frac{{{\rm{Ma}}}}{{{\mathop{\rm Re}\nolimits} }}.
\end{equation}
A larger Re number, or a smaller Kn number, usually means smaller flow structures, which should be captured by computational mesh with higher resolution. In this paper we assume the Re number is not very large so that the turbulent behavior does not occur; otherwise the turbulence model should be used for the simulation of the gas flow.

\subsection{Parametrization by the fictitious porosity model}\label{sec:formula_poro}

\begin{figure}[t]
	\centering
	\includegraphics[width=0.45\textwidth]{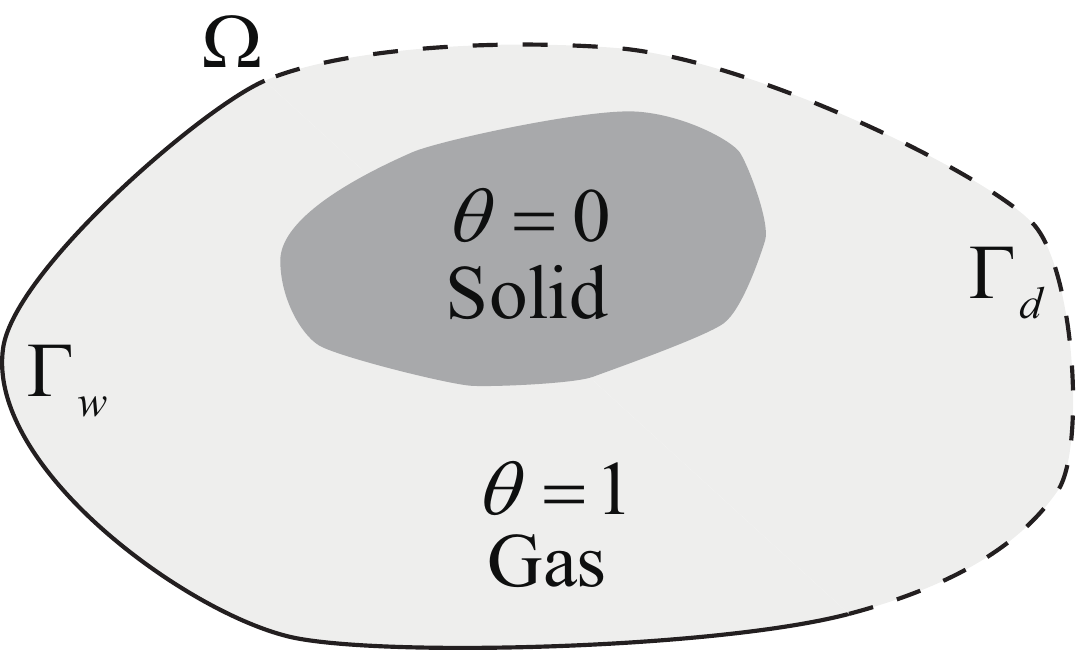}
	\caption{\label{fig:formula_domain}Illustration for the material-density-based topology optimization. $\Omega$ is the whole design domain, $\Gamma _{\rm d},\Gamma _{\rm w}$ are the domain boundaries. The gas and solid regions are represented by $\theta=1$ and 0, respectively. }
\end{figure}

A set of parameters describing the shape of solid areas to be optimized should be specified. Here, the widely-used density-based description for solid areas is adopted~\cite{bendsoe2003topology}. As shown in Fig.~\ref{fig:formula_domain}, in the design domain $\Omega$, a material density variable $\theta$ is introduced to describe the gas region ($\theta=1$) and the solid region ($\theta=0$). Realistically the material density $\theta$ should be either $1$ (gas) or $0$ (solid) and form a sharp interface between the gas and solid regions. However, in numerical treatment, it is hard to handle or maintain such a distinct interface with discontinuous distribution of $\theta$. Therefore, in the density-based method, it uses a continuous variation between the pure gas $\theta=1$ and the pure solid $\theta=0$; the gray regions with $\theta  \in (0,1)$ stand for the intermediate state between the gas and solid. To describe this continuous transformation from the gas to solid, the porous medium model is applied. For the NS equation governing the continuum gas flow, this can be accomplished by introducing a body force term similar to the term involving the permeability tensor in Darcy's law \cite{borrvall2003topology,gersborg2005topology}, and controlling the permeability by $\theta$. When $\theta=1$ the permeability is infinite and the pure fluid flow is recovered. When $\theta$ decreases, the permeability decreases accordingly and finally approaches $0$ when $\theta=0$, implying the transformation to a non-penetrating solid medium. For the gas-kinetic BGK equation~\eqref{eqn:bgk_org}, similar artificial porous force source term can be applied \cite{liu2014discrete}, but here we consider the fictitious porosity model used by Pingen \textit{et al}. in their topology optimization method for the lattice Boltzmann method~\cite{pingen2007topology,spaid1997lattice}:
\begin{equation}\label{eqn:bgk_poroP}
\frac{{\partial f}}{{\partial t}} + \bm v \cdot \nabla f = \frac{{g_{\rm M}(\rho,\bm{u}_{\theta},T) - f}}{\tau },
\end{equation}
in which the velocity $\bm{u}_{\theta}$ is calculated by the scaling
\begin{equation}\label{eqn:bgk_poroP_simp}
\bm u_\theta=\bm u_\theta ^{\rm{Pingen}} = \left[ {1 - {{\left( {1 - \theta } \right)}^q}} \right]\bm u.
\end{equation}
When $\theta=1$, i.e., in the gas region, $\bm u_\theta=\bm u$, so that the BGK equation~\eqref{eqn:bgk_org} is recovered. When $\theta=0$, it will have $\bm u_\theta=\bm 0$, then for the continuum gas flow (namely $\tau$ is very small) the gas velocity $\bm u$ will soon relax to $\bm u = \bm 0$, and thus the no-penetration no-slip boundary condition can be realized at the gas-solid interface. Note that the power-law scaling of Eq.~\eqref{eqn:bgk_poroP_simp} is similar to the material interpolation of the solid isotropic material with penalization \cite{bendsoe1999material} which is widely used in the topology optimization of solid structure. With the parameter $q>1$, this power-law scaling can avoid too severe gradient (more exactly, gradient of the objective function with respective to the material density $\theta$) when $\theta \to 1$ to prevent the gas region from transforming into the solid region during the optimization, and therefore avoid the local optima.

To extend the above fictitious porosity model to rarefied gas flow with a large Kn number (namely large $\tau$), Sato \textit{et al.}~\cite{sato2019topology} modified Eq.~\eqref{eqn:bgk_poroP} to
\begin{equation}\label{eqn:bgk_poro}
\frac{{\partial f}}{{\partial t}} + \bm v \cdot \nabla f = \frac{g_{\rm M}(\rho ,\bm u_\theta ,T_\theta ) - f}{\tau _\theta },
\end{equation}
and the material interpolation of rational approximation of material properties \cite{stolpe2001alternative}, which is also a popular interpolation method in the  structural topology optimization, was used to calculate $\bm u_\theta,T_\theta$:
\begin{equation}\label{eqn:matintp_Sw}
\bm u_\theta ^{{\rm{Sato}}} = \frac{{\theta \left( {1 + q} \right)}}{{\theta  + q}}\bm u,\quad T_\theta ^{{\rm{Sato}}} = \frac{{\theta \left( {1 + q} \right)}}{{\theta  + q}}\left( {T - {T_{\rm{s}}}} \right) + {T_{\rm{s}}},
\end{equation}
with $T_{\rm{s}}$ the temperature of the solid; meanwhile, they used the linear scaling to calculate $\tau _\theta$,
\begin{equation}\label{eqn:matintp_St}
\frac{1}{{{\tau _\theta^{{\rm{Sato}}} }}} =  \frac{\theta}{\tau } + \frac{1 - \theta }{{{\tau _{\rm{s}}^{{\rm{Sato}}}}}},
\end{equation}
where $\tau _{\rm{s}}^{{\rm{Sato}}}$ is a small parameter and set as a value corresponding to  Kn $=10^{-3}\sim 10^{-4}$~\cite{sato2019topology}. It is easy to figure out that in the solid region of $\theta=0$, Eq.~\eqref{eqn:bgk_poro} will drive the distribution function $f$ towards the Maxwellian distribution with the velocity $\bm u=\bm 0$ and temperature $T=T_{\rm s}$, and thus recover the diffuse boundary condition at the gas-solid interface~\cite{sato2019topology}.
Take moments of the porosity model equation \eqref{eqn:bgk_poro}, we can get the transport equation for the macroscopic quantities
\begin{equation}\label{eqn:macgov_poro}
\frac{{\partial \bm W}}{{\partial t}} + \nabla  \cdot {\bf{F}} = \frac{{{\bm W_\theta } - \bm W}}{{{\tau _\theta }}},
\end{equation}
in which  ${\bm W_\theta } = {\left( {\rho ,\rho {\bm u_\theta },\frac{1}{2}\rho \bm u_\theta ^2 + \frac{3}{2}\rho R{T_\theta }} \right)^T}$. The source term on the right-hand side of this  equation comes from the exchange of momentum and energy between the gas and  solid.

It is worth noting that although the model~\eqref{eqn:bgk_poro} combined with the material interpolation Eq.~\eqref{eqn:matintp_Sw} and Eq.~\eqref{eqn:matintp_St} works fine in the pure gas region and pure solid region, this model seems to encounter troubles in handling the gray region  during the optimization process for cases of large Kn, and may yield some undesirable gray areas in the final optimized material distributions. This is not only seen in the work of Sato \textit{et al.}~\cite{sato2019topology}, but also found in our early numerical experiments of airfoil optimization for drag reduction, where this gray problem seems to both
slow down the optimization process and spoil the optimized airfoil in the large-Kn-number case. According to our analysis, the mechanism of this problem can be described as followings. Suppose Kn is large and we are decreasing the material density $\theta$ at the approximate-gas region where $\theta  \to 1$, when the momentum transport of Eq.~\eqref{eqn:bgk_poro} is concerned, the reduction of $\theta$ will lead to  two effects:
\begin{enumerate}
\item Due to the decrease of $\bm u_\theta$ and $\tau_\theta$, the relaxation rate for $\bm u$ to relax to $\bm u=\bm 0$ will increase. This can be seen by investigating the momentum source term of Eq.~\eqref{eqn:macgov_poro}. This effect tends to drive the flow to static, therefore will increase the obstruction for the gas flow and cause a drag increase.
\item The decrease of $\tau_\theta$ will on the other hand increase the relaxation rate for $f$ to relax to the Maxwellian distribution $g_{\rm M}$. This effect tends to decrease the Kn number, therefore will decrease the  viscous/diffusing effect of the gas flow and cause a drag decrease.
\end{enumerate}
So there are two competing effects when $\theta$ is reduced but only the first effect is what we desired, namely a smaller $\theta$ should mean a larger obstruction for the flow. In the case of large Kn, the second effect may overwhelm the first effect in the area where $\theta  \to 1$, meaning that the gray region will not act as an obstruction but impose a positive effect on the gas flow. This occurs in our early numerical experiments of drag-reduction optimization when Kn is large, where a large area of gray region formed surrounding the solid region during the optimization process and caused an anomaly significant drag reduction due to the decrease of  Kn around the solid area. For the energy transport of Eq.~\eqref{eqn:bgk_poro}, the mechanism is similar and the reduction of $\theta$ will result in contradicting effects: the heat flux increase due to the larger relaxation rate to $T=T_{\rm s}$, and the heat flux decrease due to the decrease of  Kn; and if Kn is large the gray region may work in an undesired manner. The similar phenomenon has been seen in the work of Sato \textit{et al.}~\cite{sato2019topology} when optimizing the thermally driven pump at ${\rm Kn}=0.5$.

To alleviate this gray problem, we adopt the form of the fictitious porosity model~\eqref{eqn:bgk_poro} but modify the material interpolation to
\begin{equation}\label{eqn:matintp}
\left .
\begin{aligned}
{\bm u_\theta } = {\theta _w}\bm u,\quad {T_\theta } &= {\theta _w}T + \left( {1 - {\theta _w}} \right){T_s},\quad \frac{1}{{{\tau _\theta }}} = \frac{\theta _\tau }{\tau } + \frac{1 - {\theta _\tau }} {{{\tau _s}}}, \\
{\theta _w} &= \frac{{{q_w}\theta }}{{1 + {q_w} - \theta }},\quad {\theta _\tau } = \frac{{\left( {1 + {q_\tau }} \right)\theta }}{{{q_\tau } + \theta }} .
\end{aligned}
\right\}
\end{equation}
The idea of this modification is to make $\bm u_\theta$ and $T_\theta$ change faster than $\tau_\theta$ along with the variation of material density $\theta$ when $\theta \to 1$, and then the first effect mentioned above can surpass the second effect even when Kn is large, so that the gray region can always work in an ideal manner. The interpolation parameters are set as $q_w=10^{-2},q_\tau=5\times10^{-3}$, which work well in our present numerical tests up to ${\rm Kn}=10$. The relaxation time $\tau_{\rm s}$ is set as a value corresponding to the cell Kn number (Kn number based on the cell size of the computational mesh) of around $0.1$, namely $\tau_{\rm s}=0.1h$ in which the physical local time step $h$ is defined later in Eq.~\eqref{eqn:numerpr_physcfl}. This ensures the gas state $\bm u,T$ relax to $\bm u=0,T=T_{\rm s}$ in around one discrete cell when $\theta=0$, therefore the diffuse boundary condition can be recovered in the precision of the corresponding discretization resolution.

According to the previous studies about the density-based topology optimization, if directly taking the material density $\theta$ as the design variable, the optimization will be very unstable and may end up with undesired local optima or the so-called checkerboard pattern~\cite{bendsoe2003topology}. To solve this problem, some filters and projections can be introduced~\cite{borrvall2001topology,sigmund2013topology}.  In this paper, the design variable is set as $\vartheta  \in [0,1]$. First, a Helmholtz PDE-based filter \cite{kawamoto2011heaviside} is applied
\begin{equation}\label{eqn:intp_filter}
\left( { - {{\bar r}^2}{\nabla ^2} + 1} \right)\bar \vartheta  = \vartheta,
\end{equation}
where the filter radius $\bar r$ is set as the discrete cell size $\Delta x$. Then the filtered variable $\bar \vartheta$ is further projected to a smoothed Heaviside function \cite{kawamoto2011heaviside} as
\begin{equation}\label{eqn:intp_heaviside}
\bar {\bar \vartheta}  = \left\{ \begin{aligned}
& 0,\quad y <  - 1,\\
& \frac{1}{2} + \frac{{15}}{{16}}y - \frac{5}{8}{y^3} + \frac{3}{{16}}{y^5},\quad  - 1 \le y < 1,\\
& 1,\quad y \ge 1,
\end{aligned} \right.
\end{equation}
with $y = \left( {{2\bar \vartheta  - 1}} \right)/{{\bar w}}$. Like the previous studies \cite{guest2004achieving,sigmund2007morphology,xu2010volume}, the bandwidth $\bar w \in (0,1]$ is controlled to gradually decrease during the optimization process, see Section \ref{sec:method_optframework} below. Finally, the interpolation of rational approximation of material properties is applied to prevent severe gradient (with respect to $\theta$) when $\theta \to 1$ that obstructs the evolution of the material density,
i.e.,
\begin{equation}\label{eqn:intp_ramp}
\theta  = \frac{{\left( {1 + {q_\theta }} \right)\bar {\bar \vartheta} }}{{{q_\theta } + \bar {\bar \vartheta} }},
\end{equation}
where the parameter is set as $q_\theta=0.1$.

At the end of this section we would like to emphasize that the physical material density representing the real gas-solid distribution is $\theta$, which appears in the governing equation \eqref{eqn:bgk_poro} and is directly constrained by the volume constraint, see Eq.~\eqref{eqn:formula_opt} below. The design variable $\vartheta$ is only a mathematical parameter to define the distribution of the physical material density $\theta$ and can be scaled into any range (e.g., in Refs.~\cite{kawamoto2011heaviside,sato2019topology} it is in the range $[-1,1]$), although here it is scaled into the same range of $\theta$.

\subsection{Reduced governing equation}

The gas-kinetic equations \eqref{eqn:bgk_org} and \eqref{eqn:bgk_poro} are defined in a 3D velocity space. In this work we consider problems in 2D physical space, and there is no variable variation in the molecular velocity direction of $v_3$. Therefore, the original distribution function $f$ in 3D velocity space can be reduced \cite{chu1965kinetic,Yang1995Rarefied} to two sets of distribution functions $f_1$ and $f_2$ in 2D velocity space, saving a lot of memory and computational cost. These two reduced (or marginal) distribution functions $f_1$ and $f_2$ are defined as 
\begin{equation}\label{eqn:f_reduce}
\left.
\begin{aligned}
{f_1}(\bm x,\bm v,t) &= \int {f(\bm x,\bm v,v_3 ,t)dv_3 } \\
{f_2}(\bm x,\bm v,t) &= \int {\frac{1}{2}{v_3 ^2}f(\bm x,\bm v,v_3 ,t)dv_3 } 
\end{aligned}
\right\},
\end{equation}
where $v_3$ is the internal molecular velocity with 1 degrees of freedom. Note that in the above equation \eqref{eqn:f_reduce} and in the rest of this paper, $\bm x=(x_1,x_2)$ and $\bm v=(v_1,v_2)$ belong to a space of $2$ degrees of freedom. Then, for the fictitious porosity model equation \eqref{eqn:bgk_poro}, the corresponding reduced governing equation for $\bm f=(f_1,f_2)^T$ can be obtained by multiplying Eq.~\eqref{eqn:bgk_poro} by $(1,v_3^2/2)^T$ and integrating out $v_3$. That is
\begin{equation}\label{eqn:bgk_poro_reduce}
\frac{{\partial \bm f}}{{\partial t}} + \bm v \cdot \nabla \bm f = \frac{{{\bm g_\theta } - \bm f}}{{{\tau _\theta }}},
\end{equation}
where the equilibrium state $\bm g_\theta={\bm g_{\rm{M}}}(\rho ,{\bm u_\theta },{T_\theta })={({g_{\theta ,1}},{g_{\theta ,2}})^T}$ is
\begin{equation}
{g_{\theta ,1}} = \rho \left( 2\pi RT_\theta \right)^{-1}e^{ - \frac{{{{(\bm v - {\bm u_\theta })}^2}}}{{2R{T_\theta }}}},\quad
{g_{\theta ,2}} = \frac{{{1 }}}{2}R{T_\theta }g_{\theta ,1}.
\end{equation}
Accordingly, the relation between the macroscopic variables $\bm W$ and the reduced distribution function $\bm f$ is
\begin{equation}\label{eqn:macint_reduce}
\bm W = \int {\bf{\Psi }}    \cdot  \bm f d\Xi , 
\end{equation}
where $\bf{\Psi }$ is the moments tensor
\begin{equation}
{\bf{\Psi }} = \left( {\begin{array}{*{20}{c}}
\begin{array}{l}
1\\
\bm v\\
{\textstyle{1 \over 2}}{\bm v^2}
\end{array}&\begin{array}{l}
0\\
0\\
1
\end{array}
\end{array}} \right),
\end{equation}
and here and hereafter $d\Xi$ denotes the reduced velocity space element of $2$ dimensions.

\subsection{Optimization problem}

In this section, the formulation of topology optimization is stated. First of all, we declare that only the steady-state problem is considered in this paper, so that the time $t$ will be ignored. The design variable $\vartheta \in [0,1]$ is defined in the design domain $\Omega$. The objective is a certain functional of the macroscopic variable $\bm W$ and the material density $\theta$ defined in the design domain $\Omega$. The volume constraint is imposed on the design. Then the optimization problem is formulated as
\begin{equation}\label{eqn:formula_opt}
\left.
\begin{aligned}
&\mathop {\min }\limits_{\vartheta \in [0,1]}  \quad J = \int_\Omega  {X(\bm W,\theta )d\Omega }, \\
&{\rm{s}}{\rm{.t}}{\rm{.}}\quad Q = \int_\Omega  {\theta d\Omega }  - {V_{\max }} \le 0,
\end{aligned}
\right\}
\end{equation}
where $X(\bm W,\theta )$ is a certain function of $\bm W$ and $\theta$, and $V_{\rm max}$ is the limit of the maximum volume for the gas region, namely there is a lower limit for the solid volume.

Note that the optimization problem \eqref{eqn:formula_opt} is written in the nested form. The macroscopic variable $\bm W$ involved in the problem is essentially determined by the distribution function $\bm f$ through Eq.~\eqref{eqn:macint_reduce}. Then the two variables in the optimization problem \eqref{eqn:formula_opt}, $\bm f$ and $\theta$, satisfy the fictitious porosity model equation~\eqref{eqn:bgk_poro_reduce} and the boundary conditions:
\begin{equation}\label{eqn:formula_bgk}
\left.
\begin{aligned}
\bm v \cdot \nabla \bm f - \frac{{{\bm g_\theta } - \bm f}}{{{\tau _\theta }}} = 0 \quad {\rm{in}}\quad \Omega  \times \Xi, \\
f - {f_{\rm d}} = 0\quad  {\rm{in}}\quad {\Gamma _{\rm d}} \times {\Xi ^ - },\\
f - {g_{\rm w}} = 0\quad  {\rm{in}}\quad {\Gamma _{\rm w}} \times {\Xi ^ - },
\end{aligned}
\right\}
\end{equation}
where $\Gamma _{\rm d}$ and $\Gamma _{\rm w}$ are the boundaries of the domain $\Omega$, as shown in Fig.~\ref{fig:formula_domain}. Two types of boundary conditions are considered. On $\Gamma _{\rm d}$, the Dirichlet boundary condition is imposed with a given fixed distribution $\bm f_{\rm d}$ defined in the velocity space $\Xi ^ -$, with the definition ${\Xi ^ \pm } = \left\{ \bm v | \bm v \cdot \bm n \gtrless 0 \right\}$ representing the gas molecules flow out of or into the boundary with a outward normal unit vector $\bm n$. On $\Gamma _{\rm w}$, the diffuse boundary condition is imposed and the reflecting molecules follow the Maxwellian distribution of ${g_{\rm{w}}} = {g_{\rm{M}}}({\rho _{\rm{w}}},{\bm u_{\rm{w}}},{T_{\rm{w}}})$, where $\bm u_{\rm{w}}$ satisfying ${\bm u_{\rm{w}}} \cdot \bm n = 0$ and $T_{\rm{w}}$ are prescribed wall velocity and temperature, and $\rho _{\rm{w}}$ can be obtained from the zero-mass-flux condition as
\begin{equation}\label{eqn:formula_bgk_fdw}
{\rho _{\rm{w}}} = \sqrt {\frac{{2\pi }}{{R{T_{\rm{w}}}}}} \int_{{\Xi ^ + }} {\bm v \cdot \bm n{f_1}d\Xi } .
\end{equation}

\subsection{Sensitivity analysis}\label{sec:sens}

Since the gradient-based method is adopted to solve the optimization problem~\eqref{eqn:formula_opt}, we should calculate the derivative (or sensitivity) of the objective $J$ with respect to the design variable $\vartheta$. Here we obtain the sensitivity by the adjoint variable method \cite{bendsoe2003topology}.

To facilitate the analysis we first investigate the derivative of $J$ with respect to the material density $\theta$. From Eq.~\eqref{eqn:formula_opt} we know that $J$ is a functional of $\bm W$ and $\theta$, where $\bm W$ is essentially determined by the distribution function $\bm f$, meanwhile $\bm f$ and $\theta$ should satisfy the gas-kinetic problem \eqref{eqn:formula_bgk}. Therefore, one can introduce a set of Lagrangian multipliers (or adjoint variables) $\bm \phi ,{\bm \varphi _{\rm{d}}},{\bm \varphi _{\rm{w}}}$ to write the Lagrangian as
\begin{equation}\label{eqn:lagrangian}
\begin{aligned}[b]
{\cal L}(\bm f,\theta ) =  & J + \int_\Omega  {\int_\Xi  {\bm \phi  \cdot \left( {\bm v \cdot \nabla \bm f - \frac{{{\bm g_\theta } - \bm f}}{{{\tau _\theta }}}} \right)d\Xi } d\Omega }  + \int_{{\Gamma _{\rm d}}} {\int_{{\Xi ^ - }} {{\bm \varphi _{\rm d}} \cdot \left( {\bm f - {\bm f_{\rm d}}} \right)d\Xi } d\Gamma } \\
 &+ \int_{{\Gamma _{\rm w}}} {\int_{{\Xi ^ - }} {{\bm \varphi _{\rm w}} \cdot \left( {\bm f - {\bm g_{\rm w}}} \right)d\Xi } d\Gamma } .
\end{aligned}
\end{equation}
Note that despite the complex form of Eq.~\eqref{eqn:lagrangian}, ${\cal L}(\bm f,\theta )$ is a functional of $\bm f$ and $\theta$. Obviously, $\forall \bm \phi ,{\bm \varphi _{\rm{d}}},{\bm \varphi _{\rm{w}}}$ there is ${\cal L} \equiv J$, so the total derivatives of ${\cal L}$ and $J$ are the same. This means that we can calculate the total derivative of $\cal L$ with respect to $\theta$ instead of that of $J$, which can be finished by finding a set of adjoint variables $\bm \phi ,{\bm \varphi _{\rm{d}}},{\bm \varphi _{\rm{w}}}$ fulfilling
\begin{equation}\label{eqn:fdelta0}
d{\cal L}(\bm f;\delta \bm f) = 0
\end{equation}
to eliminate the contribution of $\delta \bm f$ to the total derivative of $\cal L$. Then the total derivative of $\cal L$ with respect to $\theta$ can be easily calculated from the explicit expression of the partial derivative $d{\cal L}(\theta ;\delta \theta )$. Thus, according to Eq.~\eqref{eqn:fdelta0}, the adjoint equation to determine the adjoint variables $\bm \phi ,{\bm \varphi _{\rm{d}}},{\bm \varphi _{\rm{w}}}$ can be established. The derivation involves the variational method and some mathematical arrangements, during which the adjoint variables ${\bm \varphi _{\rm{d}}},{\bm \varphi _{\rm{w}}}$ can be cancelled (substituted by terms with $\bm \phi$). Finally the adjoint governing equation for $\bm \phi=(\phi_1,\phi_2)^T$ along with the boundary conditions can be formulated as
\begin{equation}\label{eqn:formula_adjointbgk}
\left.
\begin{aligned}
 - \bm v \cdot \nabla \bm \phi  = \frac{{{\bm \phi _{{\rm{eq}}}} - \bm \phi }}{{{\tau _\theta }}} + {\bm \phi _\tau } + {\bm \phi _J}, \quad {\rm{in}}\quad \Omega  \times \Xi, \\
\bm \phi  = \bm 0, \quad {\rm{in}} \quad {\Gamma _{\rm d}} \times {\Xi ^ + },\\
\bm \phi  = \left( \begin{aligned}
 - \sqrt {\frac{{2\pi }}{{R{T_{\rm{w}}}}}} \int_{{\Xi ^ - }} \bm v & \cdot \bm n \bm\phi  \cdot \frac{{\partial {\bm g_{\rm{w}}}}}{{\partial {\rho _{\rm{w}}}}}d\Xi  \\
 0
\end{aligned} \right),\quad {\rm{in}}\quad {\Gamma _{\rm w}} \times {\Xi ^ + },
\end{aligned}
\right\}
\end{equation}
where the collision-related terms ${\bm \phi _{{\rm{eq}}}},{\bm \phi _\tau },{\bm \phi _J}$ are
\begin{equation}\label{eqn:adjoint_eqdefines}
{\bm \phi _{\rm eq}} = \hat {\bm W} \cdot {\bf{\Theta }} \cdot {\bf{\Psi }},\quad {\bm \phi _\tau } =  - {{\hat \rho }_\tau }\frac{1}{{{\tau _\theta }}}\frac{{\partial {\tau _\theta }}}{{\partial \bm W}} \cdot {\bf{\Psi }},\quad {\bm \phi _J} =  - \frac{{\partial X}}{{\partial \bm W}} \cdot {\bf{\Psi }},
\end{equation}
with
\begin{equation}\label{eqn:adjoint_macdefines}
\hat {\bm W} = \int_\Xi  {\bm \phi  \cdot \frac{{\partial {\bm g_\theta }}}{{\partial {\bm W_\theta }}}d\Xi } , \quad {\bf{\Theta }} = \frac{{\partial {\bm W_\theta }}}{{\partial \bm W}} , \quad {\hat \rho _\tau } = \int_\Xi  {\bm \phi  \cdot \frac{{{\bm g_\theta } - \bm f}}{{{\tau _\theta }}}d\Xi }. 
\end{equation}
It can be seen that the moments factor to get the adjoint macroscopic variable $\hat {\bm W}$ is $\partial {\bm g_\theta }/\partial {\bm W_\theta }$, by analogy with the moments tensor $\bf{\Psi }$ for the primal gas-kinetic governing equation. It is also worth noting that in the adjoint problem \eqref{eqn:formula_adjointbgk}, the only term relating to the objective is $\bm \phi _J$, which is a constant source term for the adjoint governing equation and reflects the influence of objective functions.

On obtaining the adjoint variable $\bm \phi$ from the adjoint problem \eqref{eqn:formula_adjointbgk}, directly taking the partial derivative of the Lagrangian \eqref{eqn:lagrangian} with respect to $\theta$ will yield
\begin{equation}\label{eqn:sens_theta}
\left.
\begin{aligned}
& d{\cal L}(\theta ;\delta \theta ) = \int_\Omega  {{{\cal L_\theta'} }\delta \theta d\Omega }, \\
& {{\cal L_\theta'} } = \frac{{\partial X}}{{\partial \theta }} - \frac{1}{{{\tau _\theta }}}\hat {\bm W}\frac{{\partial {\bm W_\theta }}}{{\partial \theta }}{\rm{ + }}{\rho _\tau }\frac{1}{{{\tau _\theta }}}\frac{{\partial {\tau _\theta }}}{{\partial \theta }},
\end{aligned}
\right\}
\end{equation}
which is just the derivative of the objective $J$ with respect to the material density $\theta$, i.e., ${{J_\theta'} } = {{\cal L_\theta'} }$. Then, according to Eqs.~\eqref{eqn:intp_heaviside} and \eqref{eqn:intp_ramp}, the derivative with respect to $\bar \vartheta$ can be calculated by the chain rule as
\begin{equation}\label{eqn:sens_vartheta1}
{J'_{\bar \vartheta }}{\rm{ = }}{J'_\theta }\frac{{\partial \bar {\bar \vartheta} }}{{\partial \theta }}\frac{{\partial \bar \vartheta }}{{\partial \bar {\bar \vartheta} }}.
\end{equation}
Finally, based on Eq.~\eqref{eqn:intp_filter}, it can be derived that relation between ${J'_{\bar \vartheta }}$ and ${J'_{\vartheta }}$ follows the same PDE-based filter \cite{kawamoto2011heaviside}
\begin{equation}\label{eqn:sens_vartheta2}
\left( { - {{\bar r}^2}{\nabla ^2} + 1} \right){J'_\vartheta } = {J'_{\bar \vartheta }},
\end{equation}
in which we can obtain the derivative $J'_\vartheta$ of the objective $J$ with respect to the design variable $\vartheta$.

Note that in the optimization problem \eqref{eqn:formula_opt}, the volume limit constraint $Q$ is involved.
Because $Q$ is not related to the flow variables $\bm f$ or $\bm W$, but only related to the material density $\theta$, it is easy to get the derivatives of $Q$ with respect to the design variable $\vartheta$ 
through the chain rule and the PDE-based filter:
\begin{equation}\label{eqn:sens_constraint}
\left.
\begin{aligned}
{Q'_{\bar \vartheta }}{\rm{ = }}{Q'_\theta }&\frac{{\partial \bar {\bar \vartheta} }}{{\partial \theta }}\frac{{\partial \bar \vartheta }}{{\partial \bar {\bar \vartheta} }},\\
\left(  - {{\bar r}^2}{\nabla ^2} + 1 \right)&{Q'_\vartheta} = {Q'_{\bar \vartheta }},
\end{aligned}
\right\}
\end{equation}
where $Q'_\theta=1$.

\section{Numerical method}\label{sec:method}
\subsection{Overall algorithm framework of the optimization}\label{sec:method_optframework}

The present topology optimization method follows the general procedure of the gradient-based optimization method, and three stages (namely, the calculation of the objective, the calculation of sensitivity, and the update of design variable) are involved. For the calculations of the objective and sensitivity, the analytic formulas have already been established in Section~\ref{sec:formula} and the corresponding numerical schemes will be detailed in the following sections. For the update of design variable, the method of moving asymptotes (MMA) \cite{svanberg1987method,svanberg2002class} algorithm is adopted. In our actual implementation of the MMA algorithm, the NLopt library of Johnson \cite{johnson2007nLopt} is employed, which is in fact an improved version of the original MMA algorithm \cite{svanberg1987method} and named as the globally convergent MMA algorithm detailed in Ref.~\cite{svanberg2002class}. Furthermore, to reduce the gray region while maintain a certain degree of plasticity of the gas-solid interface for guaranteeing the optimization efficiency, the bandwidth $\bar w$ of the Heaviside projection \eqref{eqn:intp_heaviside} is controlled to gradually decrease during optimization, which is similar to the strategies in Refs.~\cite{guest2004achieving,sigmund2007morphology,xu2010volume}. More specifically, the whole optimization will finish in three rounds denoted by $l=0,1,2$, during which the bandwidth $\bar w$  descends in a geometric sequence. For each round of optimization, the convergence criterion is set as that the iteration step $m \ge 5$ and after the $m$-th iteration the change of the design variable satisfies the following criterion:
\begin{equation}\label{eqn:criterion_opt}
\frac{{{{\left\| {\vartheta _i^{(l,m + 1)} - \vartheta _i^{(l,m)}} \right\|}_1}}}{{{{\left\| {\vartheta _i^{(l,m)}} \right\|}_1}}} < 5 \times {10^{ - 5}},
\end{equation}
where ${\left\| {\vartheta _i^{(l,m)}} \right\|_1} = \sum\nolimits_i {\left| {\vartheta _i^{(l,m)}} \right|} $ is the $L_1$ norm and the subscript $i$ denotes the discrete cell number.

The whole optimization procedure is listed as follows:
\begin{description}
    \item[Step 1.] Initialize the design variable $\vartheta^{(l=0,m=0)}$, set the initial value for the bandwidth to ${\bar w^{(l = 0)}} = 1$.
    
    \item[Step 2.] Based on $\vartheta^{(l,m)}$, solve the primal gas-kinetic problem \eqref{eqn:formula_bgk} and obtain the primal flow variables. Then calculate the objective $J^{(l,m)}$ and the constraint $Q^{(l,m)}$ by Eq.~\eqref{eqn:formula_opt}.
    
    \item[Step 3.] Based on $\vartheta^{(l,m)}$ and the corresponding primal flow variables, solve the adjoint problem \eqref{eqn:formula_adjointbgk} and calculate the sensitivity ${J_\vartheta'^{(l,m)} }$. Meanwhile calculate the sensitivity of the constraint ${Q_\vartheta'^{(l,m)} }$ from Eq.~\eqref{eqn:sens_constraint}.
    
    \item[Step 4.] Substitute the objective $J^{(l,m)}$, the constraint $Q^{(l,m)}$, and the corresponding sensitivities ${J_\vartheta'^{(l,m)} }$ and ${Q_\vartheta'^{(l,m)} }$ into the MMA optimizer and get the updated design variables $\vartheta^{(l,m+1)}$.
    
	\item[Step 5.] If the iteration number $m \ge 5$ and the criterion Eq.~\eqref{eqn:criterion_opt} is met, end this round of optimization and go to Step 6. Otherwise set $m=m+1$ and go to Step 2 to continue this round of optimization.
	\item[Step 6.] Set the round number $l=l+1$. If $l=3$, namely 3 rounds of sub-optimizations are complete, the whole optimization is finished. Otherwise reduce the bandwidth ${\bar w^{(l)}} = {2^{ - l/2}}$ and reset the sub-iteration number $m=0$, go to Step 2 to start the next round of optimization.
\end{description}

In the following sections, the numerical schemes to solve the primal gas-kinetic problem~\eqref{eqn:formula_bgk} and the corresponding adjoint problem \eqref{eqn:formula_adjointbgk}, which are efficient in both the continuum and rarefied gas flow regimes,  will be detailed.

\subsection{Numerical scheme for the primal equation}\label{sec:method_pr}

To get the steady-state solution of the primal problem \eqref{eqn:formula_bgk}, an implicit numerical scheme which is accurate and efficient in all flow regimes is constructed. The scheme takes the framework of the finite-volume DVM. Applying the backward Euler implicit temporal discretization, the primal gas-kinetic governing equation \eqref{eqn:bgk_poro_reduce} is discretized as
\begin{equation}\label{eqn:numerpr_gov}
\frac{{\Delta {\Omega _i}}}{{\Delta t}}\left( {\bm f_{i,k}^{n + 1} - \bm f_{i,k}^n} \right) + \sum\limits_{j \in N\left( i \right)} {{A_{ij}}{\bm v_k} \cdot {\bm n_{ij}}\bm f_{ij,k}^{n + 1}}  = \Delta {\Omega _i}\frac{{\bm g_{\theta ,i,k}^{n + 1} - \bm f_{i,k}^{n + 1}}}{{\tau _{\theta ,i}^{n + 1}}},
\end{equation}
where the super/subscripts $i,n,k$ correspond to the discretizations in physical
space, time and velocity space, respectively; $j$ denotes the neighboring cell of cell $i$ and $N\left( i \right)$ is the set of all of the neighbors of $i$; $ij$ denotes the variable at the interface between cell $i$ and $j$; $A_{ij}$ is the interface area, ${\bm n_{ij}}$ is the outward normal unit vector of interface $ij$ relative to cell $i$, and $\Delta {\Omega _i}$ is the volume of cell $i$; $\Delta t$ is the implicit time step and can be handled by various of traditional implicit time step control techniques. Note that in this and following sections the superscripts $l,m$ denoting the optimization rounds/iterations appearing in Section \ref{sec:method_optframework} will be omitted for simplicity.

In order to solve the above implicit discretized gas-kinetic equation \eqref{eqn:numerpr_gov} accurately and efficiently in all flow regimes, the interface distribution function $\bm f_{ij,k}^{n + 1}$ and the $(n+1)$-th equilibrium state $\bm g_{\theta ,i,k}^{n + 1}$ should be carefully treated, as described in the following sections.

\subsubsection{Multiscale numerical flux}\label{sec:method_pr_flux}
For the discretized gas-kinetic equation \eqref{eqn:numerpr_gov}, the numerical flux at the interface of the control volume is determined by the interface distribution function $\bm f_{ij,k}^{n}$ (suppose here we are considering the flux at the $n$-th step), and the treatment for $\bm f_{ij,k}^{n}$ determines whether the scheme is accurate in all flow regimes or not, namely whether the scheme has the so-called multiscale property. If the interface distribution $\bm f_{ij,k}^{n}$ is directly calculated by reconstructing the initial data in the control volume at the $n$-th time step, namely the advection-collision-splitting treatment is applied, the scheme will yield very dissipating result in the continuum regime when the cell size is generally much larger than the mean free path of gas molecules. This is reported in many previous studies \cite{zhu2017performance,yuan2020conservative,yuan2021novel}. To overcome this problem, the idea of the discrete unified gas-kinetic scheme \cite{guo2013discrete} is adopted in the present work to construct a multiscale numerical flux considering both the effects of molecular transport and collision. That is, the initial distribution function stored in the control volume is evolved to the interface over a time $h_{ij}$ by a temporal difference scheme of the governing equation \eqref{eqn:bgk_poro_reduce} along the molecule trajectory. This time evolution is only for the interface distribution $\bm f_{ij,k}^{n}$ to get an accurate instantaneous value to calculate the multiscale numerical flux, but not the actual time marching of the flow field.
The involving temporal difference scheme is constructed by the backward Euler formula. For more detailed discussions and derivations please refer to \cite{yuan2020conservative},
and here we finally formulate the interface distribution $\bm f_{ij,k}^{n}$ as
\begin{equation}\label{eqn:numerpr_fij}
\bm f_{ij,k}^n = \frac{{\tau _{\theta ,ij}^n}}{{\tau _{\theta ,ij}^n + {h_{ij}}}}\bm f({\bm x_{ij}} - {\bm v_k}{h_{ij}},{\bm v_k},{t^n}) + \frac{{{h_{ij}}}}{{\tau _{\theta ,ij}^n + {h_{ij}}}}{\bm g_\theta }({\bm x_{ij}},{\bm v_k},{t^n}),
\end{equation}
where the free-transport term is calculated as
\begin{equation}\label{eqn:numerpr_fij_freetrans}
\bm f({\bm x_{ij}} - {\bm v_k}{h_{ij}},{\bm v_k},{t^n}) = \left\{ 
\begin{aligned}
{\bm f_{i,k}^n + ({\bm x_{ij}} - {\bm x_i} - {\bm v_k}{h_{ij}}) \cdot {\theta _i}\nabla \bm f_{i,k}^n,\; {\bm v_k} \cdot {\bm n_{ij}} \ge 0},\\
{\bm f_{j,k}^n + ({\bm x_{ij}} - {\bm x_j} - {\bm v_k}{h_{ij}}) \cdot {\theta _j}\nabla \bm f_{j,k}^n,\; {\bm v_k} \cdot {\bm n_{ij}} < 0}.
\end{aligned}
\right.
\end{equation}
In the above expressions, the gradients $\nabla \bm f_{i,k}^n$ and $\nabla \bm f_{j,k}^n$ can be calculated by the reconstruction of the initial distribution function data. Here the 1-st order (linear) reconstruction with the weighted least-squares method is applied, where the weight is set as $0.999{\theta _i} + 0.001$. Note that in Eq.~\eqref{eqn:numerpr_fij_freetrans} these gradients are multiplied by the material density $\theta_i$ and $\theta_j$. This is because we found in our numerical experiments that adopting a $0$-th order reconstruction in the solid region of $\theta=0$ can accelerate the computation while not affect the accuracy, and more importantly can suppress the numerical error in the discontinuity region where the material density abruptly jumps form $\theta=0$ to $\theta=1$. In Eq.~\eqref{eqn:numerpr_fij}, the calculation of ${\bm g_\theta }({\bm x_{ij}},{\bm v_k},{t^n})$ involves the macroscopic variables $\bm W^n_{ij}$ and the material density $\theta_{ij}$, in which $\theta_{ij}=\max ({\theta _{i}},{\theta _{j}})$, and $\bm W^n_{ij}$ is calculated through an upwind manner \cite{xu2001gas}
\begin{equation}\label{eqn:numerpr_wij}
\bm W_{ij}^n = \int_{\bm v \cdot {\bm n_{ij}} \ge 0} {{\bf{\Psi }} \cdot {\bm g_{\rm M}}(\bm W_{ij}^{n,\rm L})d\Xi }  + \int_{\bm v \cdot {\bm n_{ij}} < 0} {{\bf{\Psi }} \cdot {\bm g_{\rm M}}(\bm W_{ij}^{n,\rm R})d\Xi } ,
\end{equation}
where the superscript $\rm L,R$ denote the interface variables on the two sides of the interface obtained by the reconstruction, i.e.
\begin{equation}\label{eqn:numerpr_wlr}
\left.
\begin{aligned}
\bm W_{ij}^{n,{\rm{L}}} = \bm W_i^n + ({\bm x_{ij}} - {\bm x_i}) \cdot {\theta _i}\nabla \bm W_i^n,\\
\bm W_{ij}^{n,{\rm{R}}} = \bm W_j^n + ({\bm x_{ij}} - {\bm x_j}) \cdot {\theta _j}\nabla \bm W_j^n.
\end{aligned}
\right\}
\end{equation}
Similar to Eq.~\eqref{eqn:numerpr_fij_freetrans}, the gradients $\nabla \bm W_i^n$ and $\nabla  \bm W_j^n$ are calculated by the weighted least-squares method and multiplied by $\theta_i,\theta_j$. $\tau _{\theta ,ij}^n$ at the interface has the form
\begin{equation}
\tau _{\theta ,ij}^n = \tau _{\theta ,ij,{\rm{physical}}}^n + \tau _{\theta ,ij,{\rm{artificial}}}^n,
\end{equation}
where $\tau _{\theta ,ij,{\rm{physical}}}^n$ is calculated by Eq.~\eqref{eqn:matintp} from the interface macroscopic variables $\bm W_{ij}^n$, and the artificial dissipation term $\tau _{\theta ,ij,{\rm{artificial}}}^n$ \cite{xu2001gas} is calculated as
\begin{equation}
\tau _{\theta ,ij,{\rm{artificial}}}^n = {\theta _{w,ij}}\frac{{\left| {p_{ij}^{\rm L} - p_{ij}^{\rm R}} \right|}}{{\left| {p_{ij}^{\rm L} + p_{ij}^{\rm R}} \right|}}{h_{ij}},
\end{equation}
in which $p_{ij}^{\rm L},p_{ij}^{\rm R}$ are the pressures calculated from $\bm W_{ij}^{n,{\rm{L}}},\bm W_{ij}^{n,{\rm{R}}}$. Finally, for the evolution time $h_{ij}$, it is often called as the physical time step \cite{Zhu2016Implicit,yuan2020conservative} and is determined by the local CFL condition as
\begin{equation}\label{eqn:numerpr_physcfl}
{h_{ij}} = \min ({h_i},{h_j}), \quad\text{with} \quad
{h_i} = \frac{{\Delta {\Omega _i}}}{{\mathop{\max }\limits_k \left( {\sum\limits_{j \in N(i)} {\left( {{\rm{H}}({\bm v_k} \cdot {\bm n_{ij}}){A_{ij}}{\bm v_k} \cdot {\bm n_{ij}}} \right)} } \right)}}  {\rm CFL }_{\rm phys},
\end{equation}
where ${\rm H}(x)$ is the Heaviside function and ${\rm CFL }_{\rm phys}$ is set as $0.9$. Once again, we emphasize that the physical time step $h_{ij}$ is only used in Eq.~\eqref{eqn:numerpr_fij} to obtain an instantaneous interface distribution function accurate in all flow regimes. 
The time-marching step is $\Delta t$ appearing in Eq.~\eqref{eqn:numerpr_gov}, which is not limited by the CFL condition due to the implicit temporal discretization of Eq.~\eqref{eqn:numerpr_gov}.

\subsubsection{Macroscopic prediction}\label{sec:method_pr_prediction}
In the implicit discretized gas-kinetic equation \eqref{eqn:numerpr_gov}, the treatment for the $(n+1)$-th equilibrium state $\bm g_{\theta ,i,k}^{n + 1}$ is the key point to ensuring the convergence efficiency in all flow regimes. If directly using $\bm g_{\theta ,i,k}^{n}$
at the old time step to approximate $\bm g_{\theta ,i,k}^{n + 1}$ at the new time step in Eq.~\eqref{eqn:numerpr_gov}, the scheme will suffer from a very slow convergence rate in the continuum flow regime where $\tau_\theta$ is very small \cite{Mieussens2000DISCRETE}. One solution for this problem is to directly calculate $\bm g^{n+1}$ from $\bm f^{n+1}$ by a linear mapping~\cite{Mieussens2000DISCRETE}; however, a large Jacobian matrix with a high computational complexity is involved. Another
treatment is the so-called macroscopic prediction method \cite{Zhu2016Implicit,yuan2020conservative,yuan2021multi}, which is adopted in the present work. The macroscopic prediction method solves an approximate prediction equation, which is normally constructed from the continuum limit of the governing equation \eqref{eqn:bgk_poro_reduce} and only involves the macroscopic variables, to obtain a set of predicted macroscopic variable $\bm W^{n+1/2}_{\theta,i}$, and then an approximate $\bm g_{\theta ,i,k}^{n + 1/2}$ can be calculated and substituted into Eq.~\eqref{eqn:numerpr_gov} to guarantee the convergence efficiency in the continuum flow regime. For simplicity here we adopt an Euler-type prediction equation, although an NS-type prediction equation could lead to a faster convergence rate~\cite{yuan2021multi}. The detailed description of the method is in the following paragraphs.

Take moments of Eq.~\eqref{eqn:numerpr_gov} for $\bf \Psi$ will yield the corresponding discretized governing equation for the macroscopic variables
\begin{equation}\label{eqn:numerpr_govmac}
\frac{{\Delta {\Omega _i}}}{{\Delta t}}\left( {\bm W_i^{n + 1/2} - \bm  W_i^n} \right) + \sum\limits_{j \in N(i)} {{A_{ij}}\bm F_{ij}^{n + 1/2}}  = \Delta {\Omega _i}\frac{{\bm W_{\theta ,i}^{n + 1/2} - \bm W_i^{n + 1/2}}}{{\tau _{\theta ,i}^{n + 1/2}}},
\end{equation}
where the superscript $n+1/2$ denotes that the variable is predicted. The source term on the right-hand side is approximately linearized as
\begin{equation}\label{eqn:numerpr_macsource}
\frac{{\bm W_{\theta ,i}^{n + 1/2} - \bm W_i^{n + 1/2}}}{{\tau _{\theta ,i}^{n + 1/2}}} \approx \frac{{\bm W_{\theta ,i}^n - \bm W_i^n}}{{\tau _{\theta ,i}^n}} + \frac{{\left( {{\bf{\Theta }}_i^n - 1} \right) \cdot \Delta \bm W_i^{n + 1/2}}}{{\tau _{\theta ,i}^n}},
\end{equation}
where $\bf{\Theta }$ is the Jacobian matrix as shown in Eq.~\eqref{eqn:adjoint_macdefines} and $\bm W_i^n$ is obtained by Eq.~\eqref{eqn:macint_reduce} through the numerical integration. The interface flux $\bm F_{ij}^{n + 1/2}$ in Eq.~\eqref{eqn:numerpr_govmac} is written as
\begin{equation}\label{eqn:numerpr_macflux}
\bm F_{ij}^{n + 1/2} = \bm F_{ij}^n + \Delta \bm F_{ij}^{n + 1/2},
\end{equation}
in which $\bm F_{ij}^n$ can be numerically integrated from the interface distribution  function $\bm f_{ij,k}^n$ (determined by Eq.~\eqref{eqn:numerpr_fij}) as
\begin{equation}\label{eqn:numerpr_intflux}
\bm F_{ij}^n = \sum\limits_k {{\bm v_k} \cdot {\bm n_{ij}}{{\bf{\Psi }}_k} \cdot \bm f_{ij,k}^n\Delta {\Xi _k}} .
\end{equation}
For the flux increment $\Delta \bm F_{ij}^{n + 1/2}$, to close the equation, it is handled by the approximate value in the continuum limit. That is, applying the Chapman-Enskog expansion \cite{chapman1990mathematical} to the governing equation \eqref{eqn:bgk_poro_reduce} and omitting all terms of $O(\tau_\theta)$, for the interface flux we will obtain the following Euler flux:
\begin{equation}\label{eqn:numerpr_euulerflux}
{\bm F_{{\rm E},ij}}(\bm W) = 
\left( 
\begin{aligned}
&\quad\quad \rho \bm u \cdot {\bm n_{ij}}\\
&\rho \bm u \bm u \cdot {\bm n_{ij}} + p{\bm n_{ij}}\\
&\left( {\rho E + p} \right)\bm u \cdot {\bm n_{ij}}
\end{aligned} 
\right).
\end{equation}
Then, similar to the traditional implicit scheme based on the Euler or NS equation \cite{luo1998fast}, $\Delta \bm F_{ij}^{n + 1/2}$ can be calculated as
\begin{equation}\label{eqn:numerpr_macfluxinc}
\Delta \bm F_{ij}^{n + 1/2} = \bm F_{{\rm R},ij}^{n + 1/2} - \bm F_{{\rm R},ij}^n,
\end{equation}
where $\bm F_{{\rm R},ij}^n$ is the approximate Roe's flux function 
\begin{equation}\label{eqn:numerpr_macfluxRoe}
\left.
\begin{aligned}
\bm F_{{\rm R},ij}^n = \frac{1}{2} & \left( {{\bm F_{{\rm E},ij}}(\bm W_i^n) + {\bm F_{{\rm E},ij}}(\bm W_j^n) + \varrho _{ij}^n\bm W_i^n - \varrho _{ij}^n\bm W_j^n} \right),\\
&\varrho _{ij}^n = \left| {\bm u_{ij}^n \cdot {\bm n_{ij}}} \right| + a_{ij}^n + 2\frac{{\mu _{ij}^n}}{{\rho _{ij}^n\left| {{\bm x_i} - {\bm x_j}} \right|}},
\end{aligned}
\right\}
\end{equation}
where $a_{ij}^n$ is the acoustic speed. Substituting Eqs.~\eqref{eqn:numerpr_macsource}, \eqref{eqn:numerpr_macflux}, \eqref{eqn:numerpr_macfluxinc} and \eqref{eqn:numerpr_macfluxRoe} into the macroscopic discretized governing equation \eqref{eqn:numerpr_govmac}, approximating $\varrho _{ij}^{n + 1/2} \approx \varrho _{ij}^n$
and noting that $\sum\nolimits_{j \in N(i)} {{A_{ij}}{\bm F_{{\rm{E}},ij}}({\bm W_i^n})}  = \bm 0$, the equation can be finally arranged as
\begin{equation}\label{eqn:numerpr_macupdate}
\begin{aligned}
& \left( {\frac{{\Delta {\Omega _i}}}{{\Delta t}} + \frac{1}{2}\sum\limits_{j \in N(i)} {\varrho _{ij}^n{A_{ij}}}  - \frac{{\Delta {\Omega _i}}}{{\tau _{\theta ,i}^n}}\left( {{\bf{\Theta }}_i^n - 1} \right)} \right) \cdot \Delta \bm W_i^{n + 1/2} \\
=  &  - \sum\limits_{j \in N(i)} {{A_{ij}}\bm F_{ij}^n}  + \frac{1}{2}\sum\limits_{j \in N(i)} {\varrho _{ij}^n{A_{ij}}\Delta \bm W_j^{n + 1/2}} \\
& - \frac{1}{2}\sum\limits_{j \in N(i)} {{A_{ij}}\left( {{\bm F_{{\rm E},ij}}(\bm W_j^{n + 1/2}) - {\bm F_{{\rm E},ij}}(\bm W_j^n)} \right)} + \Delta {\Omega _i}\frac{{\bm W_{\theta ,i}^n - \bm W_i^n}}{{\tau _\theta ^n}}.
\end{aligned}
\end{equation}
The above equation is solved by the symmetric Gauss-Seidel (SGS) method, more specifically here the point-relaxation SGS  method \cite{rogers1995comparison,yuan2002comparison} is implemented. It is worth noting that there is a more precise treatment for the last term (the source term) on the right-hand side of Eq.~\eqref{eqn:numerpr_macupdate}, i.e., applying the following substitution for the $\kappa$-th SGS iteration to update $\Delta \bm W_i^{n + 1/2,(\kappa +1)}$:
\begin{equation}
\frac{{\bm W_{\theta ,i}^n - \bm W_i^n}}{{\tau _\theta ^n}} = \frac{{\bm W_{\theta ,i}^{n + 1/2,(\kappa )} - \bm W_i^{n + 1/2,(\kappa )}}}{{\tau _{\theta ,i}^{n + 1/2,(\kappa )}}} - \frac{{\left( {{\bf{\Theta }}_i^n - 1} \right) \cdot \Delta \bm W_i^{n + 1/2,(\kappa )}}}{{\tau _{\theta ,i}^n}},
\end{equation}
which is a nonlinearization process for the linearized source term. Finally, a set of predicted macroscopic variable $\bm W_i^{n + 1/2}$ can be obtained by solving Eq.~\eqref{eqn:numerpr_macupdate}.

\subsubsection{Overall calculation procedure}
After obtaining the interface distribution $\bm f_{ij,k}^{n}$ and the predicted macroscopic variable $\bm W_i^{n + 1/2}$, the distribution function $\bm f_{i,k}^{n+1}$ for the next step can be calculated. Arranging the discretized gas-kinetic governing equation \eqref{eqn:numerpr_gov} into the incremental form :
\begin{equation}\label{eqn:numerpr_increment}
\begin{aligned}
& \left( {\frac{{\Delta {\Omega _i}}}{{\Delta t}} + \frac{{\Delta {\Omega _i}}}{{\tau _{\theta ,i}^{n + 1/2}}}} \right)\Delta \bm f_{i,k}^{n + 1} + \sum\limits_{j \in N(i)} {{A_{ij}}{\bm v_k} \cdot {\bm n_{ij}}\Delta \bm f_{ij,k}^{n + 1}}  \\
= & \Delta {\Omega _i}\frac{{\bm g_{\theta ,i,k}^{n + 1/2} - \bm f_{i,k}^n}}{{\tau _{\theta ,i}^{n + 1/2}}} - \sum\limits_{j \in N(i)} {{A_{ij}}{\bm v_k} \cdot {\bm n_{ij}}\bm f_{ij,k}^n} ,
\end{aligned}
\end{equation}
where  $\bm f_{ij,k}^n$ is calculated by Eq.~\eqref{eqn:numerpr_fij}, $\bm g_{\theta ,i,k}^{n + 1/2}$ and $ \tau _{\theta ,i}^{n + 1/2}$ can be calculated from the predicted $\bm W_i^{n + 1/2}$. The increment of the interface distribution function $\Delta \bm f_{ij,k}^{n + 1}$ is simply handled by the first-order upwind scheme, and then Eq.~\eqref{eqn:numerpr_increment} can be further arranged into
\begin{equation}\label{eqn:numerpr_update}
\begin{aligned}[b]
& \left( {\frac{{\Delta {\Omega _i}}}{{\Delta t}} + \frac{{\Delta {\Omega _i}}}{{\tau _{\theta ,i}^{n + 1/2}}} + \sum\limits_{j \in N_k^ + (i)} {{A_{ij}}{\bm v_k} \cdot {\bm n_{ij}}} } \right)\Delta \bm f_{i,k}^{n + 1} \\
=  & \Delta {\Omega _i}\frac{{\bm g_{\theta ,i,k}^{n + 1/2} - \bm f_{i,k}^n}}{{\tau _{\theta ,i}^{n + 1/2}}} - \sum\limits_{j \in N(i)} {{A_{ij}}{\bm v_k} \cdot {\bm n_{ij}}\bm f_{ij,k}^n} 
\; - \sum\limits_{j \in N_k^ - (i)} {{A_{ij}}{\bm v_k} \cdot {\bm n_{ij}}\Delta \bm f_{j,k}^{n + 1}} 
\end{aligned},
\end{equation}
where $N_k^ + (i)$ is the set of $i$'s neighbors satisfying ${\bm v_k} \cdot {\bm n_{ij}} \ge 0$, while $N_k^ - (i)$ corresponds to those satisfying ${\bm v_k} \cdot {\bm n_{ij}} < 0$. Similar to the solving of the macroscopic equation \eqref{eqn:numerpr_macupdate}, the above equation \eqref{eqn:numerpr_update} is solved by SGS iterations, and finally the distribution function $\bm f_{i,k}^{n + 1}$ for the next time step is obtained.

At the end of this section, we  give a summary of the computation procedure for the primal equation from step $n$ to step $n+1$:
\begin{description}
	\item[Step 1.] Reconstruct the data at the step $n$ and calculate the interface distribution function by Eq.~\eqref{eqn:numerpr_fij}.
	\item[Step 2.] Compute the macroscopic flux $\bm F_{ij}^n$ by the numerical integration Eq.~\eqref{eqn:numerpr_intflux}, calculate all terms of step $n$ on the right-hand side of Eq.~\eqref{eqn:numerpr_macupdate}.
	\item[Step 3.] Solve Eq.~\eqref{eqn:numerpr_macupdate} by the SGS iterations to obtain the predicted macroscopic variable $\bm W^{n+1/2}_i$.
	\item[Step 4.] Calculate $\bm g_{\theta ,i,k}^{n + 1/2}$ and $ \tau _{\theta ,i}^{n + 1/2}$ from  $\bm W^{n+1/2}_i$, and then solve Eq.~\eqref{eqn:numerpr_update} by the SGS iterations, and finally the distribution function $\bm f_{i,k}^{n + 1}$ for the next time step is obtained.
\end{description}

\subsection{Numerical scheme for the adjoint equation}\label{sec:method_ad}

After obtaining the solution of the primal gas-kinetic problem \eqref{eqn:formula_bgk}, the adjoint problem~\eqref{eqn:formula_adjointbgk} can be solved to determine the adjoint variable $\bm \phi$. Similar to the primal discretization equation \eqref{eqn:numerpr_gov}, introducing a temporal term into the left-hand side of the steady-state adjoint equation \eqref{eqn:formula_adjointbgk} and applying the backward Euler implicit temporal discretization will yield
\begin{equation}\label{eqn:numerad_gov}
 \frac{{\Delta {\Omega _i}}}{{\Delta t}}\left( {\bm \phi _{i,k}^{n + 1} - \bm \phi _{i,k}^n} \right) - \sum\limits_{j \in N\left( i \right)} {{A_{ij}}{\bm v_k} \cdot {\bm n_{ij}}\bm \phi _{ij,k}^{n + 1}}  
=  \Delta {\Omega _i}\frac{{\bm \phi _{{\rm eq},i,k}^{n + 1} - \bm \phi _{i,k}^{n + 1}}}{{\tau _{\theta ,i}}} + \Delta {\Omega _i}\bm \phi _{\tau ,i,k}^{n + 1} + \Delta {\Omega _i}\bm \phi _{J,i,k},
\end{equation}
where the super/subscripts have the same meaning of those in Eq.~\eqref{eqn:numerpr_gov}. Here it's worth noting that the time step $\Delta t=t^{n+1}-t^{n}$ can be viewed as the pseudo time step to evolve the adjoint equation to the steady state and get the solution of the steady-state adjoint problem \eqref{eqn:formula_adjointbgk}, otherwise there should be a negative sign in front of the real physical temporal term for the unsteady adjoint equation. Also note that when solving Eq.~\eqref{eqn:numerad_gov} , the source term $\bm \phi _{J,i,k}$ and terms such as $\bm f_{i,k},\bm W_i,\tau _{\theta ,i}$ which are determined by the solution  of the primal gas-kinetic equation are all constant. A numerical scheme similar to that described in Section \ref{sec:method_pr} for the primal equation is constructed to solve the adjoint equation. The key points are the treatment of $\bm \phi _{ij,k}^{n + 1}$ at the cell interface and the treatment of $\bm \phi _{{\rm eq},i,k}^{n + 1},\bm \phi _{\tau ,i}^{n + 1}$ at the $(n+1)$-th step, which will be detailed in the followings.

\subsubsection{Multiscale numerical flux}
Like the calculation of the interface distribution function $\bm f_{ij,k}^n$ described in Section \ref{sec:method_pr_flux}, the idea of the discrete unified gas-kinetic scheme \cite{guo2013discrete} is adopted to calculate $\bm \phi _{ij,k}^n$ at the cell interface to ensure the accuracy of the adjoint scheme in all flow regimes. The thought is to evolve the initial data inside the cell over a physical time step $h_{ij}$ to the interface based on the temporal difference scheme of the adjoint equation along the characteristic line. The detailed treatment is similar to the construction of Eq.~\eqref{eqn:numerpr_fij}, and finally the formula for $\bm \phi _{ij,k}^n$ at the interface can be written as
\begin{equation}\label{eqn:numerad_phiij}
\begin{aligned}
\bm \phi _{ij,k}^n =  & \frac{{\tau _{\theta ,ij}}}{{\tau _{\theta ,ij} + {h_{ij}}}}\bm \phi ({\bm x_{ij}} + {\bm v_k}{h_{ij}},{\bm v_k},{t^n}) + \frac{{{h_{ij}}}}{{\tau _{\theta ,ij} + {h_{ij}}}}{\bm \phi _{\rm eq}}({\bm x_{ij}},{\bm v_k},{t^n})\\
& + \frac{{\tau _{\theta ,ij}{h_{ij}}}}{{\tau _{\theta ,ij} + {h_{ij}}}}{\bm \phi _\tau }({\bm x_{ij}},{\bm v_k},{t^n}) + \frac{{\tau _{\theta ,ij}{h_{ij}}}}{{\tau _{\theta ,ij} + {h_{ij}}}}{\bm \phi _J}({\bm x_{ij}},{\bm v_k}),
\end{aligned}
\end{equation}
in which the free-transport term is calculated as
\begin{equation}
\bm \phi ({\bm x_{ij}} + {\bm v_k}{h_{ij}},{\bm v_k},{t^n}) = \left\{ {\begin{aligned}
&{\bm \phi _{i,k}^n + ({\bm x_{ij}} - {\bm x_i} + {\bm v_k}{h_{ij}}) \cdot {\theta _i}\nabla \bm \phi _{i,k}^n ,\; - {\bm v_k} \cdot {\bm n_{ij}} \ge 0},\\
&{\bm \phi _{j,k}^n + ({\bm x_{ij}} - {\bm x_j} + {\bm v_k}{h_{ij}}) \cdot {\theta _j}\nabla \bm \phi _{j,k}^n ,\; - {\bm v_k} \cdot {\bm n_{ij}} < 0},
\end{aligned}} \right.
\end{equation}
where the gradients $\nabla \bm \phi _{i,k}^n,\nabla \bm \phi _{j,k}^n$ are calculated by the same method of $\nabla \bm f_{i,k}^n, \nabla \bm f_{j,k}^n$ in Eq.~\eqref{eqn:numerpr_fij_freetrans}. The equilibrium state ${\phi _{\rm eq}}({x_{ij}},{v_k},{t^n})$ involves the adjoint macroscopic variable $\hat {\bm W}_{ij}^n$ (see Eq.~\eqref{eqn:adjoint_eqdefines}) which is calculated by the similar upwind idea of Eq.~\eqref{eqn:numerpr_wij} as
\begin{equation}
\hat {\bm W}_{ij}^n = \hat {\bm W}_{ij}^{n,\rm L} \cdot \frac{{\partial \left( {\int_{\bm v \cdot {\bm n_{ij}} < 0} {{\bf{\Psi }} \cdot {\bm g_{\theta ,ij}}d\Xi } } \right)}}{{\partial {\bm W}_{\theta ,ij}}} + \hat {\bm W}_{ij}^{n,\rm R} \cdot \frac{{\partial \left( {\int_{\bm v \cdot {\bm n_{ij}} \ge 0} {{\bf{\Psi }} \cdot {\bm g_{\theta ,ij}}d\Xi } } \right)}}{{\partial {\bm W}_{\theta ,ij}}},
\end{equation}
where ${\bm g_{\theta ,ij}}$ is just the term ${\bm g_\theta }({\bm x_{ij}},{\bm v_k},{t^n})$ in Eq.~\eqref{eqn:numerpr_fij}; $\hat {\bm W}_{ij}^{n,\rm L}$ and $\hat {\bm W}_{ij}^{n,\rm R}$ are adjoint macroscopic variables on the two sides of the interface, and are calculated by the same method of Eq.~\eqref{eqn:numerpr_wlr}. For ${\bm \phi _\tau }({\bm x_{ij}},{\bm v_k},{t^n})$ in Eq.~\eqref{eqn:numerad_phiij}, according to Eq.~\eqref{eqn:adjoint_eqdefines}, its calculation involves $\hat \rho _{\tau ,ij}^n$ at the interface, which is calculated by the average
\begin{equation}
\hat \rho _{\tau ,ij}^n = \frac{1}{2}\left( {\hat \rho _{\tau ,ij}^{n,\rm L} + \hat \rho _{\tau ,ij}^{n,\rm R}} \right),
\end{equation}
where $\hat \rho _{\tau ,ij}^{n,\rm L},\hat \rho _{\tau ,ij}^{n,\rm R}$ are obtained by the same reconstruction method of $\hat {\bm W}_{ij}^{n,\rm L},\hat {\bm W}_{ij}^{n,\rm R}$, based on the cell-center value of $\hat \rho _{\tau ,i}^{n}$ calculated according to Eq.~\eqref{eqn:adjoint_macdefines}. For the last term ${\bm \phi _J }({\bm x_{ij}},{\bm v_k})$ in Eq.~\eqref{eqn:numerad_phiij}, it is a constant source term (constant during the solving of the adjoint equation) defined in Eq.~\eqref{eqn:adjoint_eqdefines}, and all the interface values involved can be obtained by reconstruction without difficulty.

\subsubsection{Macroscopic prediction}
Similar to the macroscopic prediction treatment for the discretized primal equation \eqref{eqn:numerpr_gov} discussed in Section \ref{sec:method_pr_prediction}, here the terms $\bm\phi _{{\rm eq},i,k}^{n + 1},\bm \phi _{\tau ,i,k}^{n + 1}$ (for definitions see Eq.~\eqref{eqn:adjoint_eqdefines}) on the right-hand side of the discretized adjoint equation \eqref{eqn:numerad_gov} are handled by the same method. Namely, a set of predicted macroscopic variables $\hat {\bm W}_i^{n + 1/2},\hat \rho _{\tau ,i}^{n + 1/2}$ are obtained from an approximate equation constructed from the continuum limit of the adjoint equation, and then the approximate values for $\bm\phi _{{\rm eq},i,k}^{n + 1/2},\bm \phi _{\tau ,i,k}^{n + 1/2}$ are calculated and substituted into Eq.~\eqref{eqn:numerad_gov} to avoid slow convergence rate in the continuum flow regime.

It can be seen from the definition of $\hat {\bm W}$ in Eq.~\eqref{eqn:adjoint_macdefines} that, for the adjoint equation, $\partial {\bm g_\theta }/\partial {\bm W_\theta }$ is the moments factor to obtain the adjoint macroscopic variable $\hat {\bm W}$. Therefore, multiplying the discretized adjoint equation \eqref{eqn:numerad_gov} by $\partial {\bm g_\theta }/\partial {\bm W_\theta }$ and integrating in the molecular velocity space will yield the discretized governing equation for the adjoint macroscopic variable $\hat {\bm W}$:
\begin{equation}
\begin{aligned}
& \frac{{\Delta {\Omega _i}}}{{\Delta t}}\left( {\hat {\bm W}_i^{n + 1/2} - \hat {\bm W}_i^n} \right) - \int_\Xi  {\left( {\sum\limits_{j \in N\left( i \right)} {{A_{ij}}\bm v \cdot {\bm n_{ij}}\bm \phi _{ij}^{n + 1/2}}  \cdot \frac{{\partial {\bm g_{\theta ,i}}}}{{\partial {{\bm W}_\theta }}}} \right)d\Xi } \\
 =  & \frac{{\Delta {\Omega _i}}}{{\tau _{\theta ,i}}}\hat {\bm W}_i^{n + 1/2} \cdot \left( {{{\bf{\Theta }}_i} - 1} \right) - \frac{{\Delta {\Omega _i}}}{{\tau _{\theta ,i}}}\hat \rho _{\tau ,i}^{n + 1/2}\frac{{\partial \tau _{\theta ,i}}}{{\partial {\bm W}}} + \Delta {\Omega _i}\int_\Xi  {\bm \phi _{J,i} \cdot \frac{{\partial {\bm g_{\theta ,i}}}}{{\partial {{\bm W}_\theta }}}d\Xi } ,
\end{aligned}
\end{equation}
where the variables at the $(n+1)$-th step are substituted by the predicted values denoted by the superscript $n+1/2$. Arranging the above equation into the form of the increment from step $n$ to step $n+1/2$ will yield
\begin{equation}\label{eqn:numerad_macinc}
\begin{aligned}
& \frac{{\Delta {\Omega _i}}}{{\Delta t}}\Delta \hat {\bm W}_i^{n + 1/2} - \int_\Xi  {\left( {\sum\limits_{j \in N\left( i \right)} {{A_{ij}}\bm v \cdot {\bm n_{ij}}\Delta \bm \phi _{ij}^{n + 1/2}}  \cdot \frac{{\partial {\bm g_{\theta ,i}}}}{{\partial {{\bm W}_\theta }}}} \right)d\Xi } \\
 =  & \frac{{\Delta {\Omega _i}}}{{\tau _{\theta ,i}}}\Delta \hat {\bm W}_i^{n + 1/2} \cdot \left( {{{\bf{\Theta }}_i} - 1} \right) - \frac{{\Delta {\Omega _i}}}{{\tau _{\theta ,i}}}\Delta \hat \rho _{\tau ,i}^{n + 1/2}\frac{{\partial \tau _{\theta ,i}}}{{\partial {\bm W}}} + \hat {\bm R}_i^n.
\end{aligned}
\end{equation}
Here, the term $\hat {\bm R}_i^n$ is only related to the known variables at the step $n$, whose expression is
\begin{equation}\label{eqn:numerad_macrsd}
\begin{aligned}
\hat {\bm R}_i^n =  & \frac{{\Delta {\Omega _i}}}{{\tau _{\theta ,i}}}\hat {\bm W}_i^n \cdot \left( {{{\bf{\Theta }}_i} - 1} \right) - \frac{{\Delta {\Omega _i}}}{{\tau _{\theta ,i}}}\hat \rho _{\tau ,i}^n\frac{{\partial \tau _{\theta ,i}}}{{\partial {\bm W}}} + \Delta {\Omega _i}\int_\Xi  {\bm \phi _{J,i} \cdot \frac{{\partial {\bm g_{\theta ,i}}}}{{\partial {{\bm W}_\theta }}}d\Xi } \\
& + \int_\Xi  {\left( {\sum\limits_{j \in N\left( i \right)} {{A_{ij}}\bm v \cdot {\bm n_{ij}}\bm \phi _{ij}^n}  \cdot \frac{{\partial {\bm g_{\theta ,i}}}}{{\partial {{\bm W}_\theta }}}} \right)d\Xi }, 
\end{aligned}
\end{equation}
in which the integration with respect to the velocity space can be completed by numerical integration. The incremental terms $\Delta \bm \phi _{ij}^{n + 1/2},\Delta \hat \rho _{\tau ,i}^{n + 1/2}$ in Eq.~\eqref{eqn:numerad_macinc} are handled by the same treatment of the flux increment in the primal macroscopic equation (i.e. Eq.~\eqref{eqn:numerpr_macfluxinc}). Specifically, applying the Chapman-Enskog expansion \cite{chapman1990mathematical} to the adjoint equation \eqref{eqn:formula_adjointbgk} and ignoring the terms of $O(\tau_\theta)$, we will get
\begin{equation}\label{eqn:numerad_macincce}
\begin{aligned}
\Delta \bm \phi _{ij}^{n + 1/2} = \Delta \hat {\bm W}_{ij}^{n + 1/2} \cdot {\bf{\Psi }},\quad
\Delta \hat \rho _{\tau ,i}^{n + 1/2} = 0.
\end{aligned}
\end{equation}
Substituting Eq.~\eqref{eqn:numerad_macincce} into Eq.~\eqref{eqn:numerad_macinc}, it can be derived that
\begin{equation}\label{eqn:numerad_macinc2}
\begin{aligned}
\frac{{\Delta {\Omega _i}}}{{\Delta t}}\Delta \hat {\bm W}_i^{n + 1/2} - \sum\limits_{j \in N\left( i \right)} {\left( {{A_{ij}}\Delta \hat {\bm W}_{ij}^{n + 1/2} \cdot {{\bf{J}}_{{\rm E},ij,i}}} \right)} 
 =   \frac{{\Delta {\Omega _i}}}{{\tau _{\theta ,i}}}\Delta \hat {\bm W}_i^{n + 1/2} \cdot \left( {{{\bf{\Theta }}_i} - 1} \right) + \hat {\bm R}_i^n,
\end{aligned}
\end{equation}
where ${{\bf{J}}_{{\rm E},ij,i}}= \partial {\bm F_{{\rm E},ij}}({\bm W_{\theta ,i}})/\partial {\bm W_\theta }$ is just the Jacobian matrix of the Euler flux Eq.~\eqref{eqn:numerpr_euulerflux}. For $\Delta \hat {\bm W}_{ij}^{n + 1/2}$ at the interface, according to the property of the Euler flux, directly using the central average to calculate it will lead to instability. Therefore, here $\Delta \hat {\bm W}_{ij}^{n + 1/2}$ is calculated based on the idea of Roe's flux splitting, and a flux function similar to Eq.~\eqref{eqn:numerpr_macfluxRoe} is constructed as
\begin{equation}\label{eqn:numerad_macfluxRoe}
\begin{aligned}
 - \Delta \hat {\bm W}_{ij}^{n + 1/2} \cdot {{\bf{J}}_{{\rm E},ij,i}} =&  - \frac{1}{2}\Delta \hat {\bm W}_i^{n + 1/2} \cdot {\bf{J}}_{{{\rm E}},ij,i} - \frac{1}{2}\Delta \hat {\bm W}_j^{n + 1/2} \cdot {\bf{J}}_{{{\rm E}},ij,i} \\
 & + \frac{1}{2}\varrho _{ij,i}\left( {\Delta \hat {\bm W}_i^{n + 1/2} - \Delta \hat {\bm W}_j^{n + 1/2}} \right).
\end{aligned}
\end{equation}
Note that the negative signs in front of the terms with ${\bf{J}}_{{{\rm E}},ij,i}$ means that the characteristic direction of the adjoint equation is opposite to that of the primal equation. The spectral radius $\varrho _{ij,i}$ is 
\begin{equation}
\varrho _{ij,i} = \left| {{\bm u_i} \cdot {\bm n_{ij}}} \right| + {a_i} + 2\frac{{{\mu _i}}}{{{\rho _i}\left| {{\bm x_i} - {\bm x_j}} \right|}}.
\end{equation}
Finally, substituting Eq.~\eqref{eqn:numerad_macfluxRoe} into Eq.~\eqref{eqn:numerad_macinc2}, and noting that $\Delta \hat {\bm W}_i^{n + 1/2} \cdot \sum\nolimits_{j \in N(i)} {{A_{ij}}{{\bf{J}}_{{\rm E},ij,i}}}  = \bm 0$,  we will get
\begin{equation}\label{eqn:numerad_macupdate}
\begin{aligned}[b]
& \Delta \hat {\bm W}_i^{n + 1/2} \cdot \left( {\frac{{\Delta {\Omega _i}}}{{\Delta t}} - \frac{{\Delta {\Omega _i}}}{{\tau _{\theta ,i}}}\left( {{{\bf{\Theta }}_i} - 1} \right) + \frac{1}{2}\sum\limits_{j \in N\left( i \right)} {{A_{ij}}\varrho _{ij,i}} } \right)\\
 =  & \hat {\bm R}_i^n + \frac{1}{2}\sum\limits_{j \in N\left( i \right)} {\left( {{A_{ij}}\Delta \hat {\bm W}_j^{n + 1/2} \cdot {{\bf{J}}_{{\rm E},ij,i}}} \right)}  + \frac{1}{2}\sum\limits_{j \in N\left( i \right)} {{A_{ij}}\varrho _{ij,i}\Delta \hat {\bm W}_j^{n + 1/2}} .
\end{aligned}
\end{equation}
Note that the form of Eq.~\eqref{eqn:numerad_macupdate} is similar to Eq.~\eqref{eqn:numerpr_macupdate}, but the operations of left multiplications and right multiplications are reversed. Likewise, Eq.~\eqref{eqn:numerad_macupdate} is solved by the SGS method, and then the predicted macroscopic variable $\hat {\bm W}_i^{n + 1/2}$ is obtained to calculate an approximate $\bm \phi _{{\rm eq},i,k}^{n + 1/2}$. For $\bm \phi _{\tau ,i}^{n + 1/2}$, we directly set  $\bm \phi _{\tau ,i}^{n + 1/2} = \bm \phi _{\tau ,i}^n$ according to the continuum-limit approximation~\eqref{eqn:numerad_macincce}.

\subsubsection{Overall calculation procedure}
Substituting the approximate $\bm \phi _{{\rm eq},i,k}^{n + 1/2}, \bm \phi _{\tau ,i}^{n + 1/2}$ into the discretized adjoint equation \eqref{eqn:numerad_gov} and arranging the equation into the incremental form will yield
\begin{equation}
\begin{aligned}[b]
& \left( {\frac{{\Delta {\Omega _i}}}{{\Delta t}} + \frac{{\Delta {\Omega _i}}}{{\tau _{\theta ,i}}}} \right)\Delta \bm \phi _{i,k}^{n + 1} - \sum\limits_{j \in N\left( i \right)} {{A_{ij}}{\bm v_k} \cdot {\bm n_{ij}}\Delta \bm \phi _{ij,k}^{n + 1}} \\
 =  & \Delta {\Omega _i}\frac{{\bm \phi _{{\rm eq},i,k}^{n + 1/2} - \bm \phi _{i,k}^n}}{{\tau _{\theta ,i}}} + \Delta {\Omega _i}\bm \phi _{\tau ,i,k}^{n + 1/2} + \Delta {\Omega _i}\bm \phi _{J,i,k} + \sum\limits_{j \in N\left( i \right)} {{A_{ij}}{\bm v_k} \cdot {\bm n_{ij}}\bm \phi _{ij,k}^n}. 
\end{aligned}
\end{equation}
For $\phi _{ij,k}^n$ at the interface, it is calculated by Eq.~\eqref{eqn:numerad_phiij}, and its increment $\Delta \bm \phi _{ij,k}^{n + 1}$ is simply obtained by the first-order upwind scheme. Finally the equation for the increment of the adjoint variable $\Delta \bm \phi _{i,k}^{n + 1}$ can be formulated as
\begin{equation}\label{eqn:numerad_update}
\begin{aligned}[b]
& \left( {\frac{{\Delta {\Omega _i}}}{{\Delta t}} + \frac{{\Delta {\Omega _i}}}{{\tau _{\theta ,i}}} - \sum\limits_{j \in N_k^ - (i)} {{A_{ij}}{\bm v_k} \cdot {\bm n_{ij}}} } \right)\Delta \bm \phi _{i,k}^{n + 1}\\
 =  & \Delta {\Omega _i}\frac{{\bm \phi _{{\rm eq},i,k}^{n + 1/2} - \bm \phi _{i,k}^n}}{{\tau _{\theta ,i}}} + \Delta {\Omega _i}\bm \phi _{\tau ,i,k}^{n + 1/2} + \Delta {\Omega _i}\bm \phi _{J,i,k} + \sum\limits_{j \in N\left( i \right)} {{A_{ij}}{\bm v_k} \cdot {\bm n_{ij}}\bm \phi _{ij,k}^n}  \\
 &+ \sum\limits_{j \in N_k^ + (i)} {{A_{ij}}{\bm v_k} \cdot {\bm n_{ij}}\Delta \bm \phi _{j,k}^{n + 1}},
\end{aligned}
\end{equation}
which is solved by the SGS method and then the adjoint variable can be updated to  $\bm \phi _{i,k}^{n + 1}$.

Now we will give a summary about the solving of the adjoint numerical system. Suppose we have already obtained the primal variables such as $\bm f_{i,k},\bm W_i,\tau _{\theta ,i}$ from the steady-state solution of the discretized primal equation \eqref{eqn:numerpr_gov}, then the computation procedure for solving the discretized adjoint equation \eqref{eqn:numerad_gov} from step $n$ to step $n+1$ is:
\begin{description}
	\item[Step 1.] Reconstruct the data of step $n$ and calculate the interface adjoint variable $\bm \phi _{ij,k}^n$ by Eq.~\eqref{eqn:numerad_phiij}.
	\item[Step 2.] Calculate $\hat {\bm R}_i^n$ by Eq.~\eqref{eqn:numerad_macrsd}, in which all integrations with respect to the molecular velocity space are computed by numerical integrations similar to Eq.~\eqref{eqn:numerpr_intflux}.
	\item[Step 3.] Solve Eq.~\eqref{eqn:numerad_macupdate} by the SGS iterations to obtain the predicted adjoint macroscopic variable $\hat {\bm W}^{n+1/2}_i$.
	\item[Step 4.] Calculate $\bm \phi _{{\rm eq},i,k}^{n + 1/2}, \bm \phi _{\tau ,i}^{n + 1/2}$ and solve Eq.~\eqref{eqn:numerad_update} by the SGS iterations, and finally the adjoint variable $\bm \phi_{i,k}^{n + 1}$ for the next time step can be obtained.
\end{description}

After obtaining the steady-state solution of the discretized adjoint equation \eqref{eqn:numerad_gov}, the sensitivity of the objective $J$ with respect to the design variable $\vartheta_i$ can be further obtained according to Eqs.~\eqref{eqn:sens_theta}, \eqref{eqn:sens_vartheta1} and \eqref{eqn:sens_vartheta2}. Among these equations, Eq.~\eqref{eqn:sens_theta} and Eq.~\eqref{eqn:sens_vartheta1} are explicit expressions, while Eq.~\eqref{eqn:sens_vartheta2} is a simple elliptic PDE and we solve it by the finite volume method; the latter is a standard process in numerical simulations and the details are omitted for simplicity.

\section{Numerical results}\label{sec:test}
\subsection{Validation of the fictitious porosity model}\label{sec:case0}

\begin{figure}[h]
	\centering
	\subfigure[Mesh used for the fictitious porosity model, 30360 cells in total.]{\includegraphics[width=0.9\textwidth]{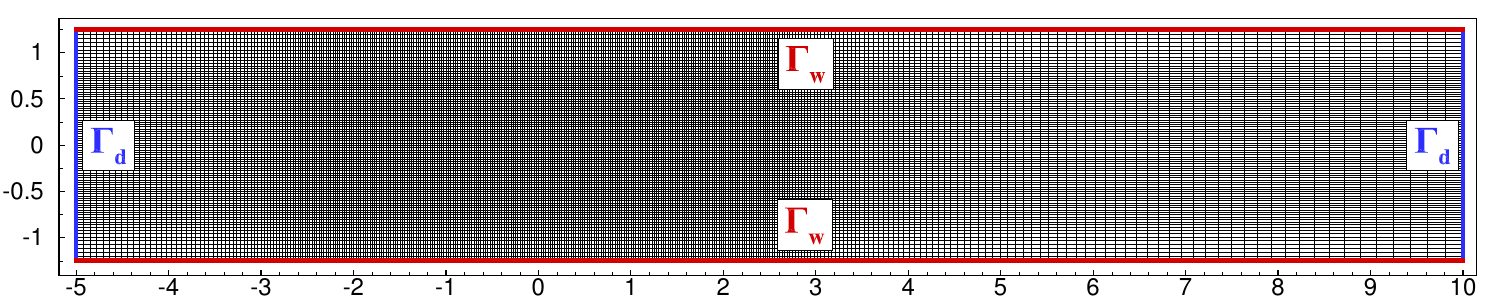}}\\
	\subfigure[Body-fitted mesh used for comparsion, 29478 cells in total.]{\includegraphics[width=0.9\textwidth]{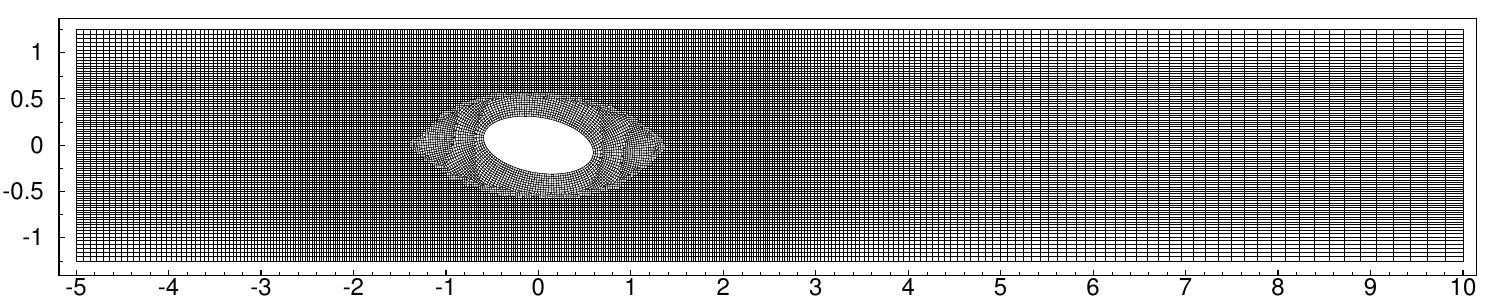}}
	\caption{\label{fig:case0_mesh}The setups of  mesh and boundary condition for the flow passing over an elliptic cylinder inside a channel.}
\end{figure}

\begin{figure}[t]
	\centering
	\includegraphics[width=0.45\textwidth]{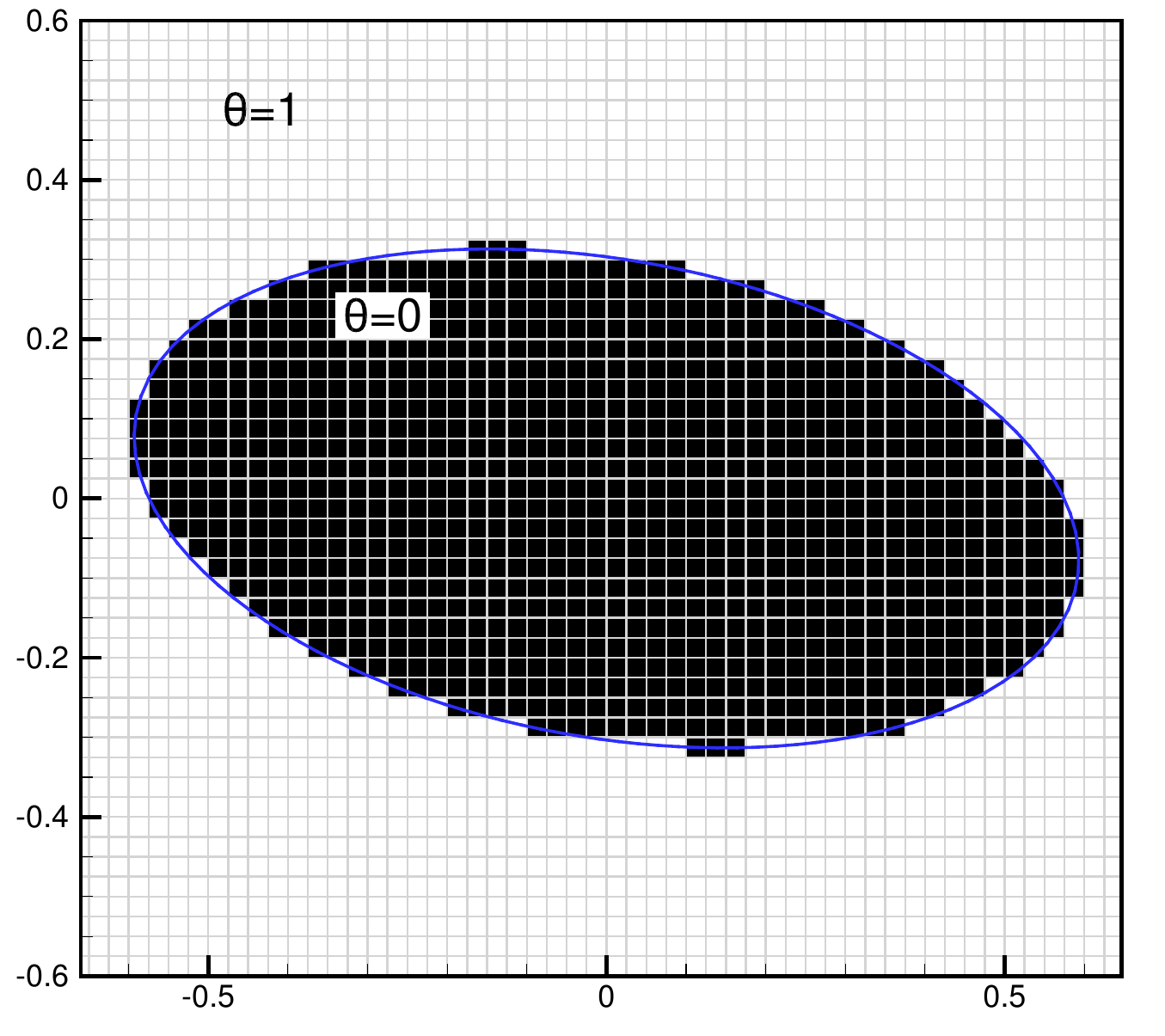}
	\caption{\label{fig:case0_mesh_theta} The distribution of material density $\theta$  for the fictitious porosity model, in the simulation of gas flow passing over an elliptic cylinder inside a channel.  }
\end{figure}

In this section, the flow passing over an elliptic cylinder inside a channel is simulated to validate the fictitious porosity model in Section \ref{sec:formula_poro} and the numerical method in Section \ref{sec:method_pr}. For comparison, the simulation under the same condition but adopting the body-fitted mesh has also been performed. The mesh and boundary condition are illustrated in Fig.~\ref{fig:case0_mesh}. As shown in the figure, the channel has a domain of $\{ \bm x|{x_1} \in [ - 5,10],{x_2} \in [ - 1.25,1.25]\} $ where the center of the cylinder is placed at $(0,0)$. The elliptic cylinder has a width of $d=1.2$ and height of $0.6$ and the angle of attack is $10^\circ$. At the inlet/outlet boundary ${\Gamma _{\rm d}}$ the following Dirichlet boundary condition is imposed:
\begin{equation}\label{eqn:bc_dirichlet}
\bm f = {\bm g_{\rm M}}({\rho _\infty },{\bm u_\infty },{T_\infty })\quad {\rm{in}}\quad {\Gamma _{\rm d}} \times {\Xi ^ - },
\end{equation}
where the state variables ${\rho _\infty },{\bm u_\infty },{T_\infty }$ correspond to a Ma number of $0.6$. At the upper/lower wall ${\Gamma _{\rm w}}$, the diffuse boundary condition as described in Eq.~\eqref{eqn:formula_bgk} and Eq.~\eqref{eqn:formula_adjointbgk} is applied, which has the condition ${\bm u_{\rm w}} = 0,{T_{\rm w}} = {T_\infty }$. For the mesh resolution, as shown in Fig.~\ref{fig:case0_mesh}, the mesh near the cylinder has been refined, and the mesh independence has been ensured (for the body-fitted mesh the variation of the drag force is less than 1\% if refining the mesh further). For the fictitious porosity model, the distribution of the material density $\theta$ adopts the nearest-neighbor interpolation to represent the elliptic cylinder, namely if the center of the cell is inside the cylinder then set $\theta_i=0$ otherwise set $\theta_i=1$, as shown in Fig.~\eqref{fig:case0_mesh_theta}. Note that here we directly give the material density $\theta$ which determines the actual gas-solid distribution, as discussed in the last paragraph of Section \ref{sec:formula_poro}. A wide range of flow regimes are considered, including ${\rm Kn}=0.5,10$ and ${\rm Re}=200$ (corresponding to ${\rm Kn}\approx 5\times 10^{-3}$), where Re and Kn are defined by Eqs.~\eqref{eqn:kndefine} and \eqref{eqn:maredefine} with the inlet  condition ${\rho _\infty },{\bm u_\infty },{T_\infty }$ and the cylinder width $d$.
 For the discretization
of the molecular velocity space, $20\times20,30\times30,60\times60$ uniform meshes in the molecular velocity range $[ - 6{a_\infty },6{a_\infty }]$ have been employed for the cases ${\rm Re}=200,{\rm Kn}=0.5, {\rm Kn}=10$ respectively, where $a_\infty$ is the acoustic speed.

\begin{figure}[t]
\centering
\subfigure[${\rm Re}=200$]{\includegraphics[width=0.4\textwidth]{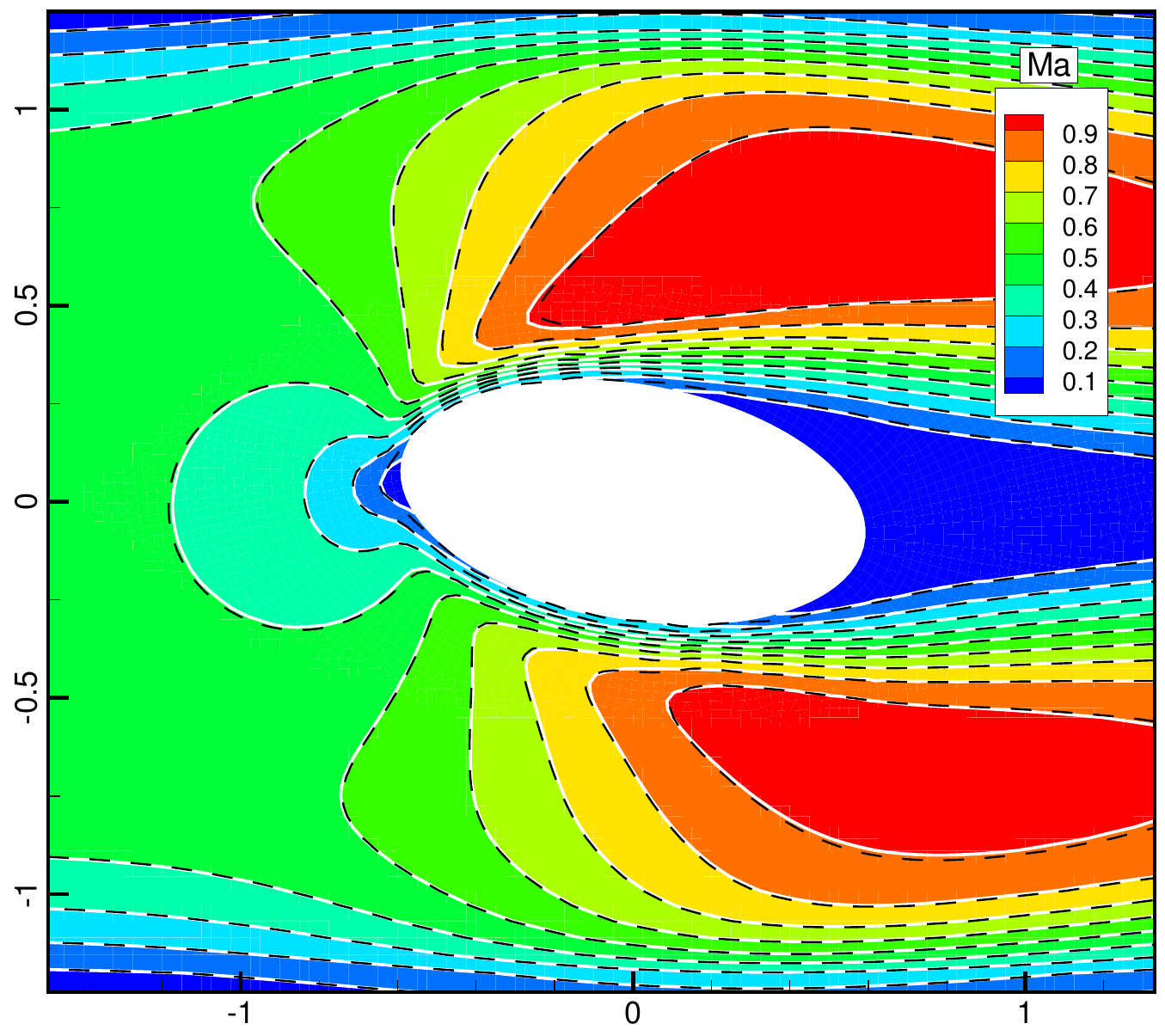}}\hspace{0.02\textwidth}
\subfigure[${\rm Kn}=0.5$]{\includegraphics[width=0.4\textwidth]{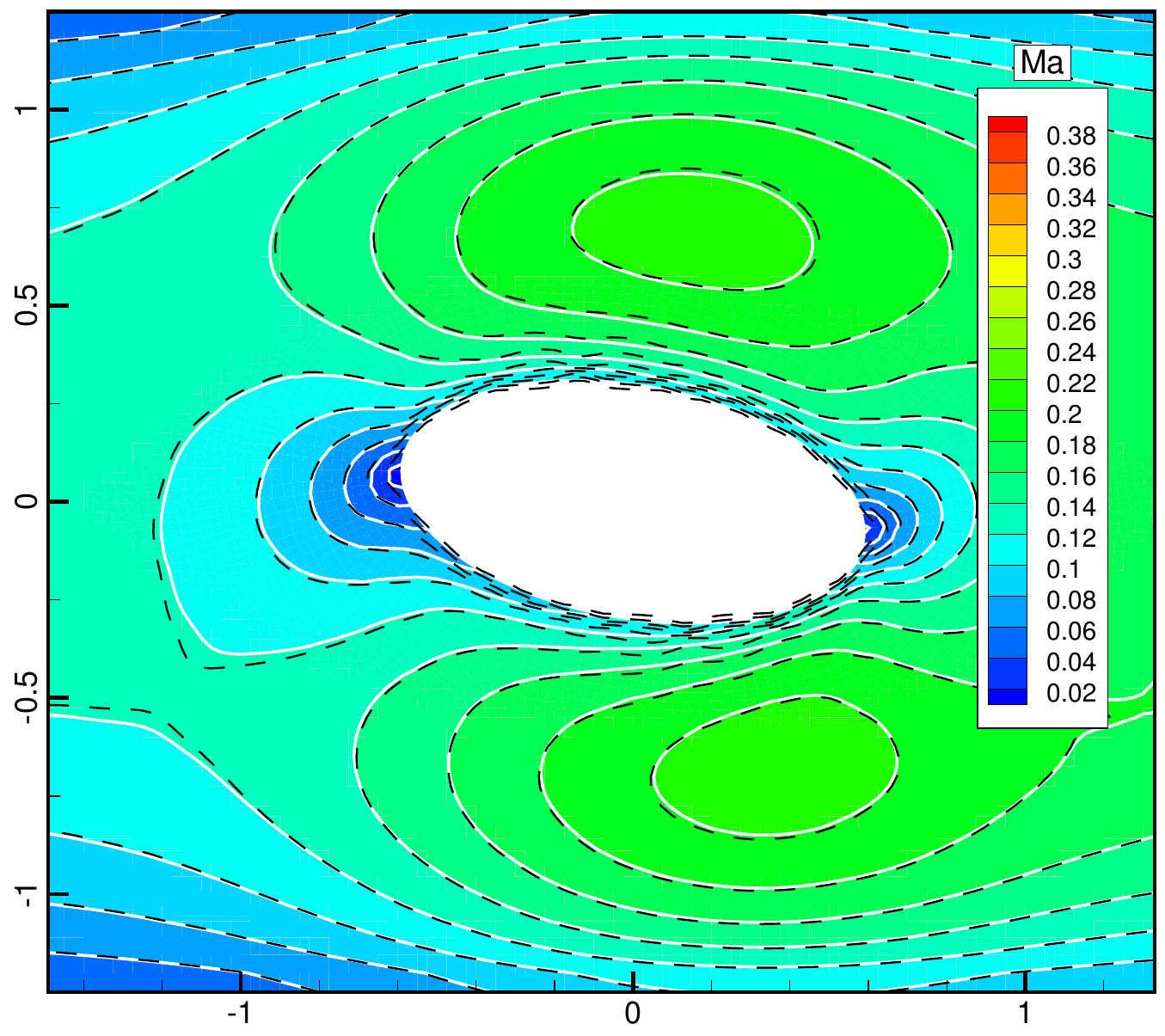}}
\\
\subfigure[${\rm Kn}=10$]{\includegraphics[width=0.4\textwidth]{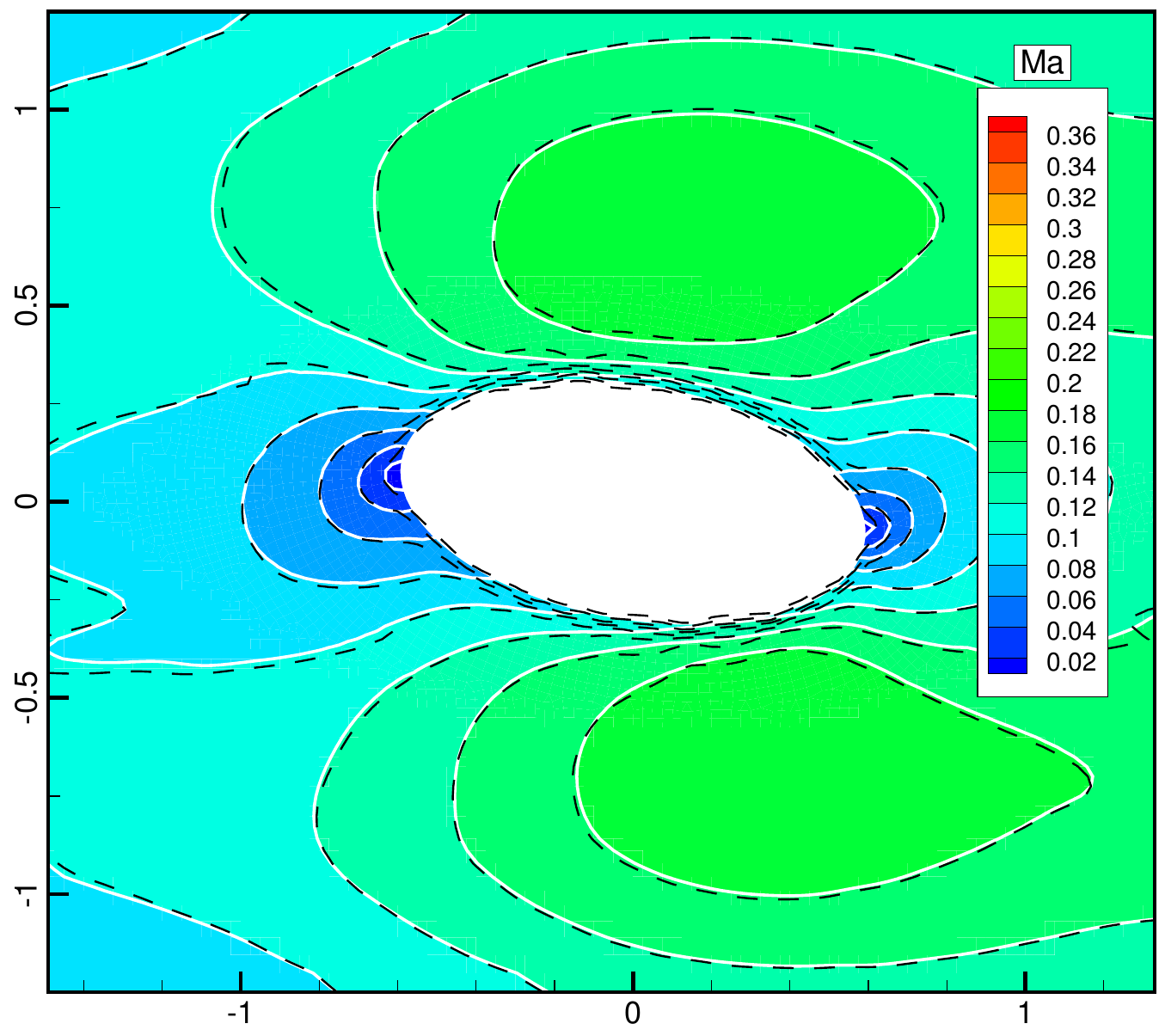}}
\caption{\label{fig:case0_ma}Mach number distribution of the flow over an elliptic cylinder inside a channel. Color bands: results of the body-fitted mesh, dashed lines: results of the fictitious porosity model.}
\end{figure}

Fig.~\ref{fig:case0_ma} shows the Mach number contours calculated by the fictitious porosity model and the body-fitted mesh (reference solution). It  is seen that the two results almost overlap with each other in different flow regimes, showing good consistence. Table \ref{tab:case0_cdcl} shows the drag coefficient $C_{\rm d}$ and the lift coefficient $C_{\rm l}$, which are defined as
\begin{equation}\label{eqn:cdcldefine}
\left(
\begin{aligned}
{C_{\rm d}}\\
{C_{\rm l}}
\end{aligned}
\right) = \frac{{\bm{\mathcal{F}}}}{{\frac{1}{2}{\rho _\infty }\bm u_\infty ^2d}}
\end{equation}
where ${\bm{\mathcal{F}}}$ is the vector of the force  exerted on the cylinder. For the fictitious porosity model, this force ${\bm{\mathcal{F}}}$ can be calculated by integrating the momentum source term on the right-hand side of the primal macroscopic equation \eqref{eqn:macgov_poro} over the design domain $\Omega$, i.e.
\begin{equation}\label{eqn:forcedefine}
\bm{\mathcal{F}} =  - \sum\limits_i {\frac{{{\rho _i}{\bm u_{\theta ,i}} - {\rho _i}{\bm u_i}}}{{{\tau _{\theta ,i}}}}\Delta {\Omega _i}},
\end{equation}
in which the negative sign comes from Newton's third law of motion. It is shown that the forces calculated by the fictitious porosity model have a maximum error of $3.2\%$ relative to the results of the body-fitted mesh. Based on the above facts, it is safe to say that the porosity model is precise to be used for optimization.

\begin{table}[]
\centering
\caption{\label{tab:case0_cdcl}Drag and lift coefficients of the flow over an elliptic cylinder inside a channel.}
\begin{tabular}{|c|ccc|ccc|}
\hline
\multirow{2}{*}{Case} & \multicolumn{3}{c|}{$C_{\rm d}$}	& \multicolumn{3}{c|}{$C_{\rm l}$}	\\ \cline{2-7} 
                      & \multicolumn{1}{c|}{Porosity model}  & \multicolumn{1}{c|}{Body-fitted}  & Error  & \multicolumn{1}{c|}{Porosity model}  & \multicolumn{1}{c|}{Body-fitted}  &  Error \\ \hline
${\rm Re}=200$	& \multicolumn{1}{c|}{1.195} & \multicolumn{1}{c|}{1.178} & 1.4\% & \multicolumn{1}{c|}{0.4096} & \multicolumn{1}{c|}{0.4220} & 2.9\% \\ \hline
${\rm Kn}=0.5$	& \multicolumn{1}{c|}{2.246} & \multicolumn{1}{c|}{2.190} & 2.6\% & \multicolumn{1}{c|}{0.2914} & \multicolumn{1}{c|}{0.2910} & 0.14\% \\ \hline
${\rm Kn}=10$	& \multicolumn{1}{c|}{2.081} & \multicolumn{1}{c|}{2.016} & 3.2\% & \multicolumn{1}{c|}{0.2973} & \multicolumn{1}{c|}{0.3016} & 1.4\% \\ \hline
\end{tabular}
\end{table}

\subsection{Validation of the sensitivity}\label{sec:case1}
In this section, the accuracy of the sensitivity calculated by the numerical method in Section \ref{sec:method_ad} based on the theory in Section \ref{sec:sens} will be verified. The setup of the test case is similar to that in Section \ref{sec:case0}, except that the material density $\theta$  is changed with a small value $\Delta \theta$ in the rectangle region of $\{ \bm x|{x_1} \in [ - 1.5,1.5],{x_2} \in [ - 0.75,0.75]\} $ for the convenience of applying the central deference (see below), as shown in Fig.~\ref{fig:case1_graysetup}. The small change $\Delta \theta$ is set as $\Delta \theta = 5 \times {10^{ - 6}},5 \times {10^{ - 4}},5 \times {10^{ - 2}}$ for the cases ${\rm Re}=200, {\rm Kn}=0.5$, and ${\rm Kn}=10$, respectively. The objective is set as the drag force exerted on the cylinder, namely $J=\mathcal{F}_1$ where $\mathcal{F}_1$ is the horizontal component of $\bm{\mathcal{F}}$ defined in Eq.~\eqref{eqn:forcedefine}. In order to keep to the point, here we only verify the sensitivity $J'_{\theta }$ with respect to the material density $\theta$ (i.e. Eq.~\eqref{eqn:sens_theta}), while the conversion from $J'_{\theta }$ to $J'_{\vartheta }$ (Eq.~\eqref{eqn:sens_vartheta1} and Eq.~\eqref{eqn:sens_vartheta2}) is a mature technology and has already been well validated \cite{sato2019topology,kawamoto2011heaviside}. The sensitivity calculated by the adjoint method is compared with that calculated by the central difference:
\begin{equation}
J_{\theta ,i}'^{\rm FDM} = \frac{{J(...,{\theta _{i - 1}},{\theta _i} + \Delta \theta ',{\theta _{i + 1}},...) - J(...,{\theta _{i - 1}},{\theta _i} - \Delta \theta ',{\theta _{i + 1}},...)}}{{2\Delta \theta '\Delta {\Omega _i}}},
\end{equation}
where the difference step size $\Delta \theta '$ is set as $\Delta \theta ' = 1 \times {10^{ - 6}},5 \times {10^{ - 5}},2 \times {10^{ - 3}}$ for the cases ${\rm Re}=200, {\rm Kn}=0.5$, and ${\rm Kn}=10$, respectively.

\begin{figure}[t]
\centering
\includegraphics[width=0.45\textwidth]{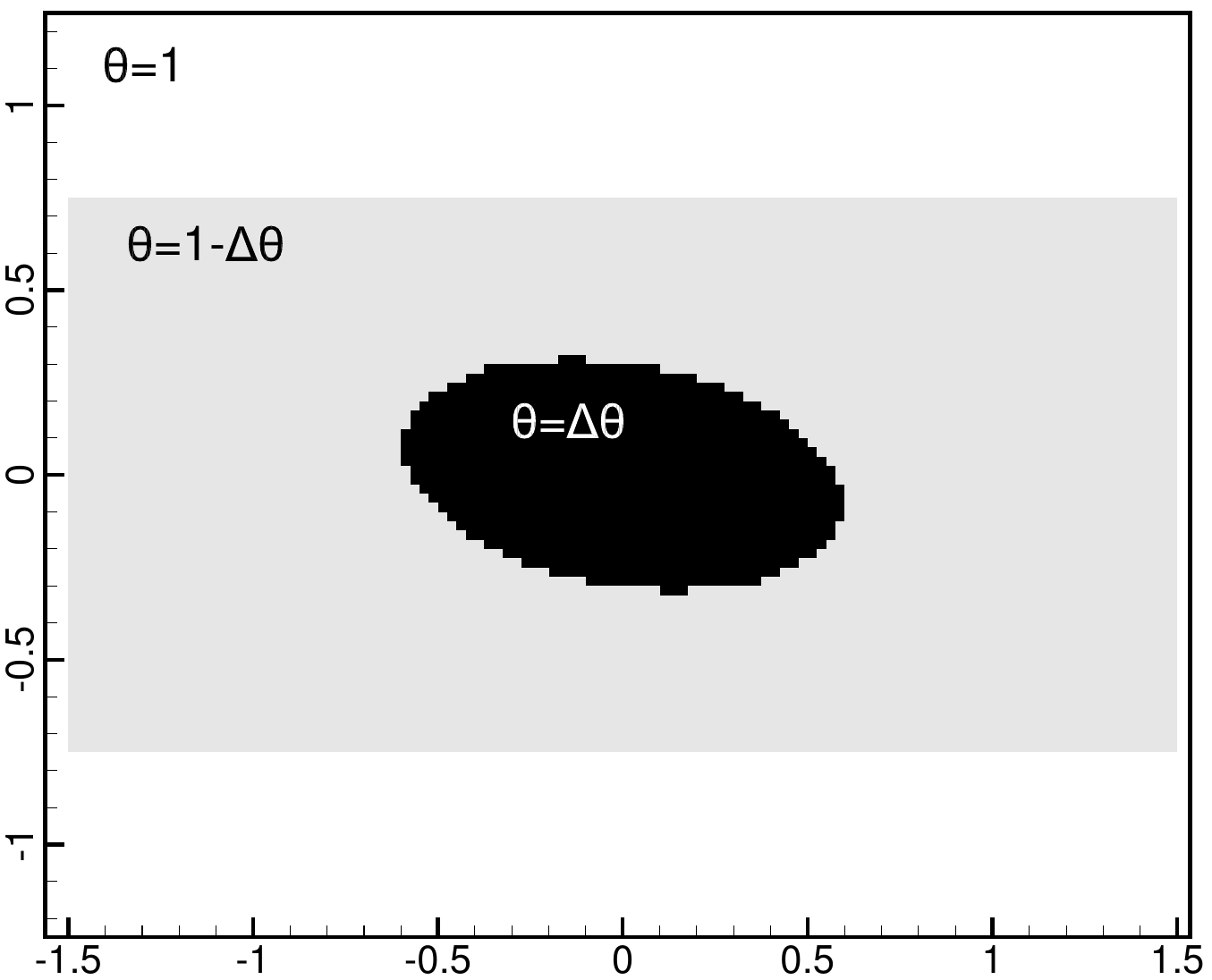}
\caption{\label{fig:case1_graysetup}The distribution of the material density $\theta$ around the elliptic cylinder in the test case for sensitivity validation.  }
\end{figure}

\begin{figure}
\centering
{\subfigure[${\rm Re}=200$]{%
\includegraphics[width=0.4\textwidth]{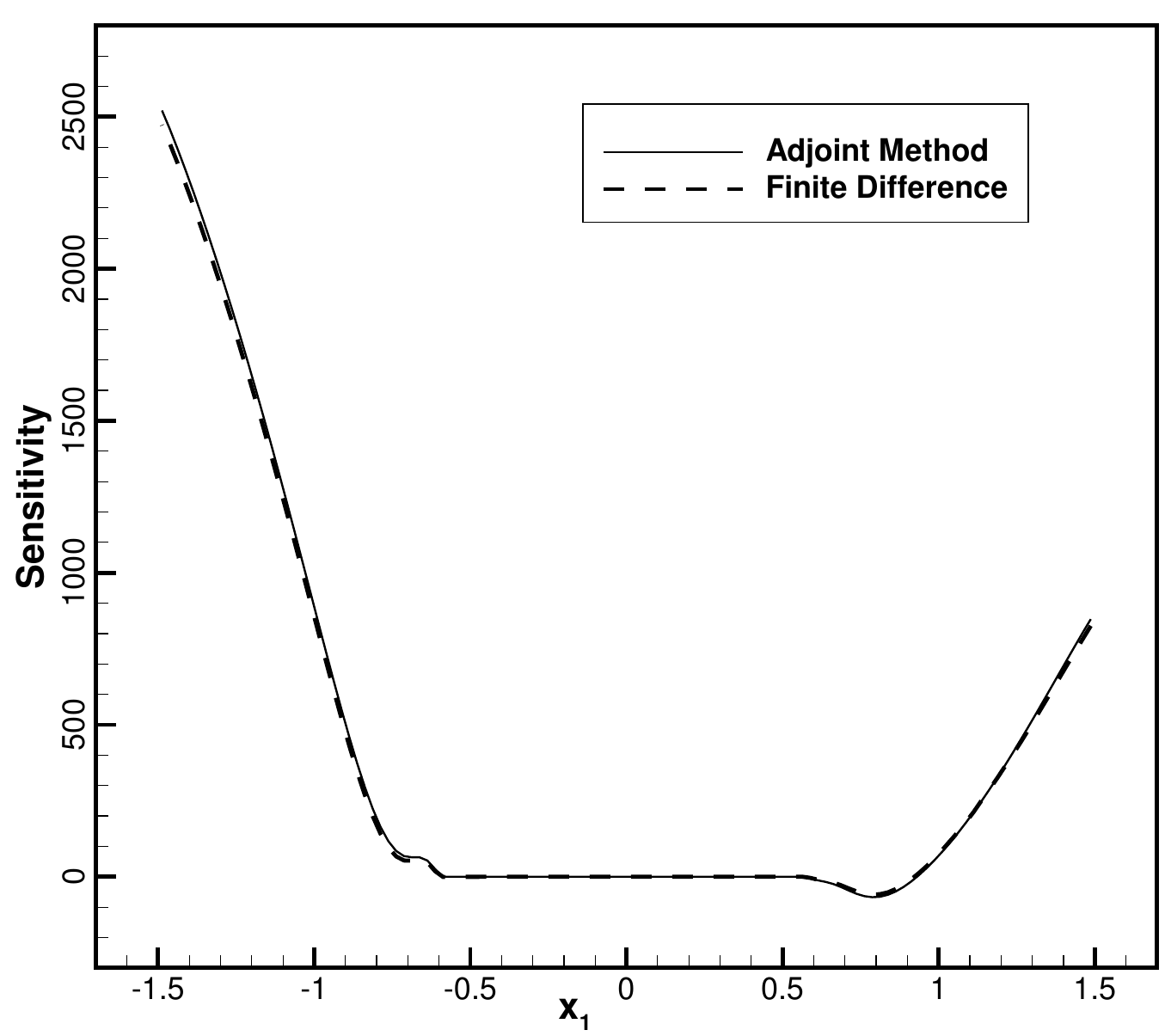}\hspace{0.02\textwidth}%
\includegraphics[width=0.4\textwidth]{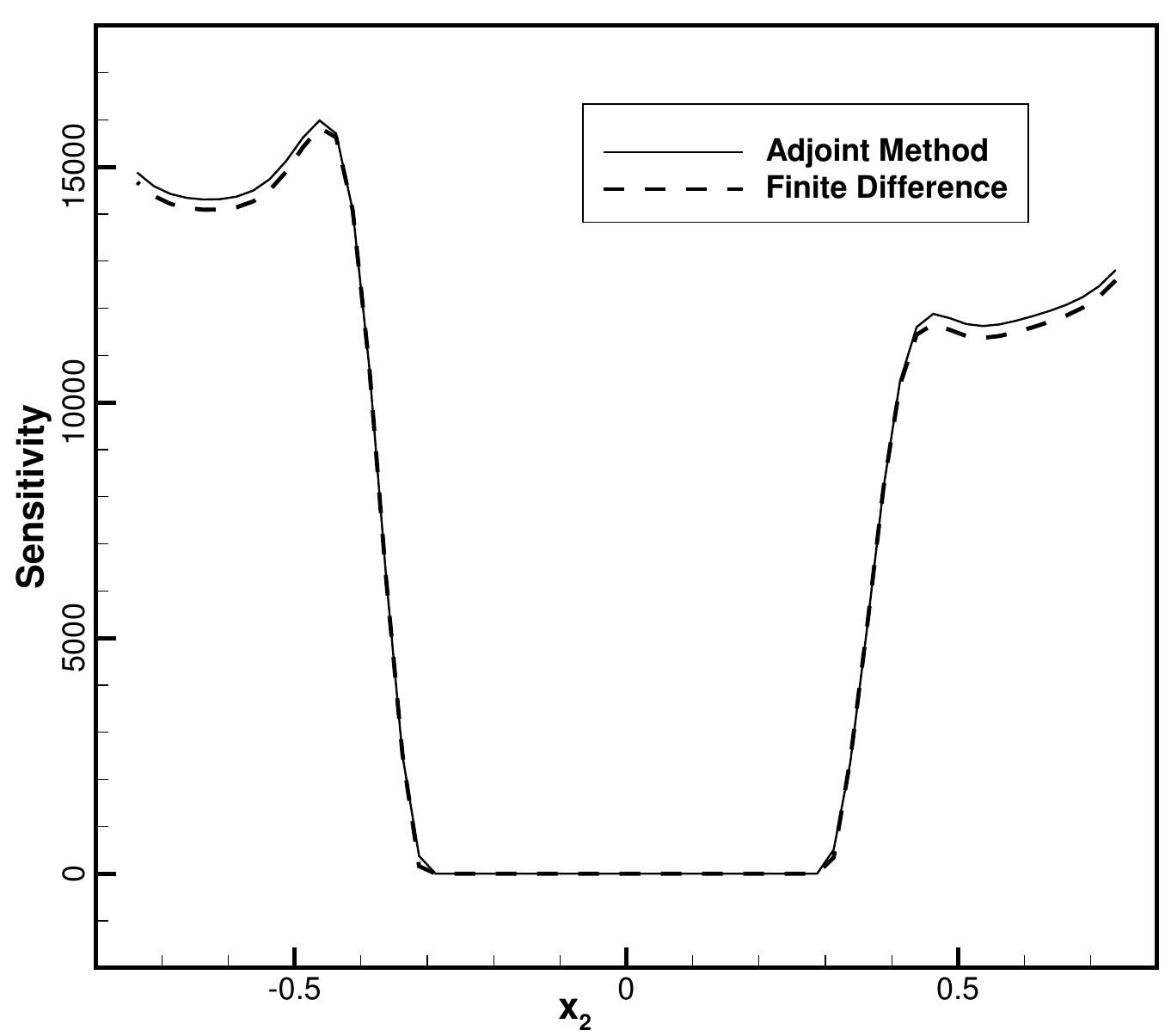}}}\\%
{\subfigure[${\rm Kn}=0.5$]{%
\includegraphics[width=0.4\textwidth]{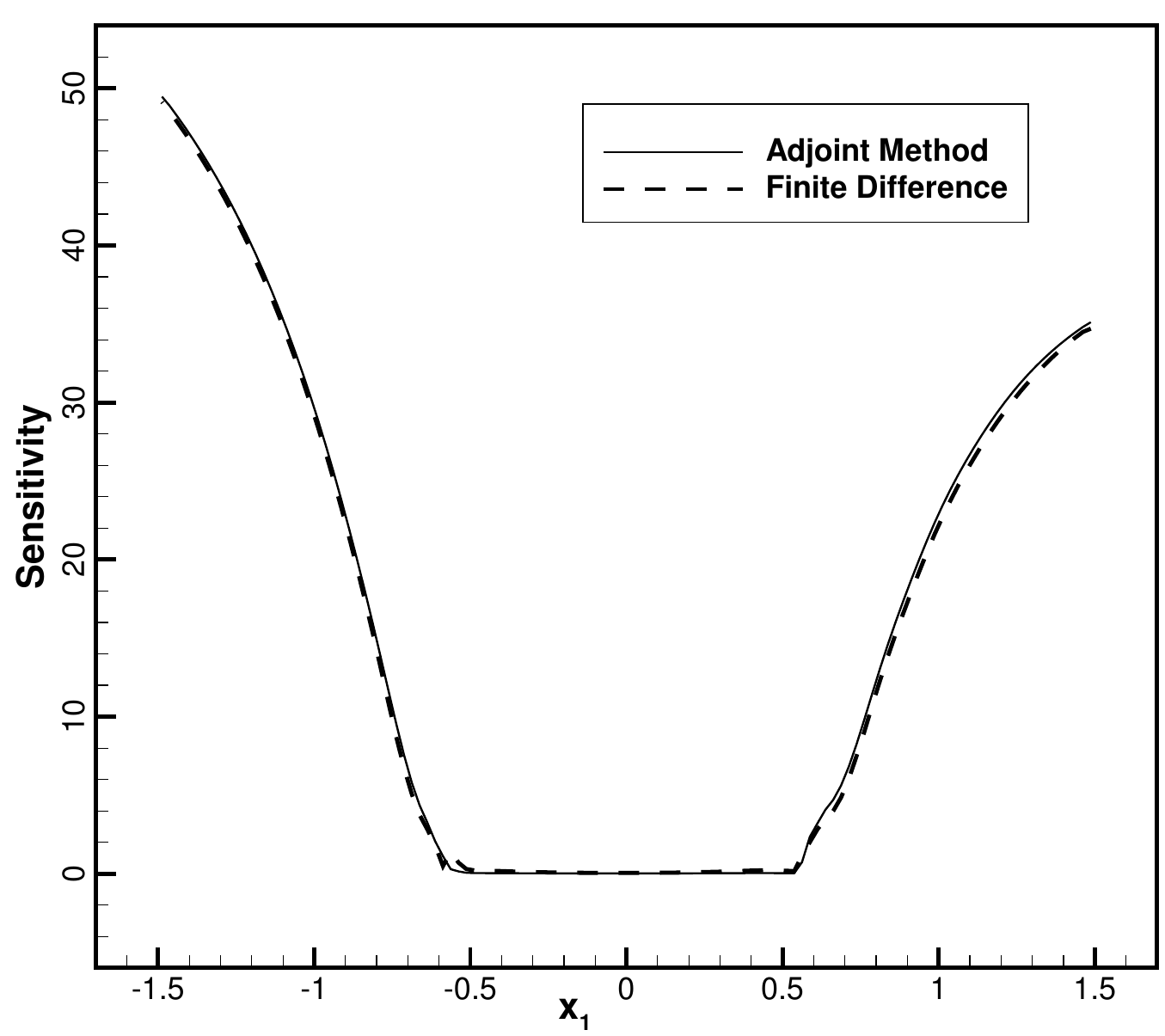}\hspace{0.02\textwidth}%
\includegraphics[width=0.4\textwidth]{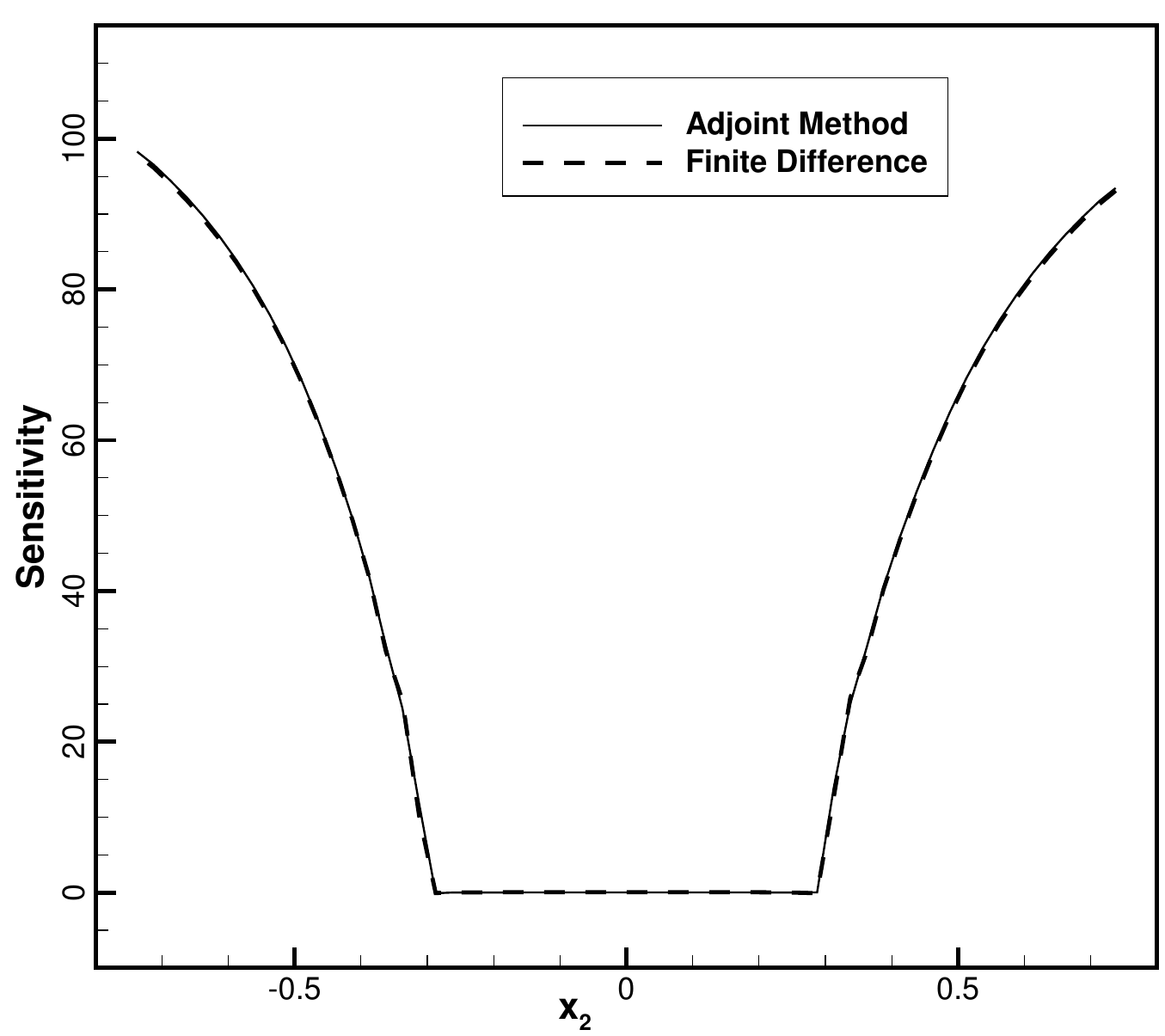}}}\\%
{\subfigure[${\rm Kn}=10$]{%
\includegraphics[width=0.4\textwidth]{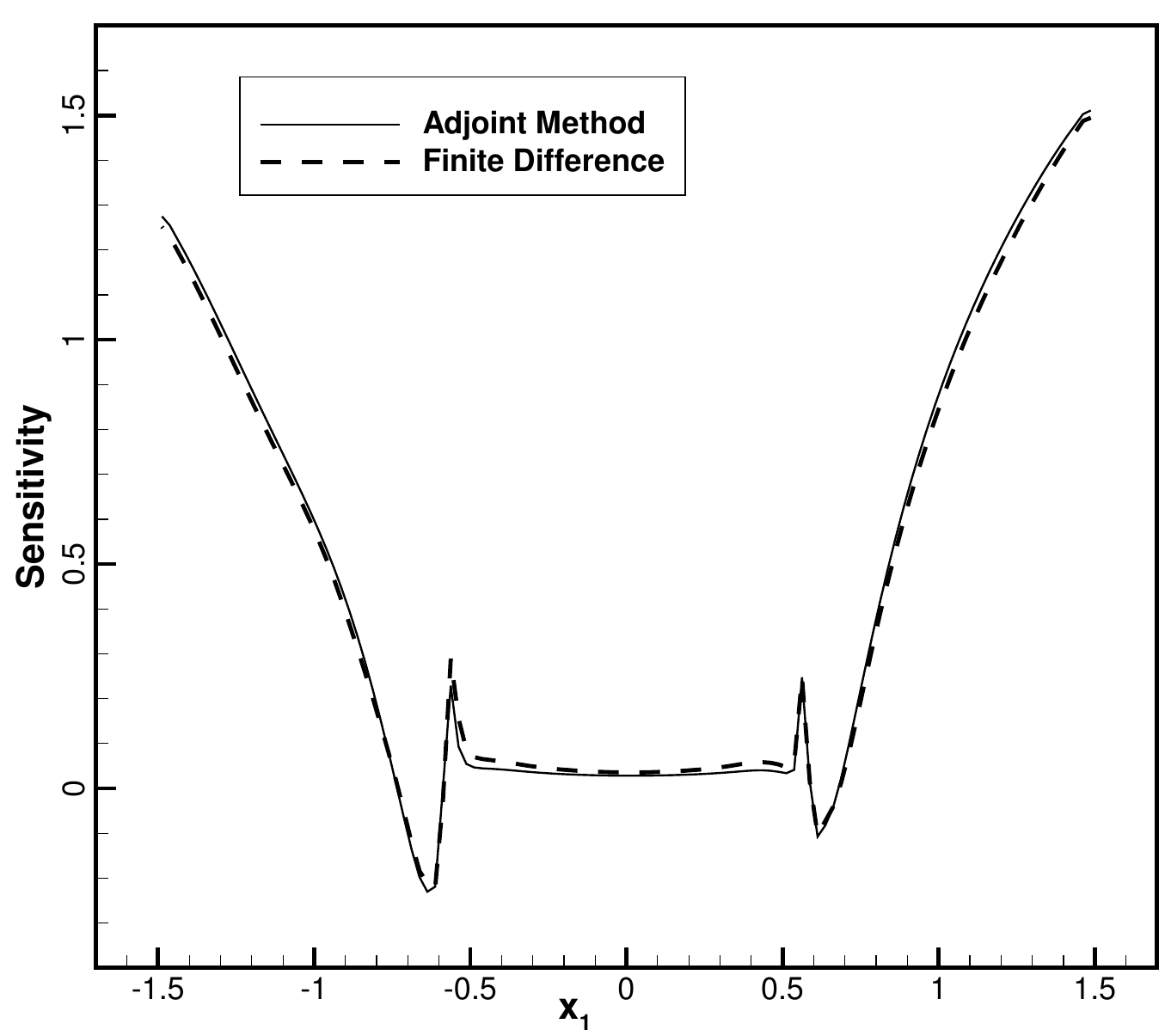}\hspace{0.02\textwidth}%
\includegraphics[width=0.4\textwidth]{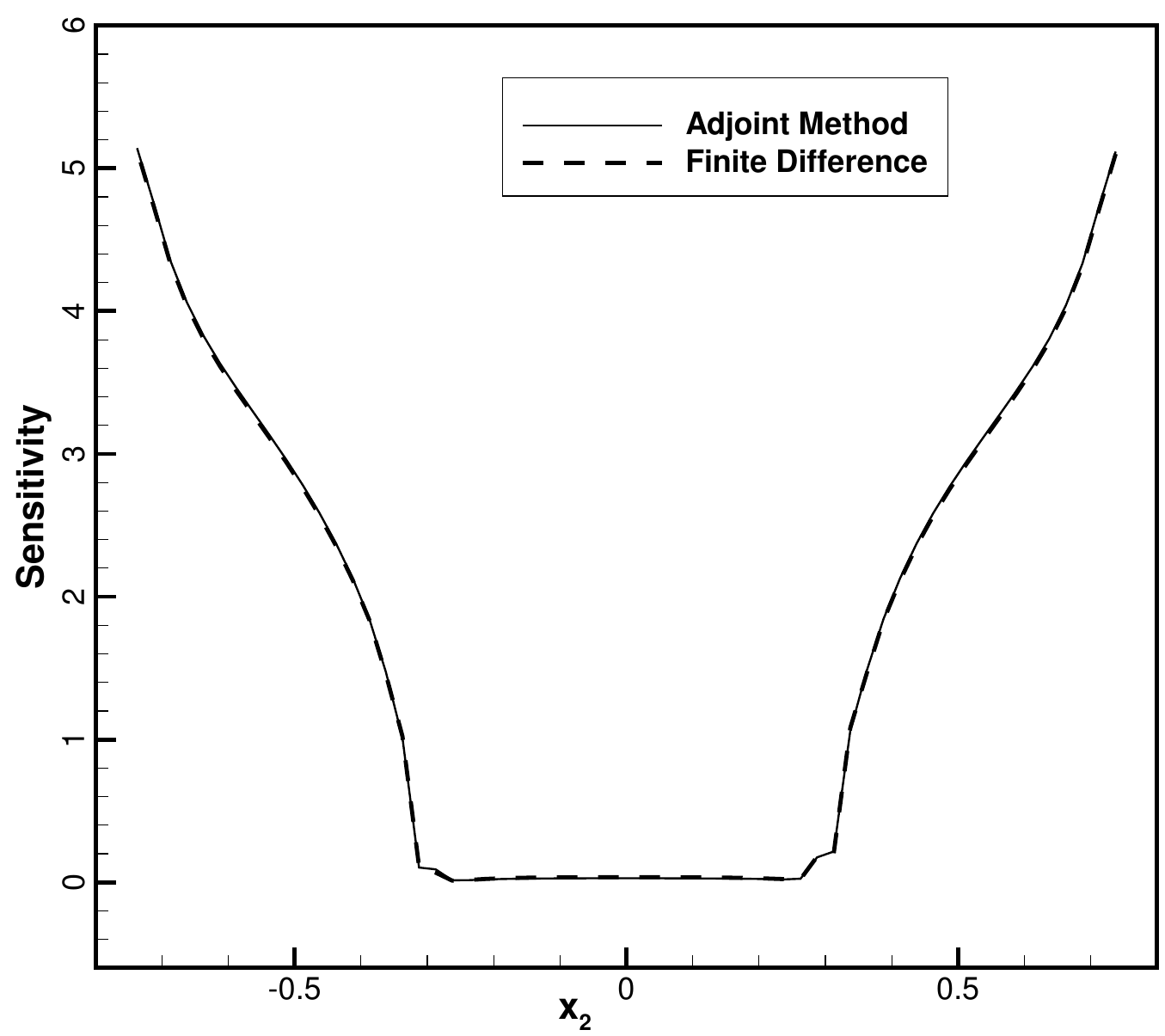}}}
{\caption{\label{fig:case1_hv}Sensitivity profiles along the horizontal line $x_2=0$ (left) and the vertical line $x_1=0$ (right) obtained by the adjoint method and the difference method in the test case for sensitivity validation.}}
\end{figure}

Fig.~\ref{fig:case1_hv} shows the comparison of the sensitivity calculated by the adjoint method and the central difference method. It can be seen that the two sets of sensitivity profiles agree perfectly well with each other, indicating that the present adjoint method has a very good accuracy.

\subsection{Optimization of the airfoil inside a channel in different flow regimes}\label{sec:case2}

\begin{figure}[p]
\centering
\subfigure[Material density distribution of the initial airfoil\label{fig:case2_inia}]{\includegraphics[width=0.47\textwidth]{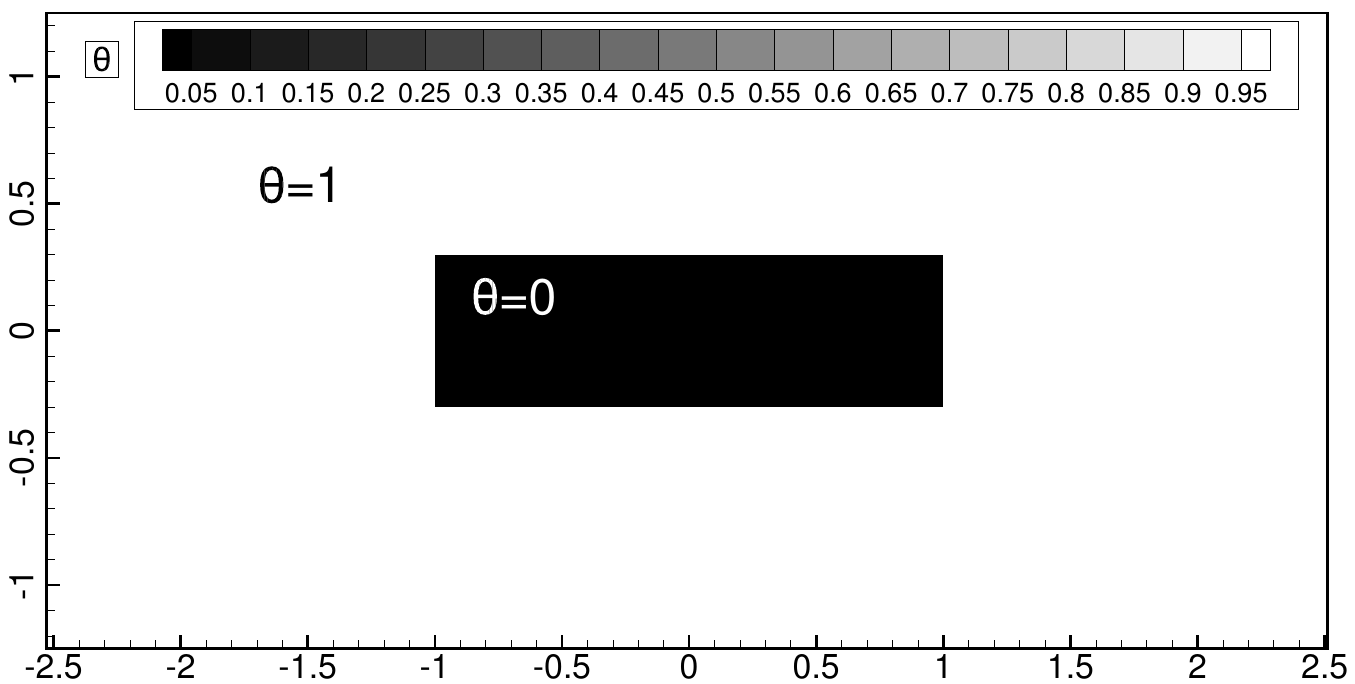}}\hspace{0.02\textwidth}
\subfigure[Material density distribution of the initialized $\vartheta^{(l=0,m=0)}$\label{fig:case2_inib}]{\includegraphics[width=0.47\textwidth]{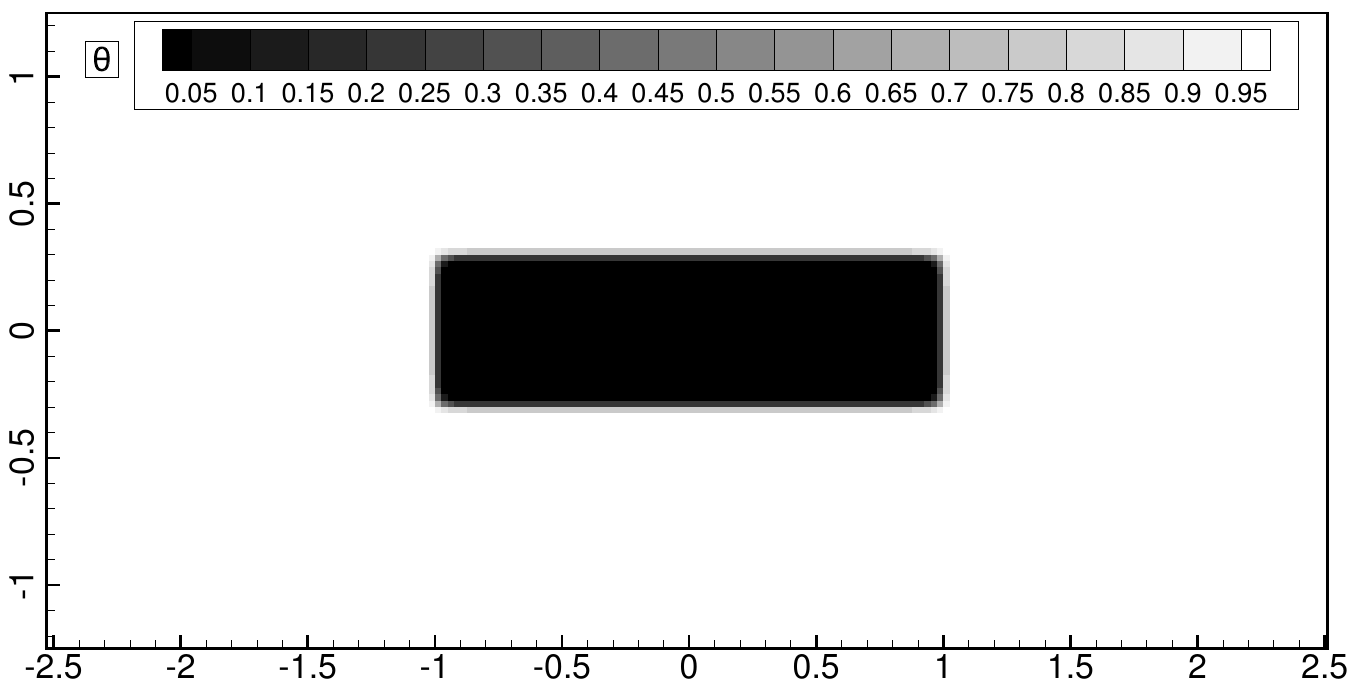}}
\caption{Optimization of the airfoil inside a channel: the initial shape of the airfoil, and the initialized material density distribution to start the optimization procedure.}
\end{figure}

In this section, the whole optimization procedure presented in Section \ref{sec:method_optframework} is verified by the optimization of an airfoil under the flow inside a channel. The setups of the channel, including the flow/boundary condition and the discretization of the physical/velocity space, are the same as that in Section \ref{sec:case0}, and accordingly the mesh independence is ensured. The objective is the same as that in Section \ref{sec:case1}, namely the drag force exerted on the airfoil $J=\mathcal{F}_1$ where $\mathcal{F}_1$ is the horizontal component of $\bm{\mathcal{F}}$ defined in Eq.~\eqref{eqn:forcedefine}. The initial shape of the airfoil is set as a rectangle of the size $2 \times 0.6$ located at $(0,0)$ as shown in Fig.~\ref{fig:case2_inia}, and the reference length of Eqs.~\eqref{eqn:kndefine},\eqref{eqn:maredefine},\eqref{eqn:cdcldefine} is set as $2$ accordingly. The volume constraint with the functional $Q$ in Eq.~\eqref{eqn:formula_opt} is imposed, where $V_{\rm max}=15\times2.5-2\times0.6=36.3$, which means that the minimum area of the airfoil is limited to the area of the initial rectangular airfoil. 

It is worth noting that, although Fig.~\ref{fig:case2_inia} gives the distribution of material density $\theta$ for the initial airfoil, we must have the initial distribution of the design variable $\vartheta$ to start the optimization. At this point, if the initial distribution of $\vartheta$ is directly set to the shape of the initial airfoil, just like the distribution of $\theta$ shown in Fig.~\ref{fig:case2_inia}, then after the processes of filtering (Eq.~\eqref{eqn:intp_filter}), Heaviside projection (Eq.~\eqref{eqn:intp_heaviside}) and material interpolation (Eq.~\eqref{eqn:intp_ramp}), the area of the airfoil obtained from the integration of $\theta$ will be no longer equal to the area of the initial airfoil, and the volume constraint $Q\le0$ may be not satisfied. On the other hand, in our practice, the performance of the MMA algorithm deteriorates when the initial point is far from the feasible region. Therefore, an initialization process is performed to obtain the $\vartheta^{(l=0,m=0)}$ distribution to start the optimization procedure of Section \ref{sec:method_optframework}. In this initialization process we first give the distribution of $\vartheta$ by the shape of the initial airfoil, then an unconstrained optimization problem with the objective $Q^2$ is solved by the MMA algorithm to obtain a solution for $\vartheta$ within the feasible region of $Q\le0$, and this solution will be taken as the initial point $\vartheta^{(l=0,m=0)}$ to start the optimization procedure presented in Section \ref{sec:method_optframework}. This initialization process is quite efficient, which can achieve sufficient accuracy ($\sqrt {{Q^2}}  \approx {10^{ - 8}}$) in less than 20 optimization steps and complete in several seconds. Fig.~\ref{fig:case2_inib} shows the distribution of $\theta^{(l=0,m=0)}$ calculated from the initialized $\vartheta^{(l=0,m=0)}$.

\begin{figure}[p]
	\centering
	\subfigure[${\rm Re}=200$]{\includegraphics[width=0.47\textwidth]{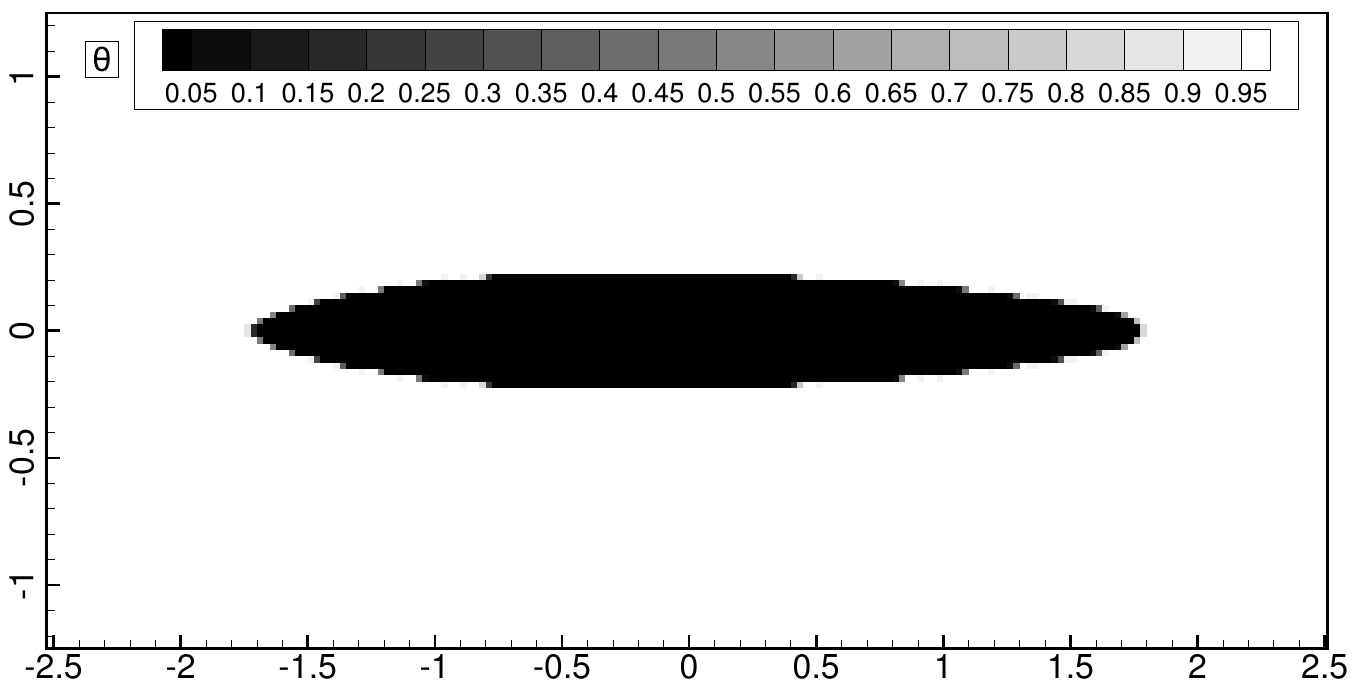}}
	\\
	\subfigure[${\rm Kn}=0.5$]{\includegraphics[width=0.47\textwidth]{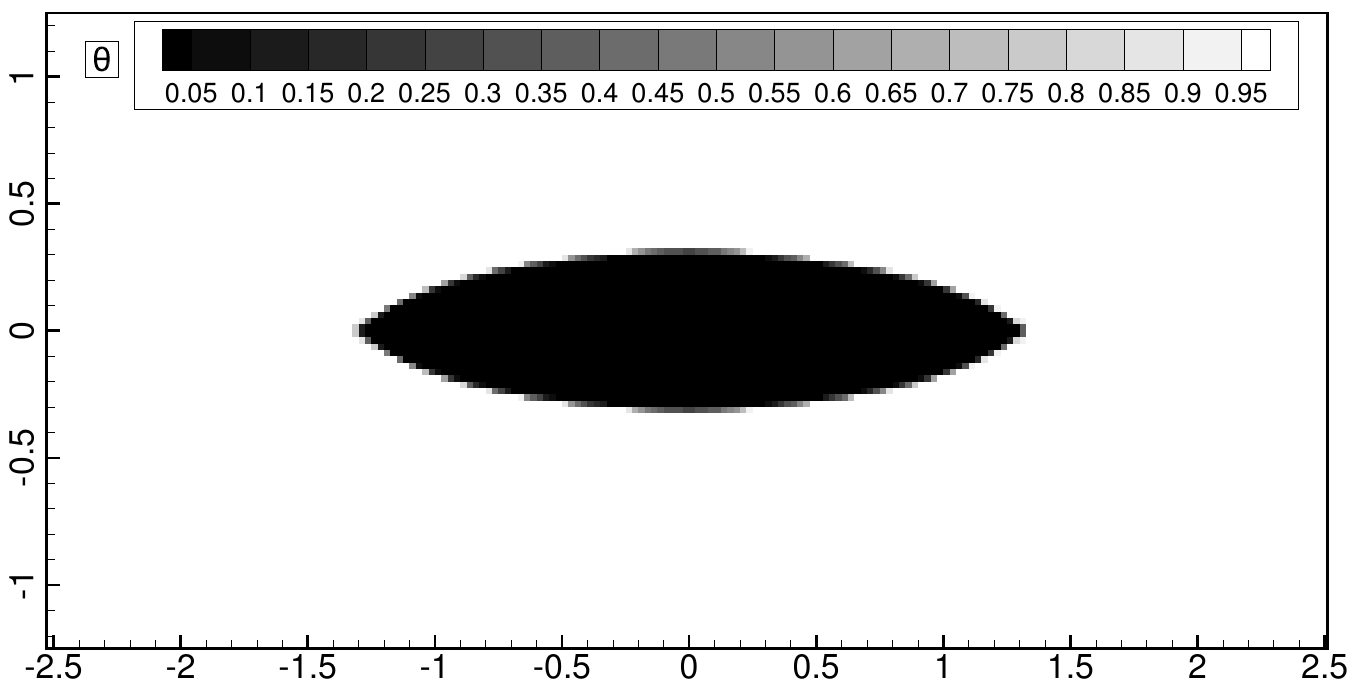}}\hspace{0.02\textwidth}
	\subfigure[${\rm Kn}=10$]{\includegraphics[width=0.47\textwidth]{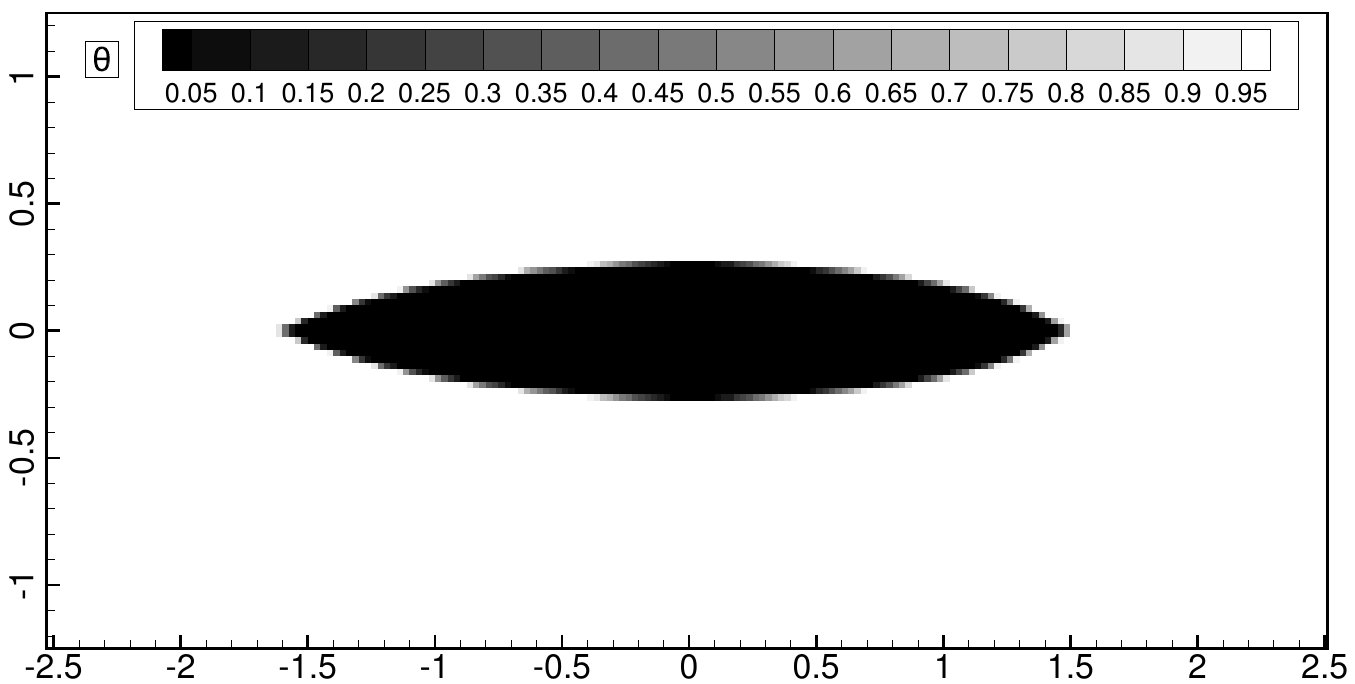}}
	\caption{\label{fig:case2_opttheta}Optimization of the airfoil inside a channel: the optimized airfoils for different flow conditions.}
\end{figure}

When $\vartheta^{(l=0,m=0)}$ is obtained, the optimization procedure of Section \ref{sec:method_optframework} is implemented. Fig.~\ref{fig:case2_opttheta} shows the optimized shapes of airfoils for different flow conditions. For the continuum flow condition ${\rm Re}=200$, the optimized result turns into the shape similar to the classical subsonic airfoil. In comparison, for the rarefied conditions ${\rm Kn}=0.5$ and 10, the optimized results tend to be the biconvex circular-arc airfoils with sharp leading and trailing edges. One interesting phenomenon is that, when Kn increases, the thickness of the optimized airfoil increases first and then decreases. We think this results from the change of the proportion between the friction drag and the pressure drag. From ${\rm Re}=200$ to ${\rm Kn}=0.5$, due to the increase of the gas viscosity, the optimized airfoil tends to become shorter and thicker to reduce its surface exposed to the gas flow to decrease the friction drag. From ${\rm Kn}=0.5$ to ${\rm Kn}=10$, due to the increase of the velocity slip on the gas-solid interface (see the discussion at the end of Section~\ref{sec:kinetictheory}), the proportion of the friction drag relative to the pressure drag decreases, so the optimized airfoil tends to become thinner and longer to decrease the pressure drag.

\begin{figure}[t]
	\centering
	{\subfigure[${\rm Re}=200$]{%
			\includegraphics[width=0.45\textwidth]{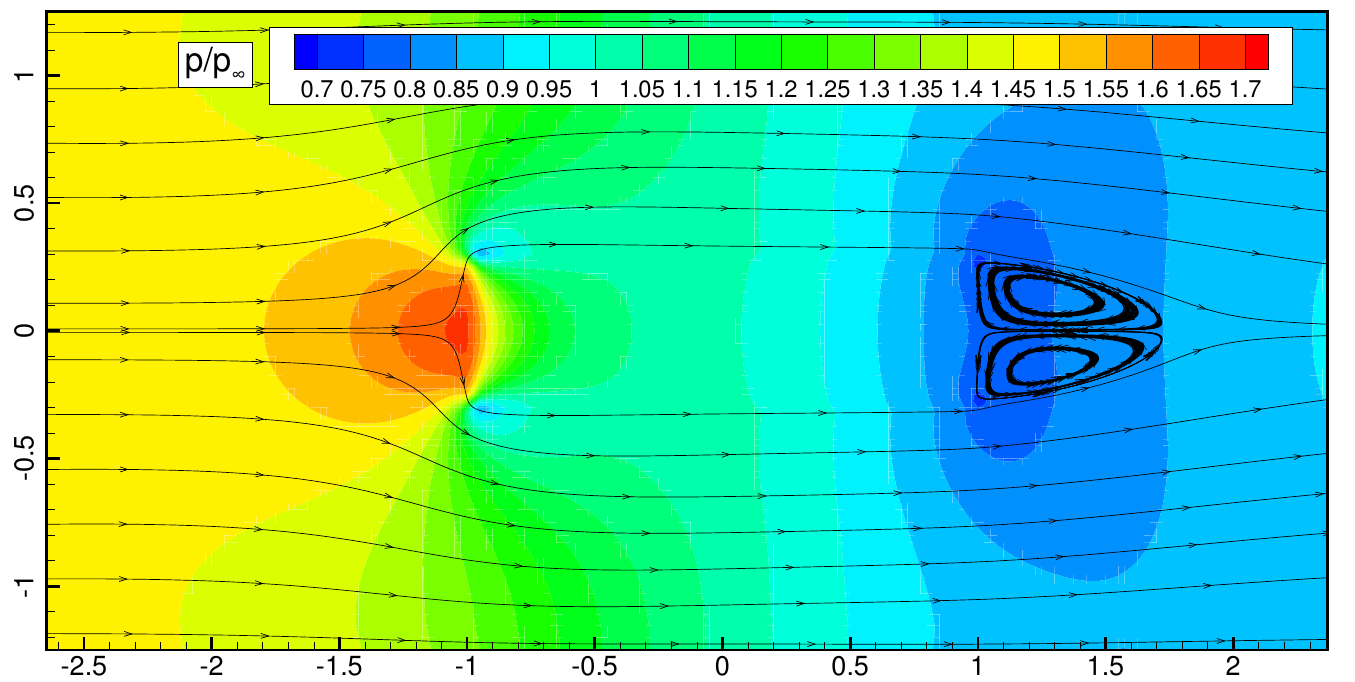}\hspace{0.02\textwidth}%
			\includegraphics[width=0.45\textwidth]{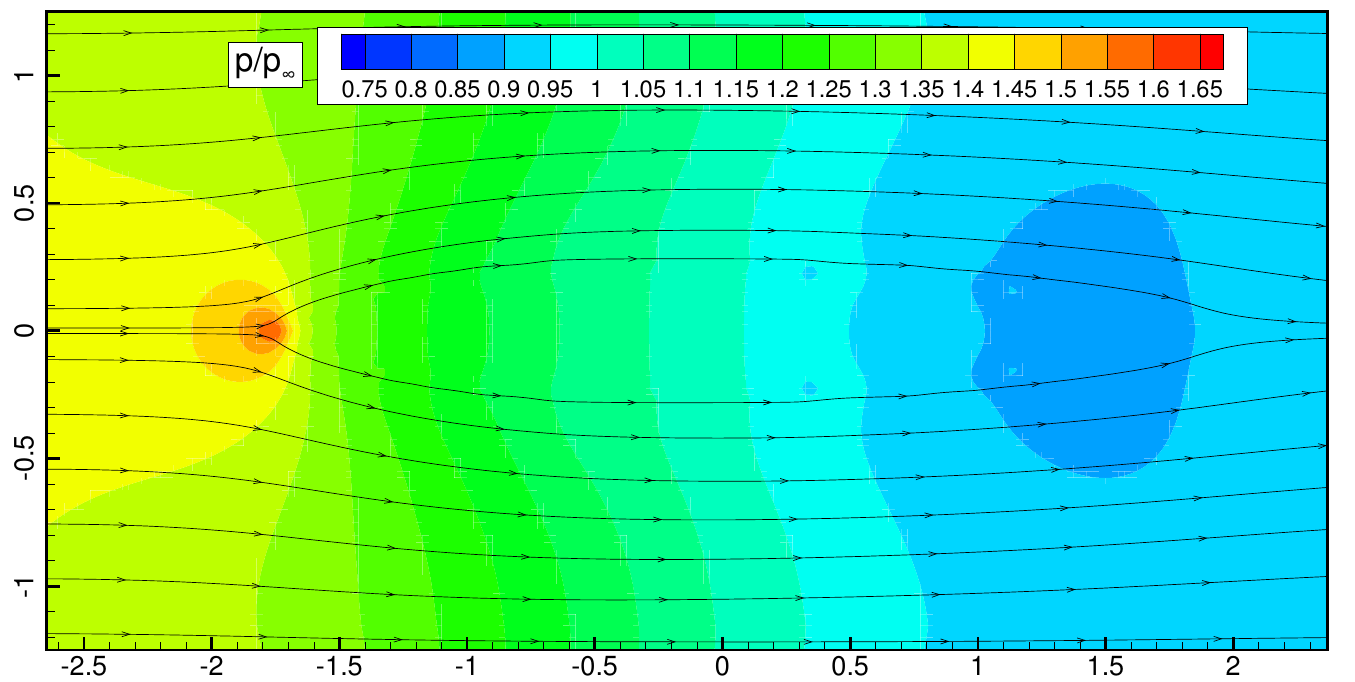}}}\\%
	{\subfigure[${\rm Kn}=0.5$]{%
			\includegraphics[width=0.45\textwidth]{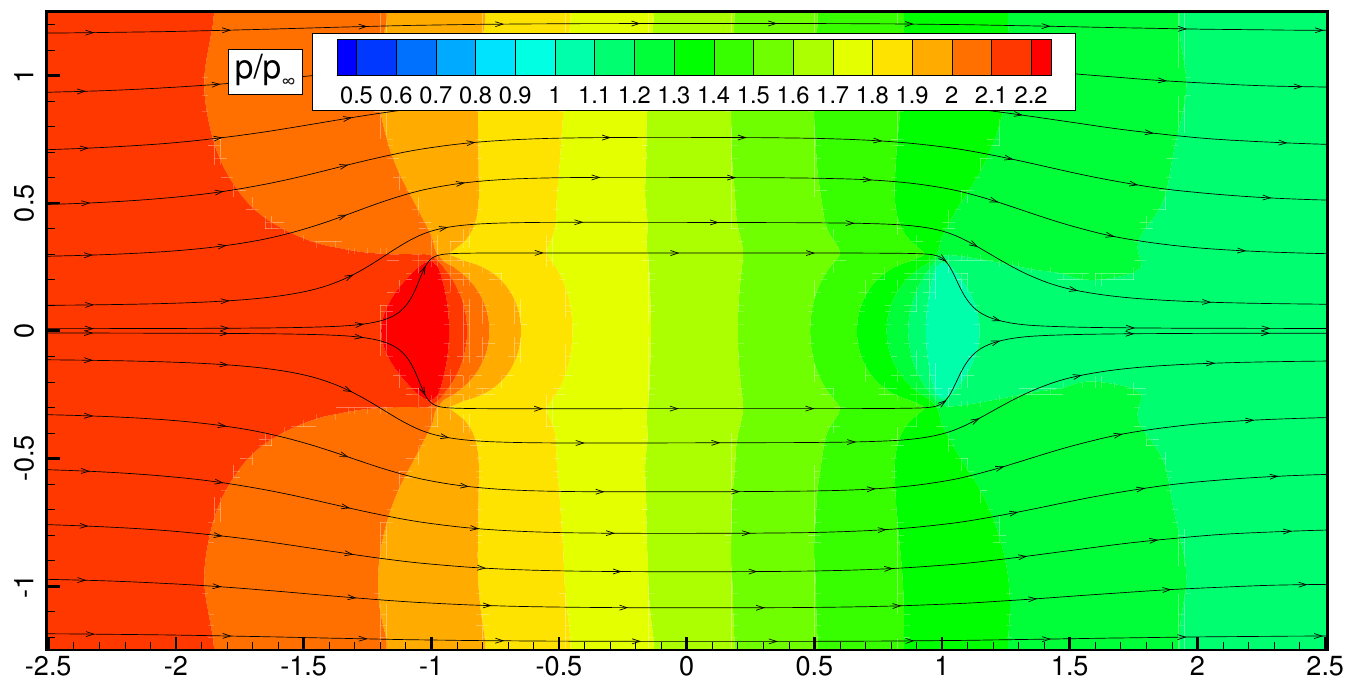}\hspace{0.02\textwidth}%
			\includegraphics[width=0.45\textwidth]{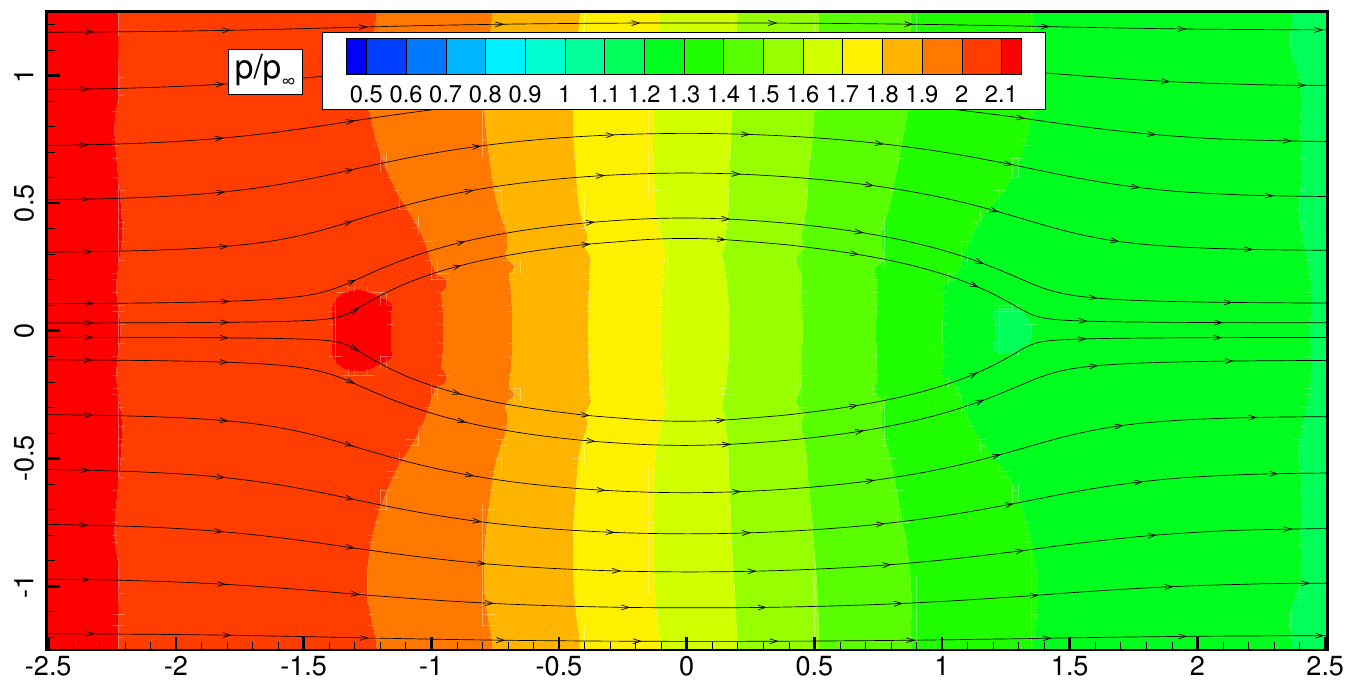}}}\\%
	{\subfigure[${\rm Kn}=10$]{%
			\includegraphics[width=0.45\textwidth]{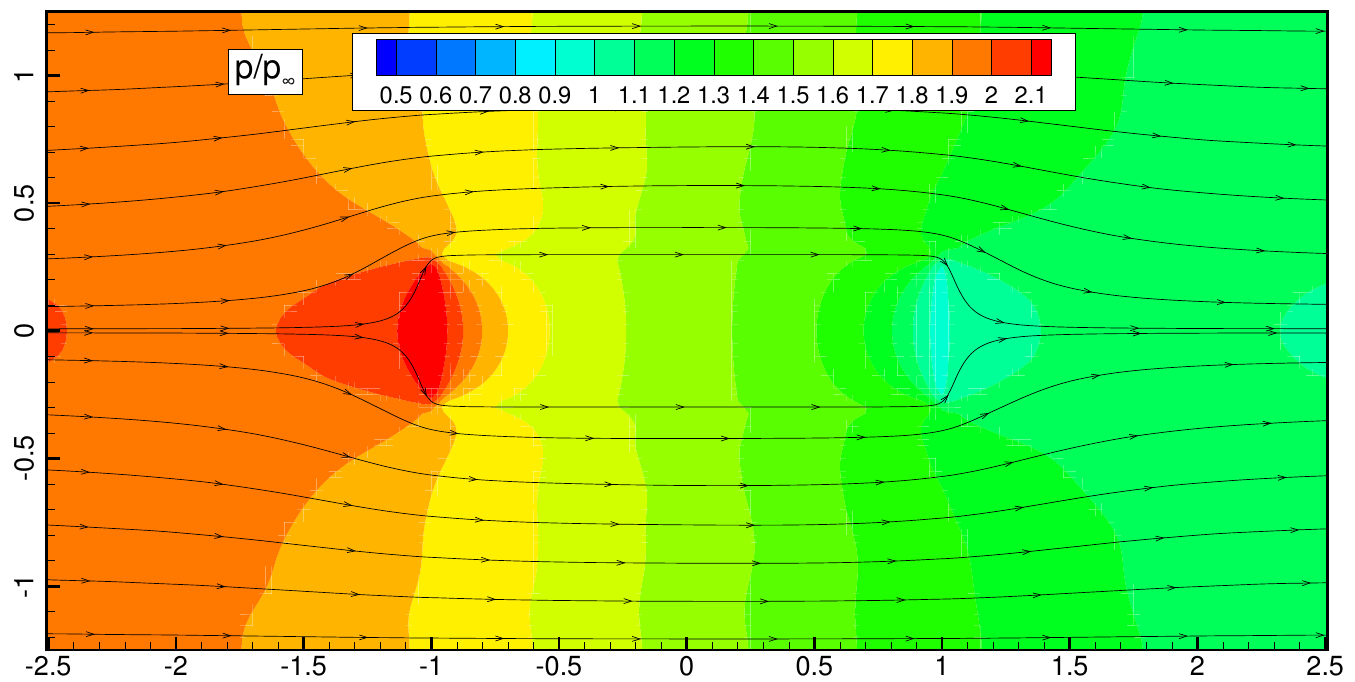}\hspace{0.02\textwidth}%
			\includegraphics[width=0.45\textwidth]{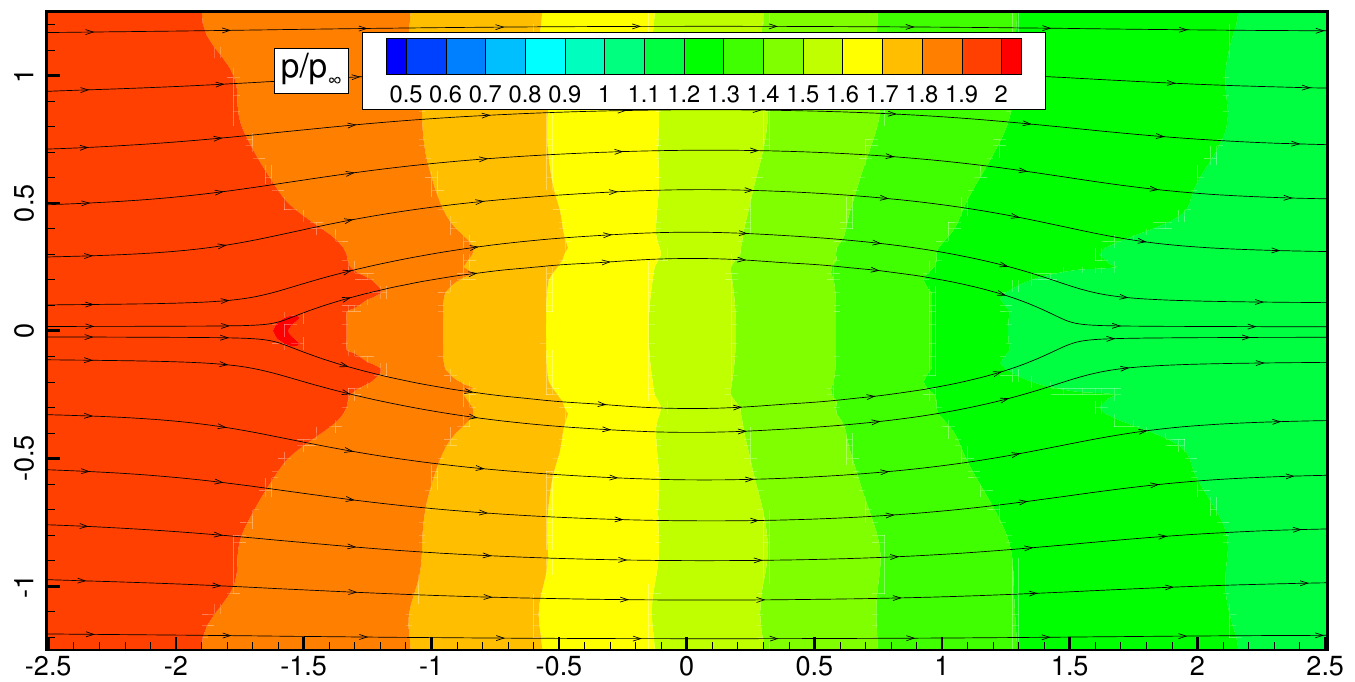}}}
	\caption{\label{fig:case2_pre}Optimization of the airfoil inside a channel: streamlines and pressure distributions before (left) and after (right) optimization.}
\end{figure}

Fig.~\ref{fig:case2_pre} shows the streamlines and pressure distributions around the initial airfoil (Fig.~\ref{fig:case2_inia}) and the optimized airfoils (Fig.~\ref{fig:case2_opttheta}). For the case ${\rm Re}=200$, the biggest difference between the flow fields before/after optimization is that, in the unoptimized flow field the gas flow separates at the trailing edge and a large pressure loss occurs, while in the optimized flow field the flow entirely attaches to the airfoil, resulting in a significant drag reduction. For the cases ${\rm Kn}=0.5,10$, the unoptimized and optimized flow fields are both entirely attached flows, but the sharp leading/trailing edges of the optimized airfoil help to suppress the high/low-pressure regions around the leading/trailing edges, and therefore cause some drag reduction.

Table \ref{tab:case2_cdeff} shows the drag coefficients $C_{\rm d}$ of the initial/optimized airfoils. Here all of the drag coefficients are calculated by Eq.~\eqref{eqn:cdcldefine} with a reference length of $2$, i.e., the chord length of the initial airfoil. When ${\rm Re}=200$ and ${\rm Kn}=10$ there are big drag reductions near 25\%, while for ${\rm Kn}=0.5$ the drag reduction is 14.6\%. Of course, it is notable that the drag reduction depends on the setup of the initial airfoil. It is also interesting that among the three conditions the maximum drag occurs at ${\rm Kn}=0.5$ for both the initial and optimized airfoils.

The optimization efficiency is also shown in Table \ref{tab:case2_cdeff}. All computations are conducted on the cluster with ``\emph{Intel(R) Xeon(R) Gold 6148 CPU @ 2.40GHz}'', the parallel scales and the computation costs are detailed in the table. It can be seen that for the cases ${\rm Re}=200$ and ${\rm Kn}=0.5$, the optimization finishes in 20-30 minutes. For the case ${\rm Kn}=10$, because a large number of velocity points are used, it takes around 1 hour. Generally speaking, the present topology optimization method achieves good performance and high efficiency from the continuum to free-molecular flow regimes.

\begin{table}[t]
	\centering
	\caption{\label{tab:case2_cdeff}Optimization of the airfoil inside a channel: drag coefficients before/after optimization and the optimization efficiency.}
	\begin{tabular}{|c|c|c|c|c|c|c|c|}
		\hline
		\multirow{2}{*}{Case} & Initial & Optimized & Drag     & Velocity & Parallel & Optimization & Time cost \\
		& $C_{\rm d}$      & $C_{\rm d}$        & decrease & points   & cores    & steps        & (seconds) \\ \hline
		${\rm Re}=200$        & 1.102   & 0.8304    & 24.6\%    & $20\times20$     & 80       & 128          & 1172.2    \\ \hline
		${\rm Kn}=0.5$        & 2.188   & 1.869     & 14.6\%    & $30\times30$     & 160      & 109          & 1866.8    \\ \hline
		${\rm Kn}=10$         & 1.995   & 1.553     & 22.2\%    & $60\times60$     & 320      & 160          & 3906.4    \\ \hline
	\end{tabular}
\end{table}

\begin{table}[t]
	\centering
	\caption{\label{tab:case2_cdmatrix}Optimization of the airfoil inside a channel: comparison of the drag coefficients of the three optimized airfoils under a certain gas condition. The minimum drag coefficient for each gas condition is shown in bold.}
	\begin{tabular}{|c|ccc|}
		\hline
		Gas       & \multicolumn{3}{c|}{Optimized airfoil for}                         \\ \cline{2-4} 
		condition & \multicolumn{1}{c|}{${\rm Re}=200$} & \multicolumn{1}{c|}{${\rm Kn}=0.5$} & ${\rm Kn}=10$  \\ \hline
		${\rm Re}=200$    & \multicolumn{1}{c|}{\textbf{0.8304}} & \multicolumn{1}{c|}{0.9371} & 0.8704 \\ \hline
		${\rm Kn}=0.5$    & \multicolumn{1}{c|}{2.009}  & \multicolumn{1}{c|}{\textbf{1.869}}  & 1.900  \\ \hline
		${\rm Kn}=10$     & \multicolumn{1}{c|}{1.624}  & \multicolumn{1}{c|}{1.593}  & \textbf{1.553}  \\ \hline
	\end{tabular}
\end{table}

Table \ref{tab:case2_cdmatrix} shows the three optimized airfoils' drag coefficients under different gas flow conditions. The minimum drag coefficient for each gas condition is just obtained from the optimized airfoil of the corresponding condition, which further validates the present optimization method. It also can be found that the optimized airfoil for ${\rm Kn}=10$ has the most stable drag performance throughout the three gas conditions from continuum to rarefied flow, while the optimized airfoil for ${\rm Re}=200$ has the biggest drag coefficients under the gas conditions ${\rm Kn}=0.5$ and ${\rm Kn}=10$, the optimized airfoil for ${\rm Kn}=0.5$ has the biggest drag coefficient under the gas condition ${\rm Re}=200$.

\subsection{Optimization of the airfoil under supersonic flow in different flow regimes}\label{sec:case3}
\begin{figure}[t]
\centering
\subfigure[Computational domain and mesh discretization (34266 cells)\label{fig:case3_mesh}]{\includegraphics[width=0.45\textwidth]{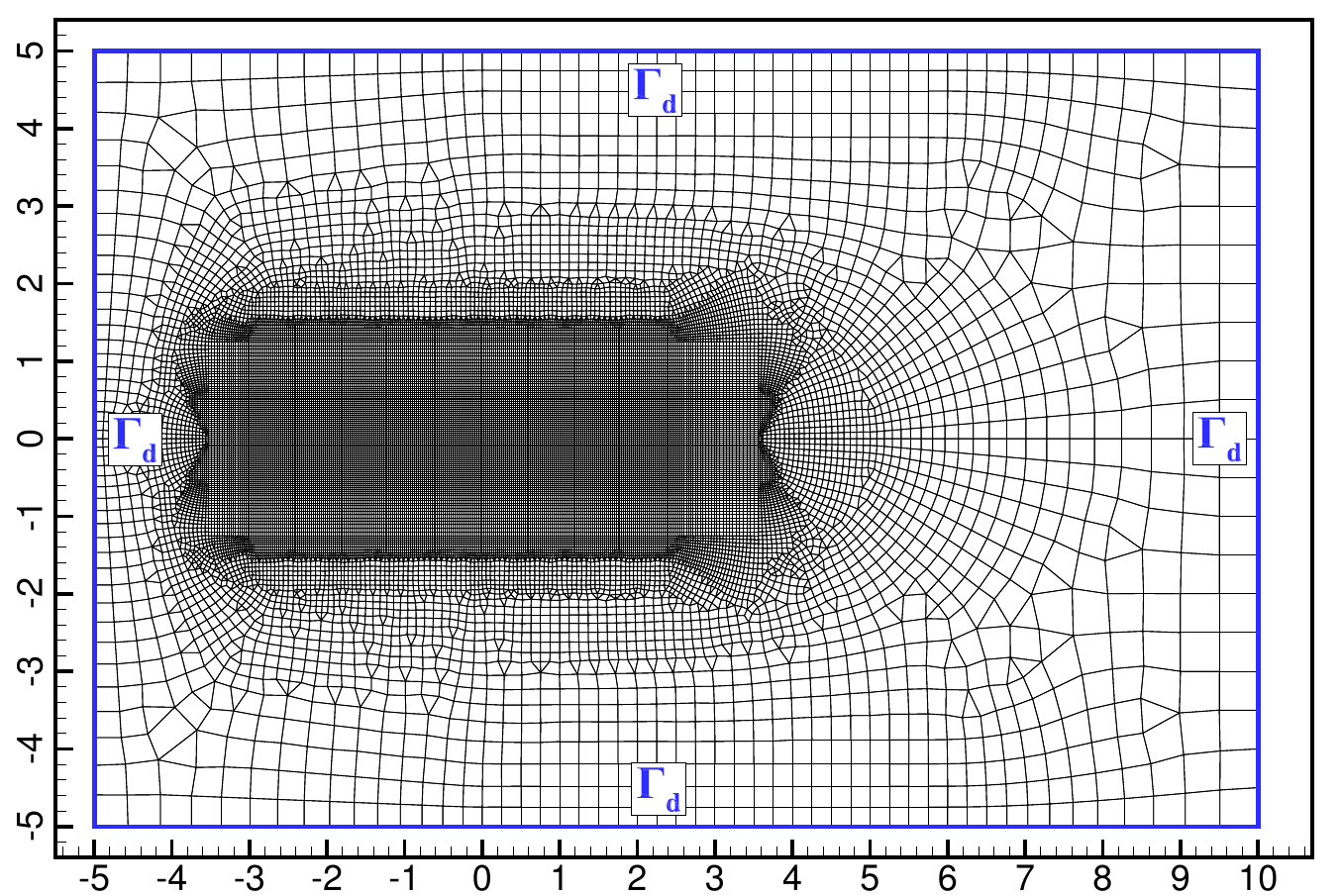}}\hspace{0.01\textwidth}
\subfigure[Initial shape of the airfoil\label{fig:case3_thetaref}]{\includegraphics[width=0.45\textwidth]{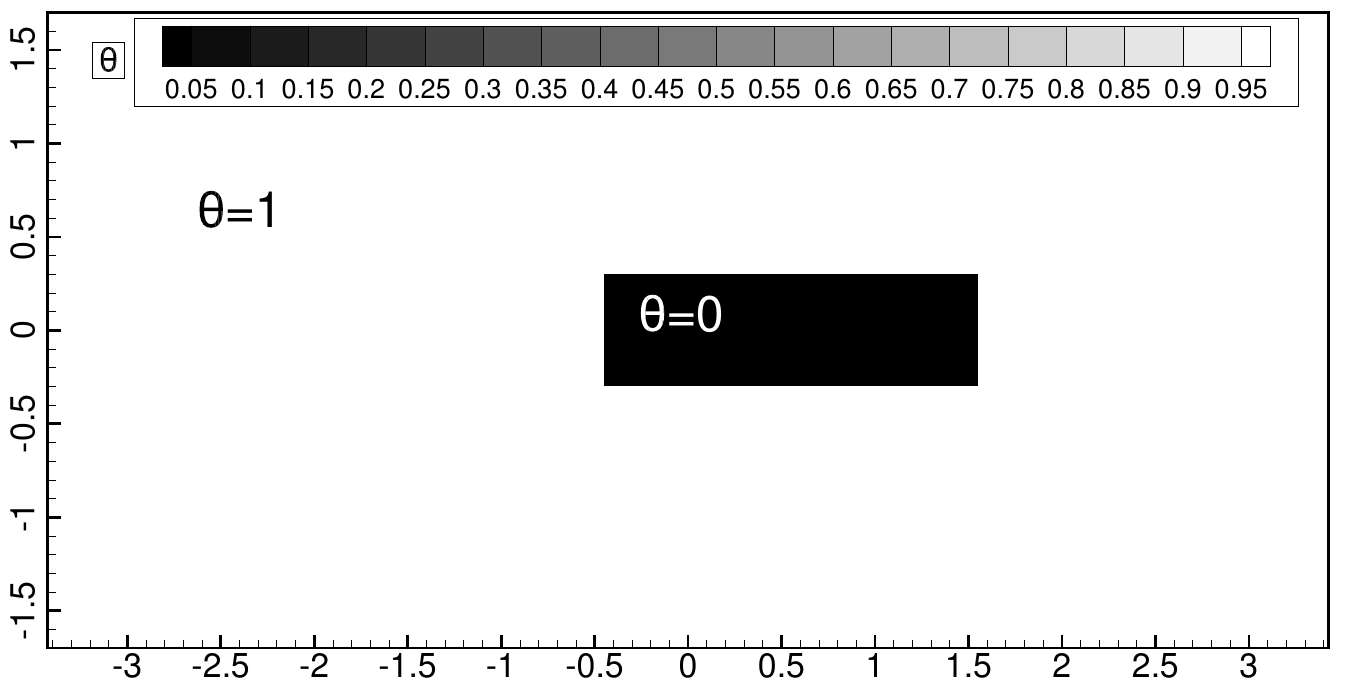}}
\caption{Setups for the optimization of the airfoil under supersonic flow.}
\end{figure}

In this section, the drag-reduction optimization of the airfoil under supersonic flow is performed to further validate the present topology optimization method. Two sets of flow conditions are considered, i.e., ${\rm Ma}=2, {\rm Re}=200$ and ${\rm Ma}=2, {\rm Kn}=0.5$. The computational domain is set as a rectangle of the size $15\times10$, and a non-uniform unstructured mesh is adopted, see Fig.~\ref{fig:case3_mesh}. In this mesh the region near the airfoil is discretized as the Cartesian grid and has the same resolution of that in Section \ref{sec:case0} where a comparable Re number is involved, so the mesh independence is ensured here. On all boundaries of the computational domain, the Dirichlet boundary condition of Eq.~\eqref{eqn:bc_dirichlet} is imposed, where the gas state corresponds to the free-stream condition of ${\rm Ma}=2$. 

For the discretization of the velocity space in the DVM solver, $28\times28$ and $40\times 40$ uniform meshes in the molecular velocity range $[ - 8{a_\infty },8{a_\infty }]$ have been employed for the cases ${\rm Re}=200$ and ${\rm Kn}=0.5$ respectively, where $a_\infty$ is the free-stream acoustic speed.

Similar to Section \ref{sec:case2}, the initial airfoil is set as a rectangle of the size $2\times0.6$ centered at $(0.55,0)$, as shown in Fig.~\ref{fig:case3_thetaref}, and the initialization process described in Section \ref{sec:case2} is performed to obtain $\vartheta^{(l=0,m=0)}$. The chord length of the initial airfoil, i.e.~a length of $2$, is used as the reference length in Eqs.~\eqref{eqn:kndefine},\eqref{eqn:maredefine} to define Re and Kn. The setups for the objective and the constraint are also similar to those in Section \ref{sec:case2}, namely the objective is the drag force obtained from Eq.~\eqref{eqn:forcedefine}, and the volume constraint is of the form in Eq.~\eqref{eqn:formula_opt} with $V_{\rm max}=15\times10-2\times0.6=148.8$, which ensures that the minimum area of the airfoil is limited to the area of the initial airfoil. 

\begin{figure}[t]
\centering
\subfigure[${\rm Re}=200$]{\includegraphics[width=0.45\textwidth]{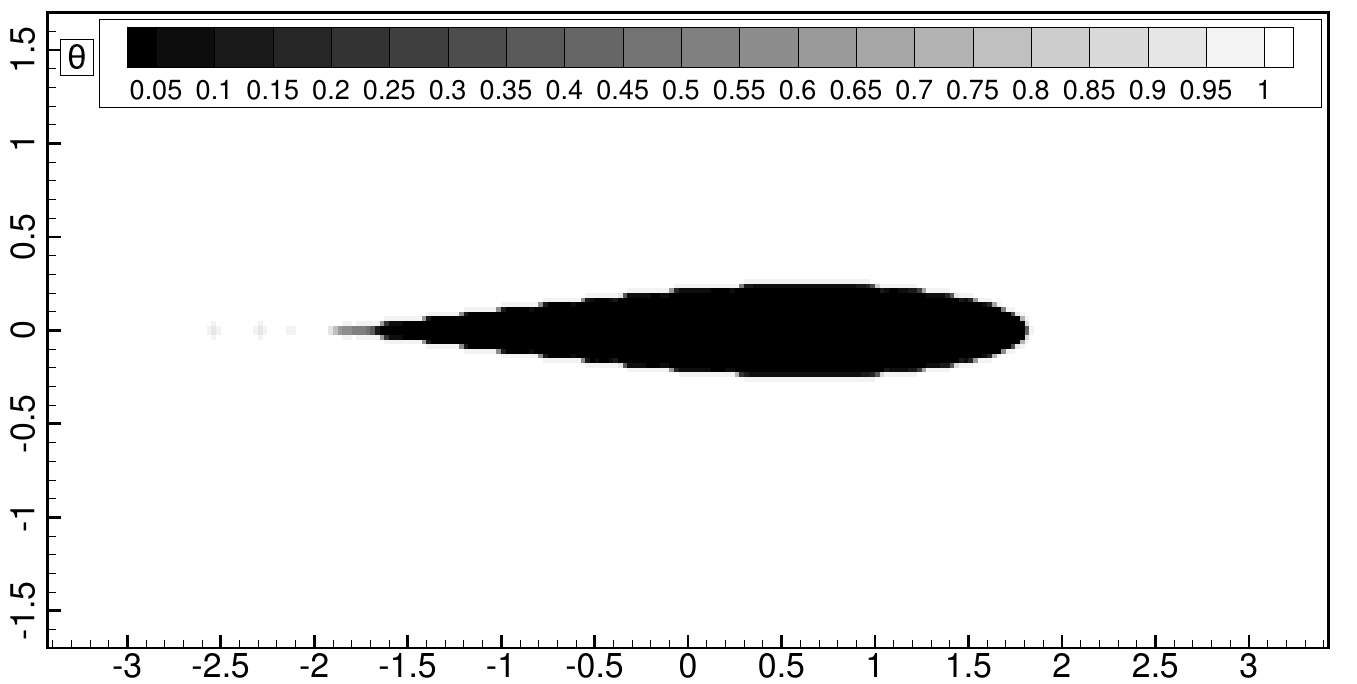}}\hspace{0.01\textwidth}
\subfigure[${\rm Kn}=0.5$]{\includegraphics[width=0.45\textwidth]{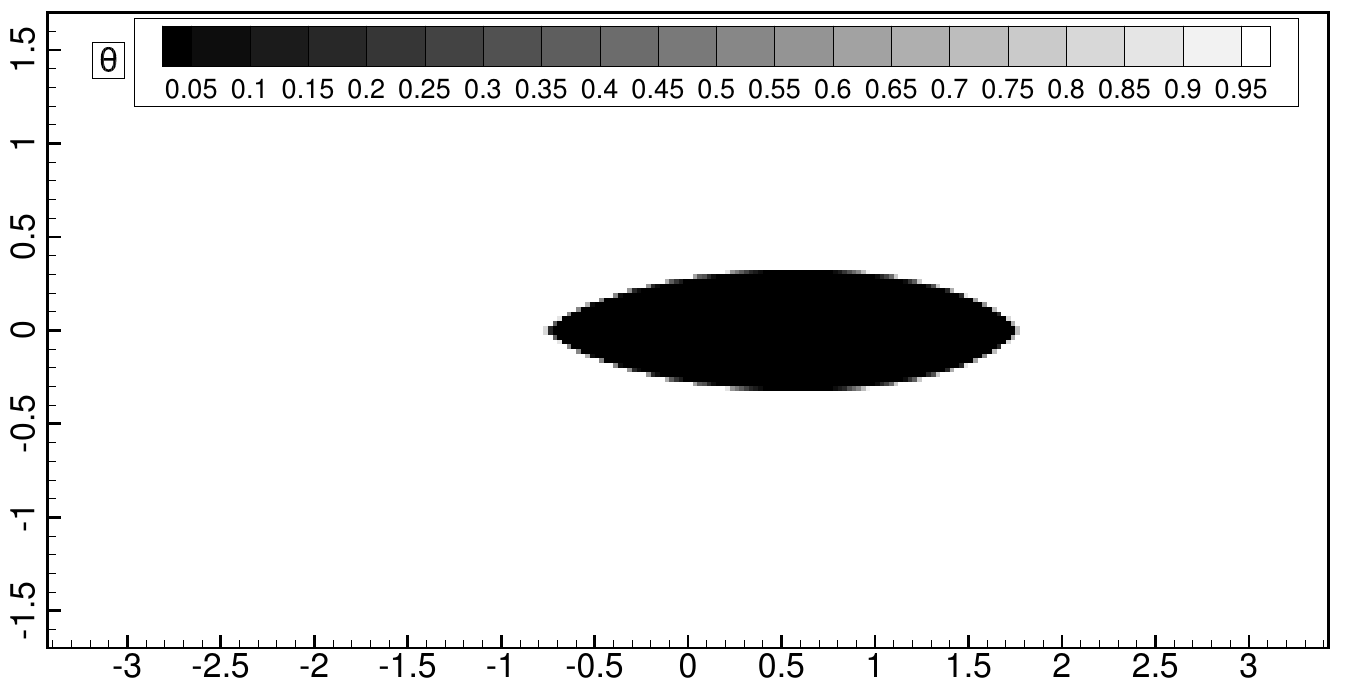}}
\caption{\label{fig:case3_thetaopt}Optimization of the airfoil under supersonic flow: the optimized airfoils for different flow conditions. Note that the leading edge of the airfoil is on the left-hand side.}
\end{figure}

The optimization follows the procedure described in Section \ref{sec:method_optframework}. Fig.~\ref{fig:case3_thetaopt} shows the material density distributions of the optimized airfoils. 
For the case ${\rm Re}=200$, the most attractive feature of the optimized airfoil is the sharp leading edge with a very long thin tip. 
For the case ${\rm Kn}=0.5$, the optimized airfoil has the shape of biconvex arc, with sharp leading/trailing edges and is leading-trailing asymmetric.

\begin{figure}
\centering
{\subfigure[Mach number distribution]{%
\includegraphics[width=0.45\textwidth]{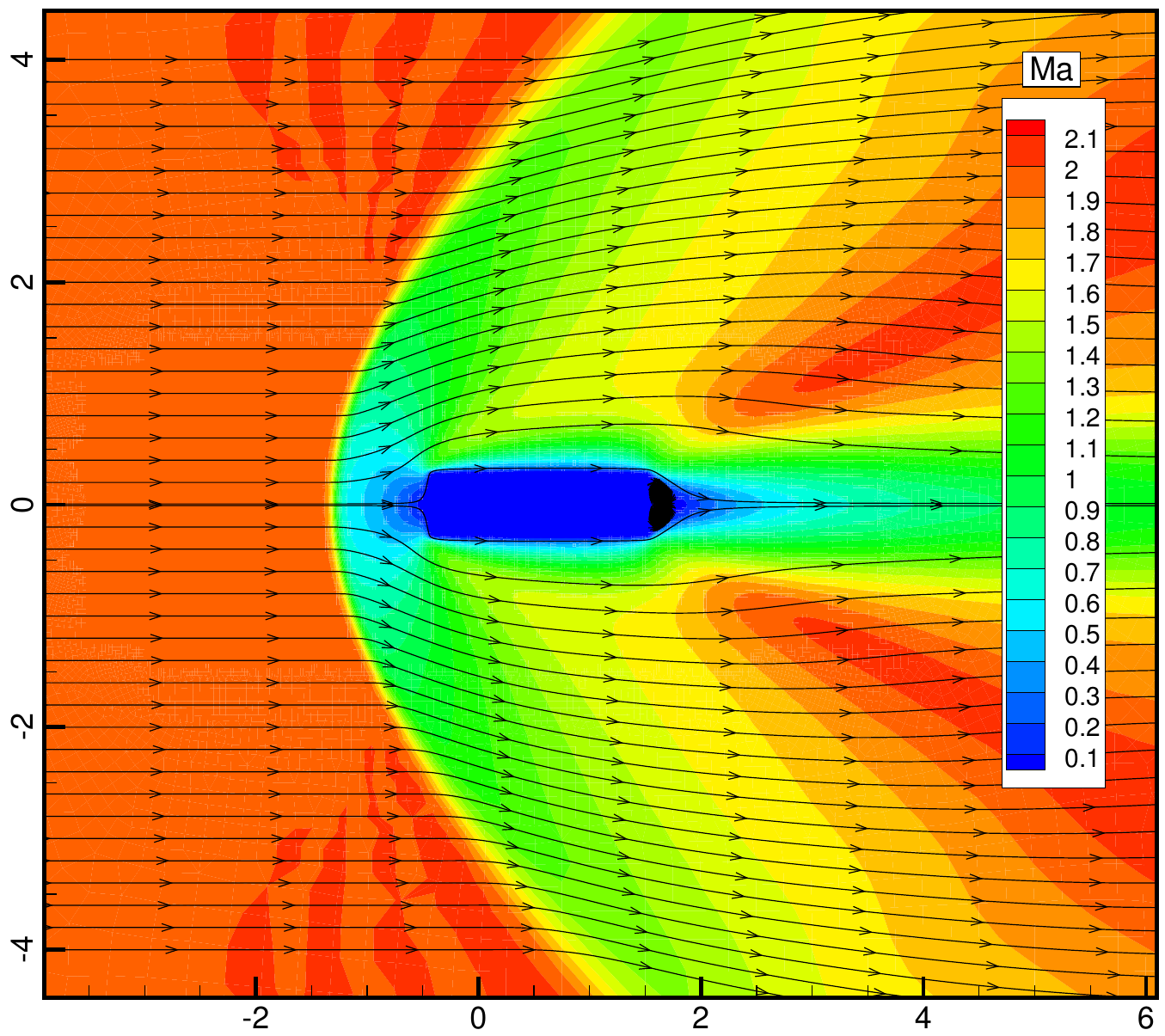}\hspace{0.02\textwidth}%
\includegraphics[width=0.45\textwidth]{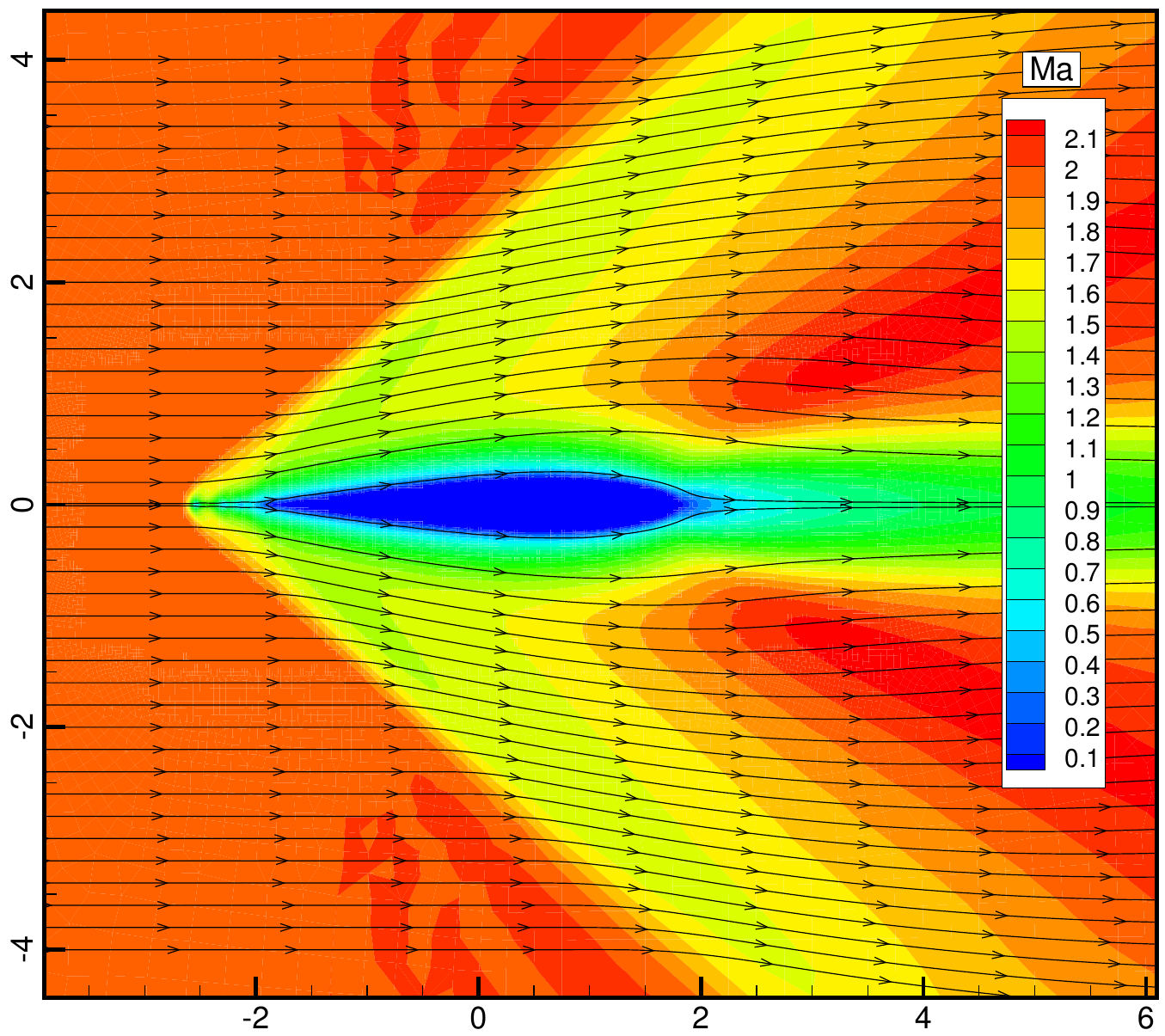}}}\\%
{\subfigure[Pressure distribution]{%
\includegraphics[width=0.45\textwidth]{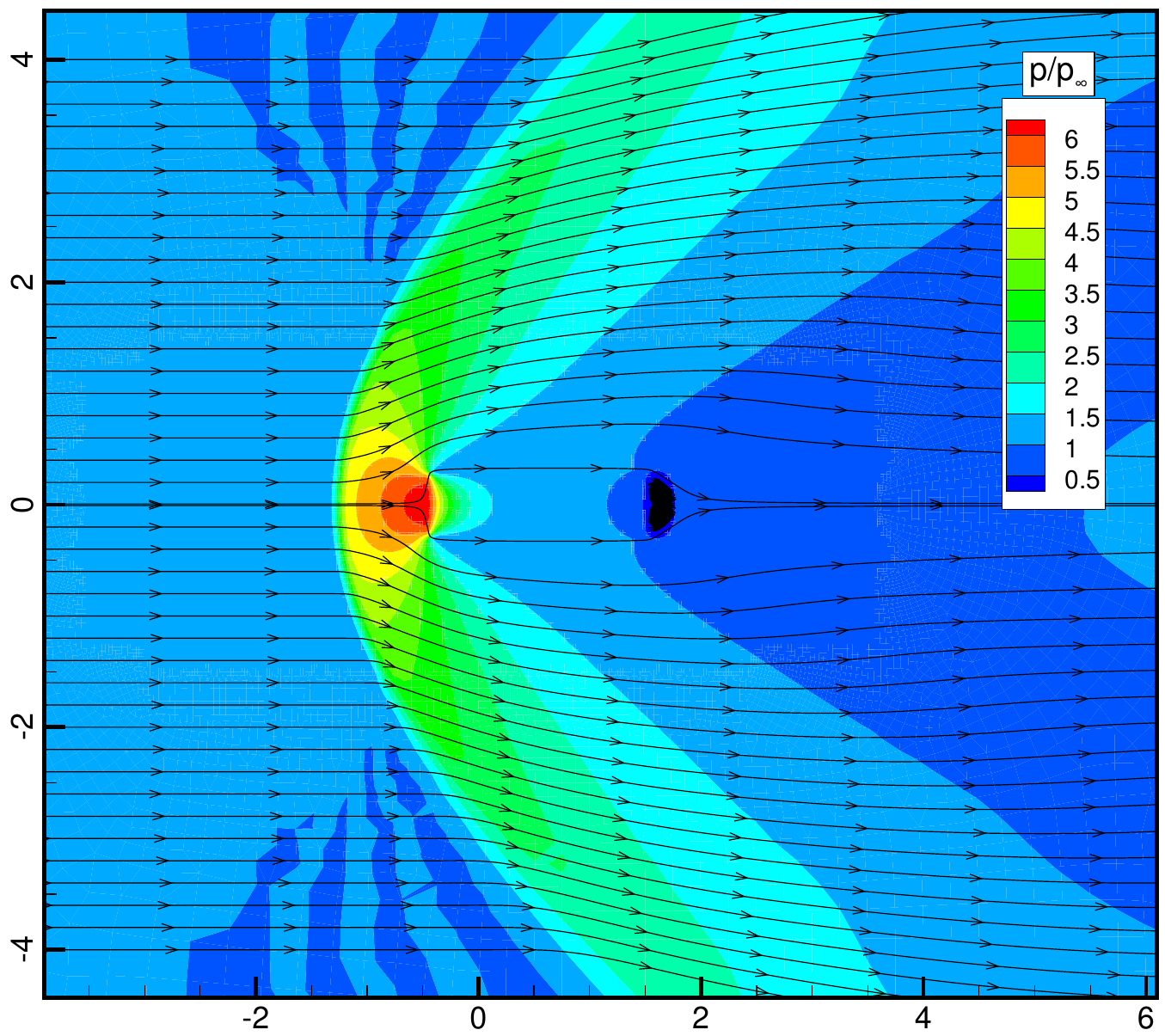}\hspace{0.02\textwidth}%
\includegraphics[width=0.45\textwidth]{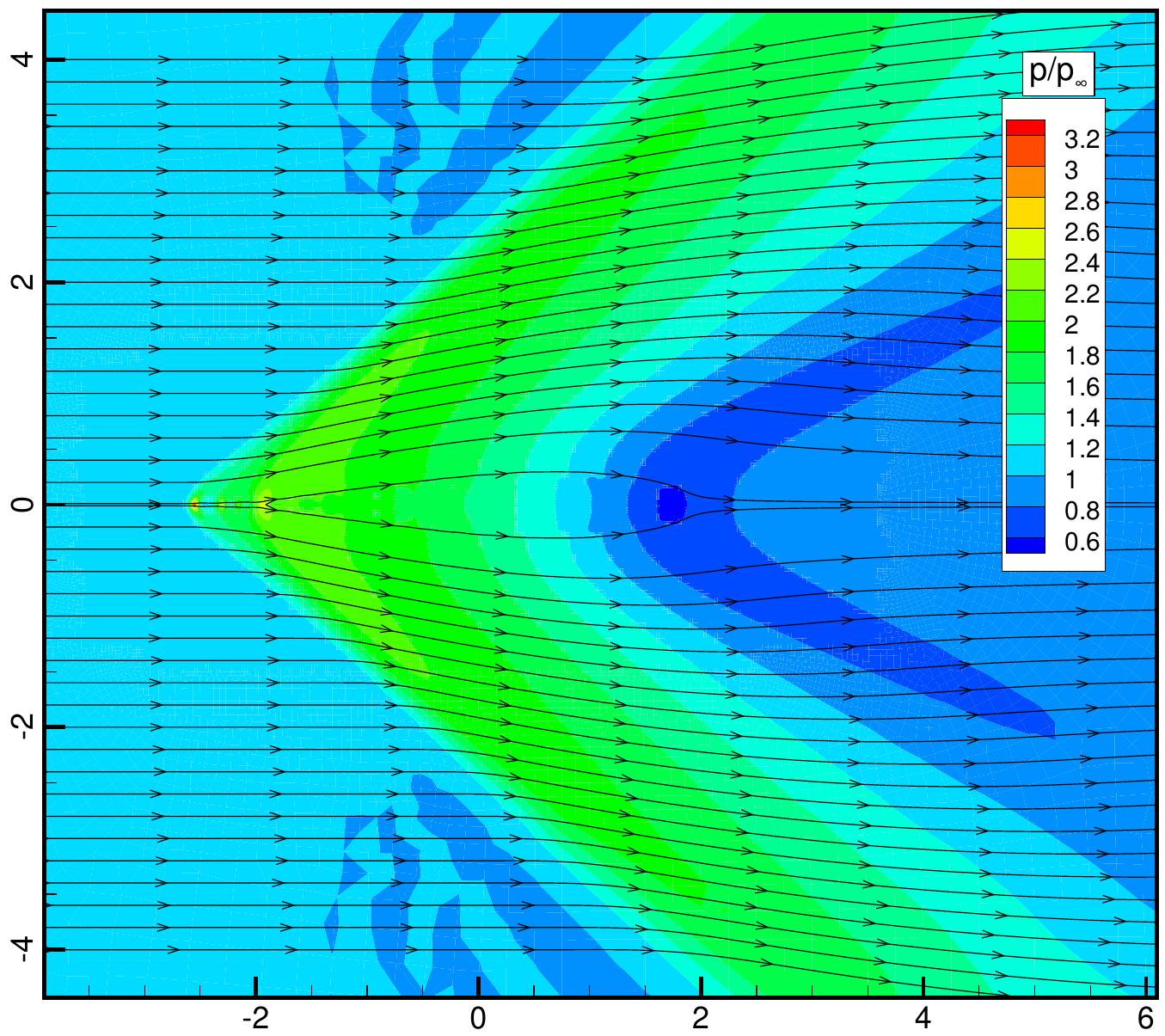}}}
\caption{\label{fig:case3_maprere200}Optimization of the airfoil under supersonic flow at ${\rm Ma}=2,{\rm Re}=200$: streamlines, Mach number and pressure distributions before (left) and after (right) optimization.}
\end{figure}

\begin{figure}[t]
\centering
{\subfigure[Mach number distribution]{%
\includegraphics[width=0.45\textwidth]{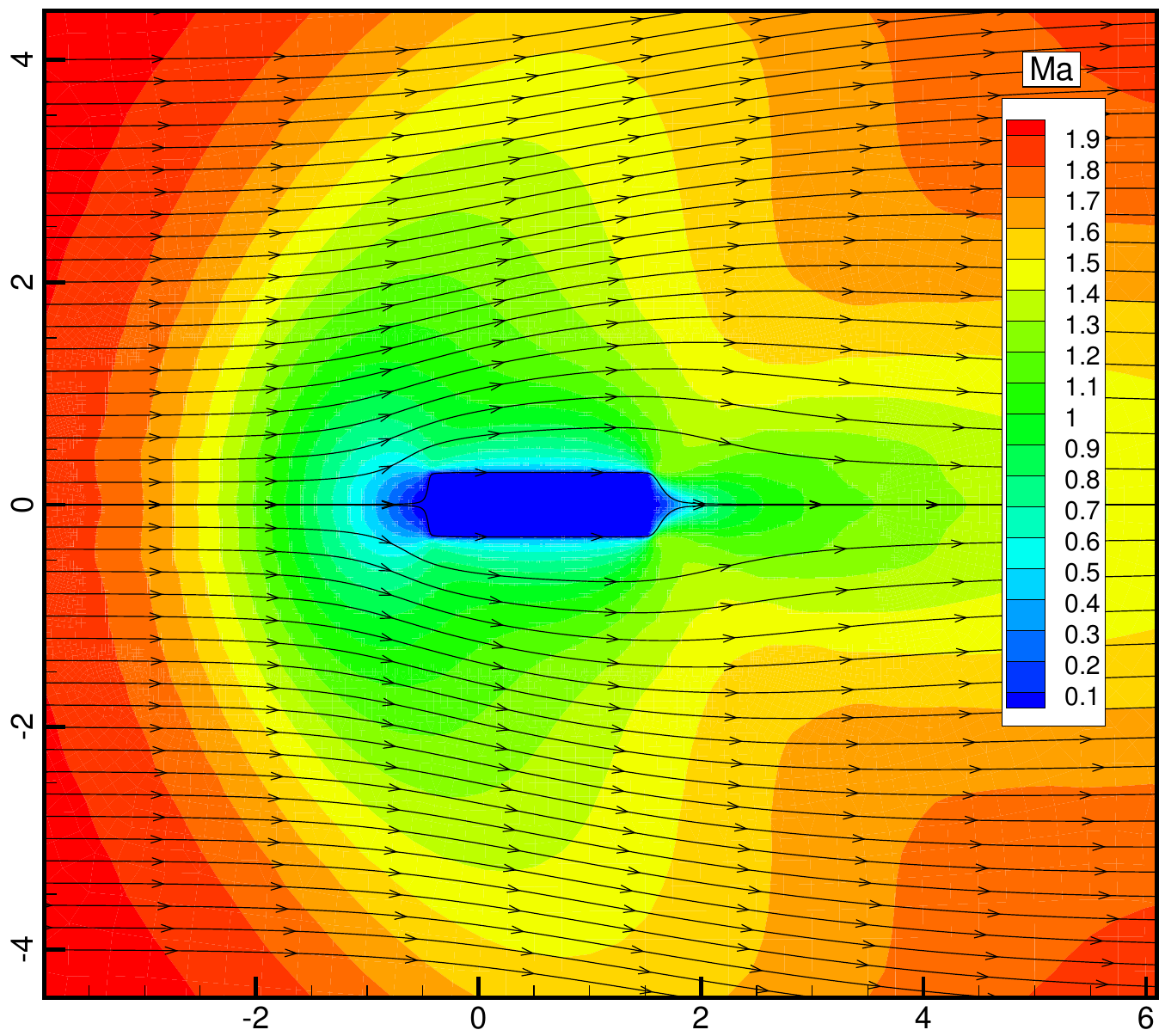}\hspace{0.02\textwidth}%
\includegraphics[width=0.45\textwidth]{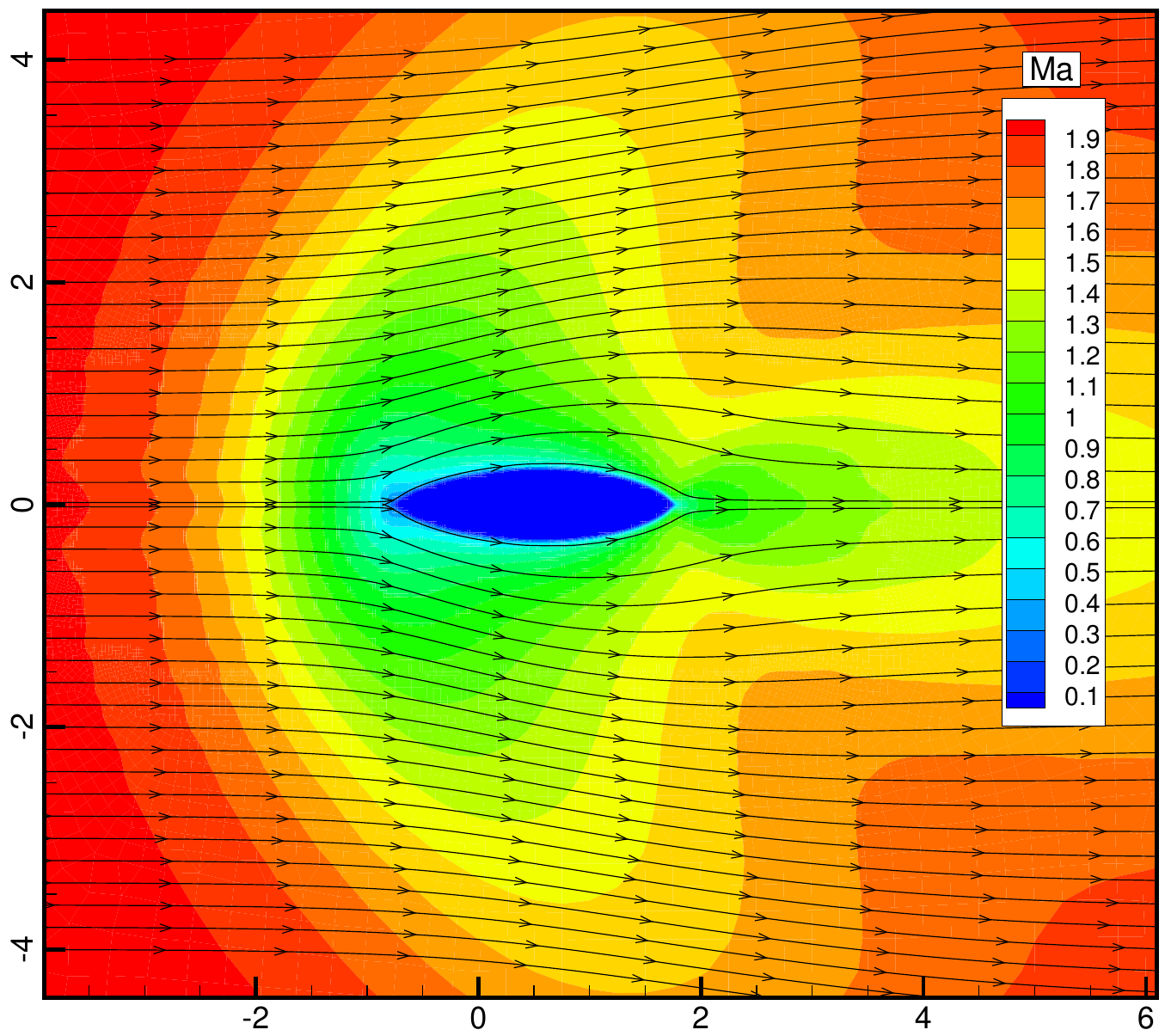}}}\\%
{\subfigure[Pressure distribution]{%
\includegraphics[width=0.45\textwidth]{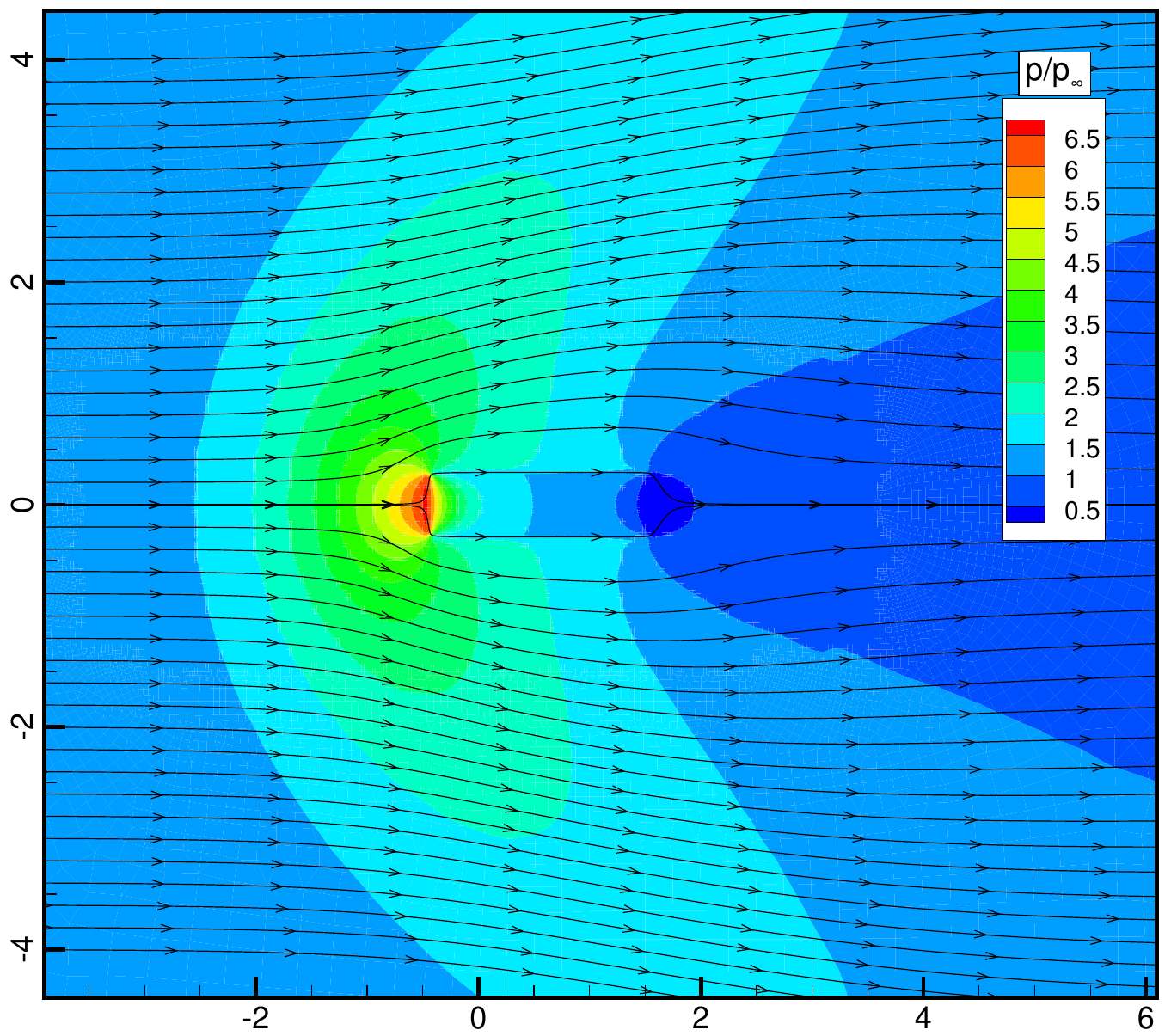}\hspace{0.02\textwidth}%
\includegraphics[width=0.45\textwidth]{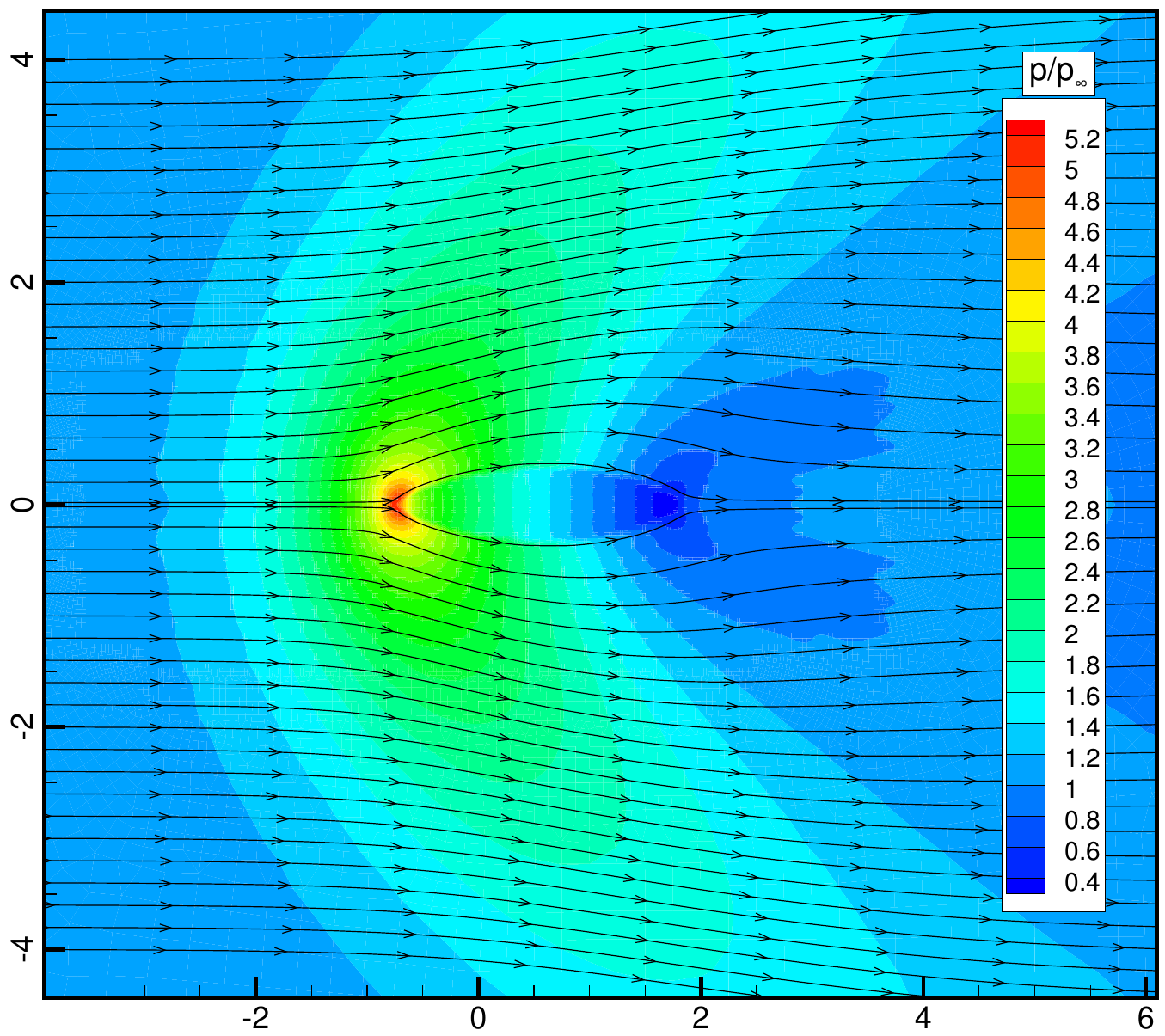}}}
\caption{\label{fig:case3_maprekn05}Optimization of the airfoil under supersonic flow at ${\rm Ma}=2,{\rm Kn}=0.5$: streamlines, Mach number and pressure distributions before (left) and after (right) optimization.}
\end{figure}

The flow fields of the initial and the optimized airfoils are shown in Fig.~\ref{fig:case3_maprere200} and Fig.~\ref{fig:case3_maprekn05}, from which the mechanism why the optimized airfoils have the shapes presented in Fig.~\ref{fig:case3_thetaopt} can be revealed. When ${\rm Re}=200$, from Fig.~\ref{fig:case3_maprere200}, it can be seen that the shock wave has the form of sharp distinct discontinuity. More specifically, for the initial rectangular airfoil, a detached bow shock forms in front of the airfoil, and a large area around the leading edge is of very high pressure, with a maximum dimensionless pressure of $6.4$, causing a large wave drag. In addition, the flow separates at the trailing edge of the airfoil, forming a pair of small vortices and an area of very low pressure. All of these make the initial airfoil suffer from a big pressure drag. After optimization, two oblique shock waves form at the long thin leading tip of the optimized airfoil, making the whole airfoil downstream of the oblique shock waves. The maximum dimensionless pressure is around $3.3$, approximately half of that before optimization, and the maximum pressure occurs only at the small leading tip of the airfoil, no large high-pressure areas. At the optimized trailing edge, no flow separation can be found. As a result, the optimized airfoil achieves a large drag reduction mainly from the decrease of the pressure drag. When ${\rm Kn}=0.5$, from Fig.~\ref{fig:case3_maprekn05} it can be seen that the shock wave turns into a thick, diffuse, continuous structure, and there is no distinct discontinuity. The general pattern of the shock wave is quite similar before and after optimization, and it seems not possible to get a big reduction of the wave drag by stretching the leading edge into a very long thin structure as what happens when ${\rm Re}=200$. Besides, there is no flow separation in both the unoptimized and the optimized flow fields. Nevertheless, the sharp leading/trailing edges of the optimized airfoil help to weaken the high/low pressure on the windward/leeward of the airfoil, and cause some reasonable drag reduction. 
It is worth noting that the optimized airfoil for ${\rm Kn}=0.5$ is shorter and thicker than that for ${\rm Re}=200$. We think this is due to the increase of the gas viscosity from ${\rm Re}=200$ to ${\rm Kn}=0.5$, by which the airfoil tends to reduce its surface exposed to the gas flow to decrease the friction drag.

The drag coefficients for the initial and the optimized airfoils are shown in Table \ref{tab:case3_cdeff}, in which all of the drag coefficients are calculated by Eq.~\eqref{eqn:cdcldefine} with a reference length of the chord length of the initial airfoil. It can be seen that the data in the table just accord with the above comparative analysis about the unoptimized and optimized flow fields in Fig.~\ref{fig:case3_maprere200} and Fig.~\ref{fig:case3_maprekn05}. In the case ${\rm Re}=200$ there is a significant drag reduction of about 36.5\% for the optimized airfoil, while when ${\rm Kn}=0.5$ the drag reduction is only 12.0\%. 

As for the optimization efficiency, the test cases in this section adopt a 160-core parallel computation with the CPU model ``\emph{Intel(R) Xeon(R) Gold 6148 CPU @ 2.40GHz}'', the optimization steps and the corresponding time costs are shown in Table \ref{tab:case3_cdeff}. It can be seen that the two sets of optimizations both finish in less than 1 hour, which further verifies the high efficiency of the present topology optimization method.

\begin{table}
\centering
\caption{\label{tab:case3_cdeff}Optimization of the airfoil under supersonic flow: drag coefficients before/after optimization and the optimization efficiency.}
\begin{tabular}{|c|c|c|c|c|c|c|c|}
\hline
\multirow{2}{*}{Case} & Initial & Optimized & Drag     & Velocity & Parallel & Optimization & Time cost \\
                      & $C_{\rm d}$      & $C_{\rm d}$        & decrease & points   & cores    & steps        & (seconds) \\ \hline
${\rm Re}=200$        & 0.6481   & 0.4117    & 36.5\%    & $28\times28$     & 160      & 191         & 3366.2    \\ \hline
${\rm Kn}=0.5$        & 1.228    & 1.081     & 12.0\%    & $40\times40$     & 160      & 97          & 2609.8    \\ \hline
\end{tabular}
\end{table}

Table \ref{tab:case3_cdmatrix} shows the two optimized airfoils' drag coefficients under different flow conditions. According to the data in the table, the minimum drag coefficient under a certain gas condition is just obtained from the airfoil optimized by the present method for that condition. This further proves the validity of the present optimization method, and also validates the analysis in the previous paragraphs about the optimization mechanism for the unoptimized/optimized flow fields.

\begin{table}[]
\centering
\caption{\label{tab:case3_cdmatrix}Optimization of the airfoil under supersonic flow: comparison of the drag coefficients of the two optimized airfoils under a certain gas condition. The minimum drag coefficient for each gas condition is shown in bold.}
\begin{tabular}{|c|cc|}
\hline
Gas       & \multicolumn{2}{c|}{Optimized airfoil for} \\ \cline{2-3} 
condition & \multicolumn{1}{c|}{Re=200}    & Kn=0.5    \\ \hline
Re=200    & \multicolumn{1}{c|}{\textbf{0.4117}}    & 0.5416    \\ \hline
Kn=0.5    & \multicolumn{1}{c|}{1.190}     & \textbf{1.081}     \\ \hline
\end{tabular}
\end{table}

\clearpage

\section{Conclusions}\label{sec:conc}
This paper presents a topology optimization method for the design problems of gas flows in all flow regimes. The present method takes the density-based parametrization approach and the design domain is governed by the BGK equation with the fictitious porosity model. The continuous adjoint method is used to perform the design sensitivity analysis and the design variable is updated by MMA.
The main innovations of the present topology optimization method are summarized as follows:
\begin{enumerate}
\item The fictitious porosity model is modified to behave well in both continuum and rarefied flow conditions, and no abnormal gray regions are generated at large Kn.
\item The primal and adjoint equations are discretized by the multiscale numerical scheme applicable to gas flows in all flow regimes.
\item The macroscopic prediction method is applied to guarantee the efficient solving of the primal and adjoint equations in all flow regimes, where the approximate macroscopic equations constructed from the continuum limit are used to accelerate the convergence of the primal and adjoint numerical systems.
\end{enumerate}

Several test cases, involving flow conditions from continuum to rarefied and from subsonic to supersonic, are performed by the present topology optimization method. From these numerical experiments, the following conclusions can be obtained for the validation of the present method:
\begin{enumerate}
\item In the test case of the flow over an elliptic cylinder inside a channel (Section \ref{sec:case0} and Section \ref{sec:case1}), the accuracy of the primal and adjoint solvers has been confirmed in all flow regimes.
\item In the drag-reduction optimization of the airfoil inside a channel (Section \ref{sec:case2}), the present optimization method achieves a drag reduction of 14.6-24.6\% with a time cost of 20-65 minutes (parallel computation with 80-320 cores) in different flow regimes. In the drag-reduction optimization of the airfoil under supersonic flow (Section \ref{sec:case3}), the drag reductions are 12.0\% and 36.5\% for the cases $\rm{Kn}=0.5$ and $\rm{Re}=200$ respectively, with an optimization time cost of around 43-56 minutes (parallel computation with 160 cores). Moreover, it is indicated that the optimized airfoils just achieve the minimum drag under the corresponding flow conditions. Therefore, the effectiveness and the high efficiency of the present topology optimization method have been fully validated.
\end{enumerate}

In addition, from the numerical results, we have come to the following interesting findings:
\begin{enumerate}
\item In the drag-reduction optimization of the airfoil inside a channel (Section \ref{sec:case2}), the optimized airfoil has the shape of the classical subsonic airfoil under the continuum condition while has the shape of biconvex circular arc under the rarefied condition. Besides, the thickness of the optimized airfoil increases first and then decreases along with the increase of Kn, which may due to the change of the proportion between the friction drag
and the pressure drag. 
\item In the drag-reduction optimization of the airfoil under supersonic flow (Section \ref{sec:case3}), for the case $\rm{Re}=200$, the optimized airfoil has a sharp leading edge with a long thin tip which forms two oblique shock waves covering the whole airfoil behind, resulting in a significant reduction of the wave drag.
For the rarefied case $\rm{Kn}=0.5$, the pattern of the shock wave is relatively insensitive to the shape of the airfoil, and the optimized airfoil tends to be shorter and thicker to reduce the friction drag.
\end{enumerate}

In conclusion, the present topology optimization method is of good performance and high efficiency in all flow regimes, can provide valuable and innovative results for the design problems involving continuum and rarefied flows, and has good potential in the related application areas.

\section*{Acknowledgments}
The authors thank Prof. Steven G. Johnson (Massachusetts Institute of Technology) for the implementation of MMA algorithm in the NLopt library. This work is supported by the National Natural Science Foundation of China (12172162) and the Guangdong-Hong Kong- Macao Joint Laboratory for Data-Driven Fluid Mechanics and Engineering Applications in China (2020B1212030001).


\bibliographystyle{yuan_arxiv}
\bibliography{2023_topo}

\begin{thebibliography}{10}

\bibitem{reed2010investigation}
E.~Reed, H.~Alkandry, J.~Codoni, J.~McDaniel, and I.~Boyd.
\newblock Investigation of the interactions of reaction control systems with
  mars science laboratory aeroshell.
\newblock In \emph{48th AIAA Aerospace Sciences Meeting Including the New
  Horizons Forum and Aerospace Exposition}, 2010 page 1558.

\bibitem{li2021kinetic}
J.~Li, D.~Jiang, X.~Geng, and J.~Chen.
\newblock Kinetic comparative study on aerodynamic characteristics of
  hypersonic reentry vehicle from near-continuous flow to free molecular flow.
\newblock \emph{Advances in Aerodynamics}, 2021.
\newblock 3:1--10.

\bibitem{hablanian1997high}
M.~H. Hablanian.
\newblock \emph{High-vacuum technology: a practical guide}.
\newblock CRC Press, 1997.

\bibitem{sharipov2005numerical}
F.~Sharipov, P.~Fahrenbach, and A.~Zipp.
\newblock {Numerical modeling of the Holweck pump}.
\newblock \emph{Journal of Vacuum Science \& Technology A}, 2005.
\newblock 23(5):1331--1339.

\bibitem{bakshi2009euv}
V.~Bakshi.
\newblock \emph{{EUV lithography}}.
\newblock SPIE press, 2009.

\bibitem{tantos2020deterministic}
C.~Tantos, S.~Varoutis, and C.~Day.
\newblock Deterministic and stochastic modeling of rarefied gas flows in fusion
  particle exhaust systems.
\newblock \emph{Journal of Vacuum Science \& Technology B}, 2020.
\newblock 38(6).

\bibitem{tsien1946superaerodynamics}
H.-S. Tsien.
\newblock Superaerodynamics, mechanics of rarefied gases.
\newblock \emph{Journal of the Aeronautical Sciences}, 1946.
\newblock 13(12):653--664.

\bibitem{Cercignanibook1988}
C.~Cercignani.
\newblock \emph{The Boltzmann Equation and its Applications}.
\newblock Springer-Verlag, New York, 1988.

\bibitem{Lei2022}
L.~Wu.
\newblock \emph{Rarefied Gas Dynamics: Kinetic Modeling and Multi-Scale
  Simulation}.
\newblock Springer, 2022.

\bibitem{bendsoe2003topology}
M.~P. Bends{\o}e and O.~Sigmund.
\newblock \emph{Topology optimization: theory, methods, and applications}.
\newblock Springer Science \& Business Media, 2003.

\bibitem{rozvany2009critical}
G.~I. Rozvany.
\newblock A critical review of established methods of structural topology
  optimization.
\newblock \emph{Structural and multidisciplinary optimization}, 2009.
\newblock 37:217--237.

\bibitem{huang2010evolutionary}
X.~Huang and M.~Xie.
\newblock \emph{Evolutionary topology optimization of continuum structures:
  methods and applications}.
\newblock John Wiley \& Sons, 2010.

\bibitem{sigmund2013topology}
O.~Sigmund and K.~Maute.
\newblock {Topology optimization approaches: A comparative review}.
\newblock \emph{Structural and Multidisciplinary Optimization}, 2013.
\newblock 48(6):1031--1055.

\bibitem{borrvall2003topology}
T.~Borrvall and J.~Petersson.
\newblock {Topology optimization of fluids in Stokes flow}.
\newblock \emph{International journal for numerical methods in fluids}, 2003.
\newblock 41(1):77--107.

\bibitem{gersborg2005topology}
A.~Gersborg-Hansen, O.~Sigmund, and R.~B. Haber.
\newblock Topology optimization of channel flow problems.
\newblock \emph{Structural and multidisciplinary optimization}, 2005.
\newblock 30:181--192.

\bibitem{kreissl2012levelset}
S.~Kreissl and K.~Maute.
\newblock Levelset based fluid topology optimization using the extended finite
  element method.
\newblock \emph{Structural and Multidisciplinary Optimization}, 2012.
\newblock 46:311--326.

\bibitem{kubo2021level}
S.~Kubo, A.~Koguchi, K.~Yaji, T.~Yamada, K.~Izui, and S.~Nishiwaki.
\newblock Level set-based topology optimization for two dimensional turbulent
  flow using an immersed boundary method.
\newblock \emph{Journal of Computational Physics}, 2021.
\newblock 446:110630.

\bibitem{pingen2007topology}
G.~Pingen, A.~Evgrafov, and K.~Maute.
\newblock {Topology optimization of flow domains using the lattice Boltzmann
  method}.
\newblock \emph{Structural and Multidisciplinary Optimization}, 2007.
\newblock 34:507--524.

\bibitem{liu2014discrete}
G.~Liu, M.~Geier, Z.~Liu, M.~Krafczyk, and T.~Chen.
\newblock {Discrete adjoint sensitivity analysis for fluid flow topology
  optimization based on the generalized lattice Boltzmann method}.
\newblock \emph{Computers \& Mathematics with Applications}, 2014.
\newblock 68(10):1374--1392.

\bibitem{kreissl2011explicit}
S.~Kreissl, G.~Pingen, and K.~Maute.
\newblock {An explicit level set approach for generalized shape optimization of
  fluids with the lattice Boltzmann method}.
\newblock \emph{International Journal for Numerical Methods in Fluids}, 2011.
\newblock 65(5):496--519.

\bibitem{norgaard2016topology}
S.~N{\o}rgaard, O.~Sigmund, and B.~Lazarov.
\newblock {Topology optimization of unsteady flow problems using the lattice
  Boltzmann method}.
\newblock \emph{Journal of Computational Physics}, 2016.
\newblock 307:291--307.

\bibitem{bhatnagar1954model}
P.~L. Bhatnagar, E.~P. Gross, and M.~Krook.
\newblock A model for collision processes in gases. {I. Small} amplitude
  processes in charged and neutral one-component systems.
\newblock \emph{Physical Review}, 1954.
\newblock 94(3):511.

\bibitem{sato2019topology}
A.~Sato, T.~Yamada, K.~Izui, S.~Nishiwaki, and S.~Takata.
\newblock {A topology optimization method in rarefied gas flow problems using
  the Boltzmann equation}.
\newblock \emph{Journal of Computational Physics}, 2019.
\newblock 395:60--84.

\bibitem{Bird1994Molecular}
G.~A. Bird.
\newblock \emph{Molecular gas dynamics and the direct simulation of gas flows}.
\newblock Clarendon Press, 1994.

\bibitem{caflisch2021adjoint}
R.~Caflisch, D.~Silantyev, and Y.~Yang.
\newblock {Adjoint DSMC for nonlinear Boltzmann equation constrained
  optimization}.
\newblock \emph{Journal of Computational Physics}, 2021.
\newblock 439:110404.

\bibitem{yang2023adjoint}
Y.~Yang, D.~Silantyev, and R.~Caflisch.
\newblock {Adjoint DSMC for nonlinear spatially-homogeneous Boltzmann equation
  with a general collision model}.
\newblock \emph{Journal of Computational Physics}, 2023.
\newblock page 112247.

\bibitem{guan2023topology}
K.~Guan, K.~Matsushima, Y.~Noguchi, and T.~Yamada.
\newblock {Topology optimization for rarefied gas flow problems using density
  method and adjoint IP-DSMC}.
\newblock \emph{Journal of Computational Physics}, 2023.
\newblock 474:111788.

\bibitem{Xu2010A}
K.~Xu and J.~C. Huang.
\newblock A unified gas-kinetic scheme for continuum and rarefied flows.
\newblock \emph{Journal of Computational Physics}, 2010.
\newblock 229(20):7747--7764.

\bibitem{guo2013discrete}
Z.~Guo, K.~Xu, and R.~Wang.
\newblock Discrete unified gas kinetic scheme for all Knudsen number flows:
  Low-speed isothermal case.
\newblock \emph{Physical Review E}, 2013.
\newblock 88(3):033305.

\bibitem{guo2015discrete}
Z.~Guo, R.~Wang, and K.~Xu.
\newblock Discrete unified gas kinetic scheme for all Knudsen number flows.
  {II}. {Thermal} compressible case.
\newblock \emph{Physical Review E}, 2015.
\newblock 91(3):033313.

\bibitem{liu2020unified}
C.~Liu, Y.~Zhu, and K.~Xu.
\newblock {Unified gas-kinetic wave-particle methods I: Continuum and rarefied
  gas flow}.
\newblock \emph{Journal of Computational Physics}, 2020.
\newblock 401:108977.

\bibitem{zhu2019unified}
Y.~Zhu, C.~Liu, C.~Zhong, and K.~Xu.
\newblock {Unified gas-kinetic wave-particle methods. II. Multiscale simulation
  on unstructured mesh}.
\newblock \emph{Physics of Fluids}, 2019.
\newblock 31(6).

\bibitem{fei2020unified}
F.~Fei, J.~Zhang, J.~Li, and Z.~Liu.
\newblock {A unified stochastic particle Bhatnagar-Gross-Krook method for
  multiscale gas flows}.
\newblock \emph{Journal of Computational Physics}, 2020.
\newblock 400:108972.

\bibitem{Zhu2016Implicit}
Y.~Zhu, C.~Zhong, and K.~Xu.
\newblock Implicit unified gas-kinetic scheme for steady state solutions in all
  flow regimes.
\newblock \emph{Journal of Computational Physics}, 2016.
\newblock 315:16--38.

\bibitem{yuan2021multi}
R.~Yuan, S.~Liu, and C.~Zhong.
\newblock A multi-prediction implicit scheme for steady state solutions of gas
  flow in all flow regimes.
\newblock \emph{Communications in Nonlinear Science and Numerical Simulation},
  2021.
\newblock 92:105470.

\bibitem{su2020can}
W.~Su, L.~Zhu, P.~Wang, Y.~Zhang, and L.~Wu.
\newblock {Can we find steady-state solutions to multiscale rarefied gas flows
  within dozens of iterations?}
\newblock \emph{Journal of Computational Physics}, 2020.
\newblock 407:109245.

\bibitem{chapman1990mathematical}
S.~Chapman and T.~G. Cowling.
\newblock \emph{The mathematical theory of non-uniform gases: an account of the
  kinetic theory of viscosity, thermal conduction and diffusion in gases}.
\newblock Cambridge university press, 1990.

\bibitem{spaid1997lattice}
M.~A. Spaid and F.~R. Phelan~Jr.
\newblock {Lattice Boltzmann methods for modeling microscale flow in fibrous
  porous media}.
\newblock \emph{Physics of fluids}, 1997.
\newblock 9(9):2468--2474.

\bibitem{bendsoe1999material}
M.~P. Bends{\o}e and O.~Sigmund.
\newblock Material interpolation schemes in topology optimization.
\newblock \emph{Archive of applied mechanics}, 1999.
\newblock 69:635--654.

\bibitem{stolpe2001alternative}
M.~Stolpe and K.~Svanberg.
\newblock An alternative interpolation scheme for minimum compliance topology
  optimization.
\newblock \emph{Structural and Multidisciplinary Optimization}, 2001.
\newblock 22(2):116--124.

\bibitem{borrvall2001topology}
T.~Borrvall.
\newblock Topology optimization of elastic continua using restriction.
\newblock \emph{Archives of Computational Methods in Engineering}, 2001.
\newblock 8:351--385.

\bibitem{kawamoto2011heaviside}
A.~Kawamoto, T.~Matsumori, S.~Yamasaki, T.~Nomura, T.~Kondoh, and S.~Nishiwaki.
\newblock {Heaviside projection based topology optimization by a PDE-filtered
  scalar function}.
\newblock \emph{Structural and Multidisciplinary Optimization}, 2011.
\newblock 44:19--24.

\bibitem{guest2004achieving}
J.~K. Guest, J.~H. Pr{\'e}vost, and T.~Belytschko.
\newblock Achieving minimum length scale in topology optimization using nodal
  design variables and projection functions.
\newblock \emph{International journal for numerical methods in engineering},
  2004.
\newblock 61(2):238--254.

\bibitem{sigmund2007morphology}
O.~Sigmund.
\newblock Morphology-based black and white filters for topology optimization.
\newblock \emph{Structural and Multidisciplinary Optimization}, 2007.
\newblock 33:401--424.

\bibitem{xu2010volume}
S.~Xu, Y.~Cai, and G.~Cheng.
\newblock Volume preserving nonlinear density filter based on heaviside
  functions.
\newblock \emph{Structural and Multidisciplinary Optimization}, 2010.
\newblock 41:495--505.

\bibitem{chu1965kinetic}
C.~Chu.
\newblock Kinetic-theoretic description of the formation of a shock wave.
\newblock \emph{The Physics of Fluids}, 1965.
\newblock 8(1):12--22.

\bibitem{Yang1995Rarefied}
J.~Y. Yang and J.~C. Huang.
\newblock Rarefied flow computations using nonlinear model {Boltzmann}
  equations.
\newblock \emph{Journal of Computational Physics}, 1995.
\newblock 120(2):323--339.

\bibitem{svanberg1987method}
K.~Svanberg.
\newblock The method of moving asymptotes---a new method for structural
  optimization.
\newblock \emph{International journal for numerical methods in engineering},
  1987.
\newblock 24(2):359--373.

\bibitem{svanberg2002class}
K.~Svanberg.
\newblock A class of globally convergent optimization methods based on
  conservative convex separable approximations.
\newblock \emph{SIAM journal on optimization}, 2002.
\newblock 12(2):555--573.

\bibitem{johnson2007nLopt}
S.~G. Johnson.
\newblock The {NLopt} nonlinear-optimization package.
\newblock \url{https://github.com/stevengj/nlopt}, 2007.

\bibitem{zhu2017performance}
L.~Zhu, P.~Wang, and Z.~Guo.
\newblock {Performance evaluation of the general characteristics based
  off-lattice Boltzmann scheme and DUGKS for low speed continuum flows}.
\newblock \emph{Journal of Computational Physics}, 2017.
\newblock 333:227--246.

\bibitem{yuan2020conservative}
R.~Yuan and C.~Zhong.
\newblock A conservative implicit scheme for steady state solutions of diatomic
  gas flow in all flow regimes.
\newblock \emph{Computer Physics Communications}, 2020.
\newblock 247:106972.

\bibitem{yuan2021novel}
R.~Yuan, S.~Liu, and C.~Zhong.
\newblock A novel multiscale discrete velocity method for model kinetic
  equations.
\newblock \emph{Communications in Nonlinear Science and Numerical Simulation},
  2021.
\newblock 92:105473.

\bibitem{xu2001gas}
K.~Xu.
\newblock A gas-kinetic {BGK} scheme for the {Navier-Stokes} equations and its
  connection with artificial dissipation and {Godunov} method.
\newblock \emph{Journal of Computational Physics}, 2001.
\newblock 171(1):289--335.

\bibitem{Mieussens2000DISCRETE}
L.~Mieussens.
\newblock Discrete velocity model and implicit scheme for the {BGK} equation of
  rarefied gas dynamics.
\newblock \emph{Mathematical Models and Methods in Applied Sciences}, 2000.
\newblock 10(08):1121--1149.

\bibitem{luo1998fast}
H.~Luo, J.~D. Baum, and R.~L{\"o}hner.
\newblock A fast, matrix-free implicit method for compressible flows on
  unstructured grids.
\newblock \emph{Journal of Computational Physics}, 1998.
\newblock 146(2):664--690.

\bibitem{rogers1995comparison}
S.~E. Rogers.
\newblock {Comparison of implicit schemes for the incompressible Navier-Stokes
  equations}.
\newblock \emph{AIAA journal}, 1995.
\newblock 33(11):2066--2072.

\bibitem{yuan2002comparison}
L.~Yuan.
\newblock Comparison of implicit multigrid schemes for three-dimensional
  incompressible flows.
\newblock \emph{Journal of Computational Physics}, 2002.
\newblock 177(1):134--155.

\end{thebibliography}

\renewcommand{\multirowsetup}{\centering}

\end{document}